\documentclass[usenatbib,usegraphicx,useAMS, referee]{mn2e}
\usepackage{amssymb,url,times}

\newcommand{\fracb}[2]{\left(\frac{#1}{#2}\right)}

\def\be{\begin{equation}}
\def\ee{\end{equation}}

\title[From SGRs to ``weak field magnetars'']{Magnetic field decay in neutron stars: from Soft Gamma Repeaters to ``weak field magnetars''}

\author[Dall'Osso et al.]{S. Dall'Osso$^{1}$\footnote{sim.dall@gmail.com}, J. Granot$^{1,2,3}$ and T. Piran$^{1}$\\
$^1$  Racah Institute of Physics, The Hebrew University, Jerusalem 91904,
  Israel \\
$^2$ Raymond and Beverly Sackler School of Physics \& Astronomy, Tel Aviv
  University, Tel Aviv 69978, Israel \\
$^3$ Centre for Astrophysics Research, University of Hertfordshire, College Lane, Hatfield, AL10 9AB, UK}
\date{Submitted: Revised:  Accepted:}
\pagerange{\pageref{firstpage}--\pageref{lastpage}} \pubyear{2011}

\begin{document}
\label{firstpage}
\maketitle

\begin{abstract}

The  recent discovery of the ``weak field, old magnetar", the soft gamma
repeater SGR~0418+5729 
, whose  dipole magnetic field, $B_{\rm dip}$, is less than   $7.5 \times
10^{12}\;$G 
, has raised perplexing questions:  How can the  neutron star produce SGR-like
bursts with such a low magnetic field?  What powers the observed X-ray
emission when neither the rotational energy nor the magnetic dipole energy
are sufficient? These  observations, that suggest either a much larger energy
reservoir or a much younger true  age (or both), have renewed the interest in
the evolutionary sequence of magnetars.  We examine, here, a phenomenological
model for the magnetic field decay:   $\dot{B}_{\rm dip} \propto B_{\rm
  dip}^{1+\alpha}$ and  compare its predictions with the observed period, $P$,
the period derivative, $\dot{P}$, and  the X-ray luminosity, $L_X$, of
magnetar candidates. We find a strong evidence for a dipole field decay on a
timescale of $\sim 10^3\;$yr for the strongest ($B_{\rm dip}\sim 10^{15}\;$G)
field objects, with a  decay index within  the range $1 \leq \alpha < 2$ and more likely within $1.5\lesssim\alpha\lesssim 1.8$. The decaying field implies a younger age than what is implied by $P/2 \dot P$. Surprisingly, even with the younger age, the energy released in the dipole field  decay is insufficient to power the X-ray emission,  suggesting  the existence of  a stronger internal field, $B_{\rm int}$.  Examining several models for the  internal magnetic field decay we find that 
it  must have a very large ($\gtrsim 10^{16}\;$G) initial value. Our findings suggest two clear distinct evolutionary tracks -- the SGR/AXP branch and the transient branch, with a possible third branch involving high-field radio pulsars that age into low luminosity X-ray dim isolated neutron stars.
\end{abstract}


\section{Introduction}
\label{sec:intro}
Soft Gamma Repeaters (SGRs) and Anomalous X-ray Pulsars (AXPs) are two
classes of pulsating X-ray sources thought to be highly magnetized
Neutron Stars (NSs) whose high-energy emission is powered by the
dissipation of their magnetic field. Hence they are deemed
magnetars. They have large rotation periods ($5\;{\rm s}\lesssim
P\lesssim 12\;$s) with relatively large time derivatives ($\dot{P}\sim
10^{-12}- 10^{-10}\;{\rm s\;s^{-1}}$), implying (through magnetic
dipole breaking) large surface dipole field strengths ($B_{\rm
dip}\gtrsim 3\times 10^{14}\;$G or energy $E_{\rm B_{\rm dip}}\gtrsim
2\times 10^{46}\;$erg) and relatively young spindown ages ($\tau_{\rm
c}\sim 10^3-10^5\;$yr). Thus, the decay of their dipole fields on the
timescale of their spindown ages appears to be capable of accounting
for their large persistent X-ray luminosities, of $L_{\rm X} \sim
10^{35}$ erg s$^{-1}$ \citep{TD95}.  On the other hand,
rotational energy losses are 1-2 orders of magnitude too low to
account for their measured $L_{\rm X}$, and the absence of binary
companions strongly argues against accretion as a power source.

Both SGRs and AXPs can emit sporadic, sub-second ($\sim 0.1\;$s)
bursts of hard X-rays to soft $\gamma$-rays, which release $\sim
10^{38}- 10^{41}\;$erg at typically super-Eddington luminosities.
Evolving stresses from a decaying magnetic field $> 10^{14}\;$G can
in principle stress the rigid lattice of the NS crust beyond its
yielding point, leading to sudden release of stored magnetic energy,
which is believed to trigger these bursts \citep{TD96, Perpons+2011}.

AXPs tend to be less burst-active, and have somewhat
weaker dipole fields and larger spindown ages compared to SGRs. This
led to the suggestion that SGRs evolve into AXPs as they age and
their magnetic field decays \citep{Kouve+1998}.  The effect of the
decay of the dipole magnetic field on the spin evolution of magnetars
was previously considered \citep{Colpi+2000}, but its
ultimate implications for their expected X-ray emission
have not yet been fully explored.

In addition to classical SGRs and AXPs, the magnetar family includes
also transient sources. First discovered in 2003 (XTE J1810-197; \citealt{Ibra+2004}), the group of transients is now the largest among magnetar
candidates and is rapidly growing, thanks to the improved detection 
capabalities of the Fermi Gamma-ray Burst Monitor (GBM). The distinctive property of transients 
is that they are discovered only thanks to outbursts, when they 
emit typical magnetar-like bursts accompanied by a very large increase ($\sim$ two orders
of magnitude) of their persistent emission. Their timing parameters can only 
be measured in outburst and they match well those of persistent
SGRs/AXPs. The implied rotational energy losses are much lower than the
outburst-enhanced persistent emission. However, when the much weaker quiescent
emission of transients could eventually be measured, it was found to be
below the level of rotational energy losses.

Typical magnetar-like bursts and outburst was also detected from the 0.3 s,
allegedly  rotation-powered X-ray pulsar PSR J1846-0258, with a dipole field
of $4.8 \times 10^{13}$ G \citep{Gavriil+2008, KumarSafi+2008}, and
the radio pulsar PSR 1622-4950, with a spin period of 4.3 s and a dipole field 
of $3\times 10^{14}$ G, displayed a flaring radio emission \citep{Levin2010}, with properties very
similar to those of two transient radio magnetars \citep{Camilo+2006, Camilo+2007}. 
On the other hand, there are radio pulsars with dipole fields
comparable to the weaker field magnetars, and larger than PSR J1846-0258, but 
showing no sign of peculiar behaviour. 
Thus, a strong dipole field does not seem a sufficient condition for powering 
magnetar-like emission \citep{Kaspi+2010}.

The greatest surprise, however, came from the recent discovery
of SGR~0418+5729 by the Fermi Gamma-ray Burst Monitor . 
This source, previously undetected in X-rays, was
discovered on 5 June, 2009, through the emission of two distinct,
sub-second magnetar-like bursts of low luminosity ($\lesssim 10^{38}$
erg s$^{-1}$) and total energy $\sim 2 {\rm ~and}~4 \times 10^{37}$
erg, \citep{vderh+2010}.  Following the bursts it became
detectable as a bright, pulsating X-ray source with a period of
$\sim$9.1$\;$s and a luminosity of $L_{\rm X}
\approx 10^{34}$ erg s$^{-1}$ that decayed by 2 orders of magnitude
during the following 6 months \citep{Esposito+2010}. However, only an
upper limit could be put on its period derivative, $\dot{P}< 6\times
10^{-15}\;{\rm s\;s^{-1}}$, translating to an upper limit of
$B_{\rm dip} < 7.5\times 10^{12}\;$G on its dipole field strength and a  
spindown age  $\tau_{\rm c} > 24\;$Myr
\citep{Rea+2010}. This demonstrates that magnetar-like activity can be
present in NSs with rather standard dipole magnetic fields. This
finding represents a breakthrough in our understanding of magnetars,
as it was unexpected that bursts could be produced in such a low-field
object, if they are associated with sudden crustal fractures.

Moreover, it is clear that the dipole field of SGR~0418+5729
cannot power its X-ray emission if it decays on the timescale
of its spindown age (of $\tau_{\rm c} > 24\;$Myr). The latter would
require a much stronger field, of $\gtrsim 5\times 10^{14}\;$G, in
order to power the weakest level of X-ray emission at which this
source was found, 1.5 yr after the outburst \citep{Rea+2010}. That is, 
either its total magnetic energy is $>$5000 times larger than that the 
dipole \citep{Turolla+2011}, or its dipole field decay time is $>$5000 times 
smaller than its spindown age (or some combination of the two).

Bearing on the above arguments,  we
address, in this work  the power
source of magnetar candidates and their spin evolution as their dipole
fields decay. The latter are jointly compared with the observed properties
of the full sample of magnetar candidates, thus enabling us to draw robust
conclusions. In \S~\ref{sec:source-classes} we describe the main properties of the
different classes of objects of interest and our observed sample. In
\S~\ref{sec:dipolebraking} we provide a simple analytic formalism for
the spin evolution of NSs whose dipole field decays as $\dot{B}_{\rm
dip} = -B_{\rm dip}/\tau_{\rm d} \propto B_{\rm dip}^{1+\alpha}$, and
discuss its general properties.  A summary of the mechanisms for
magnetic field decay in NSs is provided in \S~\ref{summary-fieldecay}.
In \S~\ref{sec:two} we show that there is indeed a strong evidence for
effective decay of the dipole fields of these sources, which
is $\sim 10^3\;$yr timescale for the strongest fields of SGRs
and AXPs, and provide quantitative constraints on the most likely
decay mechanisms (we show that $1\lesssim\alpha<2$ is required).
These constraints are made tighter ($1.5\lesssim\alpha\lesssim 1.8$)
in \S~\ref{sec:explaining}, where we  show that decay of the
dipole component alone is insufficient to power the X-ray emission
and that
a stronger power source is required, presumably a stronger internal
magnetic field. We examine  two models for internal magnetic field decay  in \S~\ref{sec:internal-field}.
Comparing  with observations we derive basic
constraints on the initial values and decay properties of such a
component. Our conclusions are summarized in
\S~\ref{sec:conclusions}. Finally, in \S~\ref{sec:evol} we build on
our conclusions to outline a self-consistent evolutionary
scenario for magnetar candidates and speculate about their relation to
other classes of objects.

\section{Source classes and the observational sample}
\label{sec:source-classes}
 In order to verify the role of field decay in NSs with  strong magnetic fields
 we  consider here the X-ray and timing properties of magnetar candidates.
 These are usually identified by their measured values of $P$ and $\dot{P}$
(hence the \textit{inferred} dipole field, $B_{\rm dip}$, and spin-down age,
$\tau_c$), detection of peculiar burst/outburst activity and persistent X-ray 
luminosities that exceed rotational energy losses.
Sources are divided into two main groups, largely for historical reasons.  
SGRs are typically identified by the emission of trains of sporadic, 
sub-second bursts of X-to-$\gamma$-ray radiation with super-Eddington 
luminosities ($\sim 10^{39}- 10^{42}$ erg s$^{-1}$).
More rarely they show much more powerful events, the Giant Flares (GFs). These are 
typically initiated by a spike of $\gamma$-ray emission lasting half a second 
and releasing $\sim 10^{44} - 10^{46}$ erg at luminosities $\sim 10^{44} - 10^{47}$ erg
s$^{-1}$ (assuming isotropic emission), which are followed by minute-long, yet
super-Eddington, tails of radiation ($\sim 10^{41} - 10^{42}$ erg
s$^{-1}$), markedly pulsed at the NS spin \citep{Mazets+1979, Cline+1998,
Hurley+1999, Terasawa+2005, Palm+2005}.

The highly super-Eddington luminosities of bursts and GFs definitely 
rules out accretion as a viable power source. Indeed, it first hinted
  to the key role of superstrong magnetic fields\footnote{It is customary to
    adopt, just as a reference value for the strength of magnetar fields, the
  critical field $B_{\rm QED} \approx 4.4 \times 10^{13}\;$G, at which
the energy of the first excited Landau level of electrons equals their
rest-mass energy} which reduce the electron scattering cross-section much below the
Thomson value. In particular, $ B \gtrsim 3\times 10^{14}$ G
were required to explain luminosities up to $10^{42}$ erg s$^{-1}$ \citep{Pac+1992}. 
The several minute-long pulsating tails are very similar in all three GFs detected so
  far. They suggest that a remarkably standard amount of energy, 
$\sim {\rm a~few} \times 10^{44}$ erg, remains trapped in a small-size, 
closed field line region in the magnetosphere ($\sim$ a few stellar radii),
likely in the form of a pair-photon plasma, and slowly radiated away
(the ``trapped fireball'' model of \citealt{TD95}). A
magnetospheric field $\gtrsim {\rm a~few} \times 10^{14}$ G would
naturally account for this.

Anomalous X-ray Pulsars were typically identified through the 
peculiar properties of their persistent emission \citep{Merestella+1995}. 
However, the distinction between these two classes has
come into question since SGR-like bursting activity has been detected
in many APXs \citep{Gavriil+2002, WooThom+2006, Mereghetti+2008}.
To date, AXP have never displayed a GF and are, in general, characterized
by less frequent, and somewhat less energetic, burst activity.

There are persistent and transient sources both in the SGR and the
AXP classes. The emission from the persistent sources is known to be 
pulsed at the NS spin period and to be characterized by a blackbody (BB) 
component, with $k T \sim 0.4\,$--$\,0.7\;$keV, plus a power-law (PL) tail 
extending up to $\sim 10\;$keV (\cite{Mereghetti+2008} and references therein). In
recent years, a new power-law component extending up to $\sim$ 150 keV
was discovered with the INTEGRAL satellite (\cite{Kuiper+2006}; for
recent reviews \citet{WooThom+2006}, \citealt{Mereghetti+2008}). This
component has a very flat $\nu F_\nu$ spectrum and is thus clearly
distinct from the lower energy PL.  A seizable fraction of the
bolometric emission from SGRs/AXPs is emitted in the 10 - 150$\;$keV
energy range, and it is likely comparable to that at lower
energies, below $10\;$keV \citep{Mereghetti+2008}. 
The persistent emission of these sources gets temporarily enhanced at bursts
(SGRs/AXPs) and Flares (SGRs only), but such variations are usually moderate --
normally at most by a factor of a few.

Transient sources, on the other hand, are characterized by a much
weaker persistent X-ray emission, which in some cases still remains
undetected. When detected, it is generally found to be at a lower level than rotational
energy losses ($L_{\rm X}<|\dot{E}_{\rm
rot}|$). Transient sources are distinctively characterized by X-ray
outbursts during which, in addition
to the emission of sub-second long bursts (similar to the bursts of
persistent sources), their persistent X-ray emission is subject to
very large enhancements, by $\sim 2\,$--$\,3$ orders of magnitude, making
them temporarily as bright as the persistent sources. In fact, most of these
objects have been discovered thanks to such events.
In outburst, their luminosities
largely exceed rotational energy losses ($L_{\rm X,outburst}\gg
|\dot{E}_{\rm rot}|$),
and gradually return to
the much lower quiescent level over timescales of $\sim$ a few years
(cfr. \cite{Rea+2011} 
for a recent review).Data for persistent SGRs/AXPs and transient sources were collected
from the McGill catalog\footnote{URL: http://www.physics.mcgill.ca/~pulsar/magnetar/main.html}.

In addition to these ``classical'' magnetar candidates, we consider the
properties of X-ray Dim Isolated NSs (XDINs; \cite{Kaplan+2008, Mereghetti+2011}, \citealt{Kaplan+2011}). These are a class of currently seven, $\sim 10^6$ yr old X-ray sources, 
which are considered as neat examples of isolated, cooling NSs. They have
stable and nearly purely thermal persistent emission (but see,
e.g. \citealt{Turolla+2009}), 
typically described as a single BB with $kT \sim
0.05\,$--$\,0.1\;$keV. 
Their spin periods are in the same range as SGRs/AXPs and their inferred dipole fields
are all $>10^{13}$ G.  As such, a possible link with magnetar candidates has long been suspected.
Data collected for all sources are summarized in Table~\ref{table}.

\begin{table}
\begin{center}
{\scriptsize
\begin{tabular}{|c|c|c|c|c|c|c|c|c|c|c|}
\hline 
          & $P$ & $\dot{P}$                    & $L_X$ & $kT$                          & $\tau_c$ &
$|\dot{E}_{\rm rot}|$          & $B_{\rm dip}$ & $a\equiv$ & $k\equiv$ & $f_A\equiv$ \\ 
\rm{Name} & (s) & ($10^{-11}\,{\rm s/s}$) & ($10^{35}\,{\rm erg/s}$) & (keV) & (kyr) &
($10^{33}\,{\rm erg/s}$) & ($10^{14}\,$G)           & $L_X\tau_c/E_{\rm dip}$ &
$L_X/|\dot{E}_{\rm rot}|$ & $\frac{L_X}{4\pi R^2\sigma T^4}$ \\
\hline
\hline
SGR$\,$1086$-$20   & 7.6022(7)     & 75(4)     & 1.6            &
$0.6^{+0.2}_{-0.1}$ & 0.16     & 67         & 24       & 0.00083 & 2.4  &
0.096  \\ \hline
SGR$\,$0526$-$66   & 8.0544(2)     & 3.8(1)    & 1.4            &               & 3.4      & 2.9        & 5.6      & 0.28    & 49   & 0.32  \\ \hline
SGR$\,$1900$+$14   & 5.19987(7)    & 9.2(4)    & 0.83--1.3      & 0.47(2)             & 0.90     & 26         & 7.0      & 0.037   & 4.1  & $\sim$0.17  \\ \hline
SGR$\,$1627$-$41   & 2.594578(9)   & 1.9(4)    & $\sim$0.025    &              & 2.2      & 43         & 2.2      & 0.020   & 0.058 &  \\ \hline
SGR$\,$0418$+$5729 & 9.07838827(4) & $<$0.0006 & 0.00062        & 0.67(11)            & $>$24000 & $<$0.00032 & $<$0.075 & $>$5050 & $>$196 & 0.000024 \\ \hline
SGR$\,$1833$-$0832 & 7.565408(4)   & 0.439(43)   & 
     &     & 27       & 0.40       & 1.8     &  &  & \\ 
\hline  
\hline
1E$\,$1547$-$5408  & 2.06983302(4)    & 2.318(5)    & $\sim$0.0058 & 0.43(4)    & 1.4  & 100   & 2.2  & 0.0032 & 0.0056 & 0.0013 \\ \hline
XTE$\,$J1810$-$197 & 5.5403537(2)     & 0.777(3)    & $\sim$0.0019 & 0.14--0.30 & 11   & 1.8   & 2.1  & 0.0092 & 0.11   & 0.0063 \\ \hline
1E$\,$1048$-$5937  & 6.45207658(54)   & $\sim$2.70  & 0.054        & 0.623(6)   & 3.8  & 3.9   & 4.2  & 0.022  & 1.4    & 0.0028 \\ \hline
1E$\,$2259$+$586   & 6.9789484460(39) & 0.048430(8) & 0.18         & 0.411(4)   & 230  & 0.056 & 0.59 & 225    & 320    & 0.049 \\ \hline
CXOU$\,$J010043.1$-$72134 & 8.020392(6) & 1.88(8)   & $\sim$0.78   & 0.38(2)    & 6.8  & 1.4   & 3.9  & 0.65   & 54     & 0.29 \\ \hline
4U$\,$0142$+$61    & 8.68832973(8)    & 0.1960(2)   & $>$0.53      & 0.395(5)   & 70   & 0.12  & 1.3  & $>$40  & $>$449 & $>$0.17 \\ \hline
CXO$\,$J164710.2$-$455216 & 10.6107(1) & 0.083(2)    & $\sim$0.0044   & 0.49(1)    & 200   & 0.027 & 0.95  & 1.7     & 16.3    & 0.0006 \\ \hline
1RXS$\,$J170849.0$-$400910 & 10.9990355(6) & 1.945(2) & $\sim$1.9  & 0.456(9)   & 9.0  & 0.57  & 4.7  & 1.47   & 329    & 0.34 \\ \hline
1E$\,$1841$-$045   & 11.7750542(1)    & 4.1551(14)  & $\sim$2.2    & 0.44(2)    & 4.5  & 0.99  & 7.1  & 0.37   & 219    & 0.45 \\ \hline
PSR$\,$J1622$-$4950 & 4.3261(1)       & 1.7(1)      & $\sim$0.0063 & $\sim$0.4  & 4.0  & 8.5   & 2.8  & 0.0064 & 0.076  & 0.0019 \\ \hline
CXOU$\,$J171405.7$-$381031 & 3.82535(5) & 6.40(14)  & $\sim$2.2    & 0.38(8)    & 0.95 & 45    & 5.0  & 0.16   & 4.9    & 0.82 \\ 
\hline
\hline
RXJ$\,$1856 & 7.05  & 0.003      & 0.00017               & 0.063    & 3700      & 0.0034  & 0.15    & 55        & 5.0       & 0.084 \\ \hline
RXJ$\,$0720 & 8.39  & 0.00698(2) & 0.00337               & 0.090(5) & 1900      & 0.0047  & 0.25    & 202       & 72        & 0.40 \\ \hline
RXJ$\,$1605 & 6.88  &     & 0.00011$d_{0.1}^2$  & 0.096    &    &  &  &    &    & 0.010 \\ \hline
RXJ$\,$0806 & 11.37 & 0.006      & 0.00033               & 0.096    & 3000      & 0.0016  & 0.26    & 27        & 20        & 0.030 \\ \hline
RXJ$\,$1308 & 10.31 & 0.011      & 0.000051$d_{0.1}^2$ & 0.102    & 1500      & 0.0040  & 0.34    & 1.2$d_{0.1}^2$ & 1.3$d_{0.1}^2$ & 0.0036$d_{0.1}^2$ \\ \hline
RXJ$\,$2143 & 9.44  & 0.004      & $\geq$0.00069         & 0.100    & 3700      & 0.0019  & 0.20    & $\geq$126 & $\geq$37  & $\geq$0.053 \\ \hline
RXJ$\,$0420 & 3.45  & 0.00028    & 0.00032               & 0.044    & 2000    & 0.027 & 0.1  & 115.2 & 1.18 & 0.66 \\ \hline
\end{tabular}}
\end{center}
\caption{Summary of the salient timing and X-ray spectral properties of SGRs, AXPs and XDINs. Data are collected from references cited in the text.}
\label{table}
\end{table}

Although timing data are homogeneuos and 
can be easily compared for all classes above,
significantly different values of $\dot{P}$ have been measured at different
epochs in SGR 1806-20
and 1900+14, leading to different values of the inferred dipole fields. We
show in all plots the maximum and minimum value of the inferred $B_{\rm dip}$, 
for either object, joined by a dotted line.
In the following discussion we consider, however, the lowest value as a more
likely indication of the dipole 
field (cfr. \cite{TLK+2002}; \cite{WooThom+2006} for
discussion of magnetic torque changes).

Data on the X-ray emission from different classes, on the other hand, must be 
compared with caution. Important spectral differences exist and, in some
cases, marked temporal variability makes the sample much less homogeneous.
We adopt the X-ray luminosity in the $(2\,$--$\,10$) keV range,
 as reported in the McGill catalog, as a measure of the
thermal emission from persistent sources, which we call $L_{\rm X}$. 
Given their typical BB temperatures, bolometric corrections are expected to
be $\sim\,2\,$--$\,4$ and will be ignored. Note that the
contribution from the PL component in this energy range is not negligible
either, which at least partially balances for the neglect of bolometric 
corrections.  
Also note that GFs might provide a non-negligible contribution to $L_{\rm
  tot}$ in SGRs\footnote{This contribution might even be dominant in
  SGR 1806-20.}, and that the hard tails up to $\gtrsim 150$ keV also
contribute significantly to the latter.

The quiescent emission of transients is less clearly understood. In most of
them it appears dominated by a BB component with temperature ($\sim 0.4$ keV) 
and an emitting area much smaller than the NS surface. The resulting
luminosities are generally $\lesssim 10^{33}$ erg s$^{-1}$, and the 
bolometric corrections to their ($2-10$) keV emission are thus similar to
those of persistent sources. 

In XTE J1810, on the other hand, the BB component which dominates the quiescent 
emission has a much lower temperature, $kT \simeq 0.18$ keV and an emitting
area  consistent with the NS 
surface \citep{Bernardo+2009}. This implies a total 
luminosity $\lesssim 10^{33}$ erg s$^{-1}$, but a relatively large bolometric 
correction to the $(2-10)$ keV flux. 

As we see later the "weak field magnetar" SGR 0418+5729, with its extreme
parameters, plays a very important role in constraining various models. 
Particularly important is its \textit{total} X-ray luminosity, $L_{\rm tot}$. 
The current observed value of $\simeq 6\times 10^{31}$ erg s$^{-1}$ in the
$(0.5 -10)$ keV range \citep{Rea+2010} might well be a post flare remnant emission larger than 
its real quiescent level, which is unknown.  It is interesting 
to use the energy budget of its recent outburst to estimate a minimal 
average power needed.
The decaying X-ray flux following the outburst was monitored from June
2009 to September 2010 \citep{Esposito+2010}. Results of this monitoring 
allow to estimate a total fluence in the $0.5\,$--$\,10\;$keV energy range 
corresponding to\footnote{This corresponds to an average luminosity of $\sim
10^{33}\;{\rm erg\; s^{-1}}$ during the outburst.} $\Delta E
\simeq 3 \times 10^{40}\;$erg, for a distance of 2$\;$kpc to this source. 
Even if the outburst recurrence time was $T_{\rm rec}\sim 100\;$yr,
this would still correspond to an average luminosity of $\langle L_{\rm outb}
\rangle \sim
10^{31}\;{\rm erg\;s^{-1}}$ due only to such outbursts, not much below 
the current upper limit.

Does such a recurrence time match what we know about outbursts from transient
sources? If $T_{\rm age}$ is the age of SGR~0418+5729 and $T_{\rm birth}$ is 
the average time between the birth of magnetars in our Galaxy, one expects a 
rate of outbursts of $R_{\rm outb} = 10\,T_{\rm age, Myr} (T_{\rm birth,kyr}~T_{\rm
rec,100})^{-1}\;{\rm yr^{-1}}$ in the Galaxy. The age of SGR~0418+5729 could be
as small as $\approx 0.1 \;$Myr (based on the upper limit on its quiescent X-ray
luminosity, as we discuss in next sections) but, on the
other hand, their actual birth rate is likely to be a few times larger than
our adopted lower limit of 1 per kyr \citep{Gaensler+1999}.
The ratio $T_{\rm age}/T_{\rm birth}$ is thus not expected to vary much,
  in any case. Comparing with observations, we note that outbursts from 4
different sources were detected, in one year of operation, by the
Fermi satellite. Two of these sources were previously unknown and
could not have been detected if they had been further than a few kpc,
and are thus detectable only from a small fraction of our Galaxy
\citep{vderh+2010}. This matches reasonably well our estimate of
 $R_{\rm outb}$, thus supporting our estimate of $\langle L_{\rm outb} \rangle$.

 We finally note that uncertain distance determinations can affect the inferred
luminosities. For most objects, different estimates
 agree to within factors $< 1.5$.  Although non-negligible, this typically
 translates to uncertainties of a factor
$\lesssim 2$ on $L_X$. The AXP 1E~1048-5937 represents, however, a
notable exception. \cite{Durant+2006} find a distance of $9.0 \pm 1.7\;$kpc 
for this source, based on a detailed study of reddening of red giant stars
in the field of this AXP. This is significantly larger
than the $2.7 \pm 0.1\;$kpc 
reported in the McGill catalogue \citep{Gaensler+2005}. The persistent X-ray 
luminosity of 1E~1048-5937 would accordingly increase by a factor $\sim 11$, 
reaching the value $L_X \sim 6\times 10^{34}\;{\rm erg\;s^{-1}}$ instead of 
$\sim 5 \times 10^{33}\;{\rm erg\;s^{-1}}$ as reported in the McGill catalog. 
We consider both values for this source.

\section{Magnetic dipole braking and field decay}
\label{sec:dipolebraking}

The energy loss rate of a magnetic dipole rotating in vacuum is:
\begin{equation}
\label{eq:L_vac}
L_{\rm vac} = \frac{2}{3}\frac{\mu^2\Omega^4}{c^3}\sin^2\theta_{\rm B}\ ,
\end{equation}
where $\theta_{\rm B}$ is the angle between the dipole and rotation axes.
\begin{equation}
\mu = B_{\rm eq}R^3 = \frac{1}{2}B_{\rm pol}R^3\ ,
\end{equation}
is the magnetic moment, where $R$ is the stellar radius while $B_{\rm
eq}$ and $B_{\rm pol}$ are the surface magnetic field strengths at the
dipole equator and pole, respectively.
More realistically, however, a magnetized rotating neutron star is
surrounded by plasma rather than vacuum. Three dimensional (3D)
force-free numerical simulation \citep{Spit+06}
have shown that this slightly increases the energy loss rate to:
\begin{equation}
\label{spitk}
L_{\rm pls} = \frac{\mu^2\Omega^4}{c^3}(1+\sin^2\theta_{\rm B})\ .
\end{equation}

The evolution of the spin period $P$ is governed by $L_{\rm
pls} = -\dot{E}_{\rm rot} = -I\Omega\dot{\Omega} = I(2\pi)^2\dot{P}/P^3$, or:
\begin{equation}\label{eq:P_sq_dot}
\frac{d}{dt}\left(P^2\right) = 2P\dot{P} = 
\frac{8\pi^2\mu^2}{Ic^3}(1+\sin^2\theta_{\rm B}) = 
\frac{8\pi^2B_{\rm eq}^2R^6}{Ic^3}(1+\sin^2\theta_{\rm B}) 
\equiv f\,\frac{8\pi^2 B_{\rm eq}^2R^6}{Ic^3}\ .
\end{equation}
Values of $B_{\rm eq}$ for isolated, spinning down NSs are usually
estimated through this formula, using the measured values of the spin
period, $P$, and its first derivative, $\dot{P}$. Generally the formula
for an orthogonal rotator in vacuum is used
(Eq.~[\ref{eq:L_vac}], with $\theta_{\rm B} = 90^{\circ}$). That
expression is formally recovered from the more realistic one given in
Eq.~(\ref{spitk}) by taking\footnote{This is just a formal choice,
made for comparison with published data, since the angular factor in
Eq. \ref{spitk} can never be smaller than unity.} $f=2/3$. In what
follows we will maintain the explicit dependence on $f$ for all
quantities but always specialize to the case $f=2/3$ when making
numerical estimates.
Defining a characteristic spindown age, $\tau_{\rm c} = P/(2
\dot{P})$, Eq.~(\ref{eq:P_sq_dot}) can be rearranged to read:
\begin{equation}
 \label{inferred-dipole} B_{\rm eq} = \left(\frac{I c^3}{f 8 \pi^2
 R^6}\right)^{1/2} \frac{P} {\tau^{1/2}_{\rm c}}\ .
\end{equation}
%

We shall now allow the dipole magnetic field to evolve in time in the
above equations and use its value at the equator as our reference
field strength from here on, $B_{\rm eq} (t) \equiv B_{\rm dip}(t)$.
For simplicity, however, the angle $\theta_{\rm B}$ and thus also the
factor $f = 1+\sin^2\theta_{\rm B}$ will be taken to be constant in
time. The exact solution for the spin period becomes:
%
%
\begin{equation}
\label{eq:P-evol}
P^2(t) = P^2(t_0) +f\,\frac{8\pi^2R^6}{Ic^3}
\int_{t_0}^tB_{\rm dip}^2(t')dt'\ .
\end{equation}
The spindown depends in a critical way on the time-dependence of magnetic
dipole energy, $E_{\rm dip} \propto B_{\rm dip}^2$, since this
determines the behaviour of the integral on the right-hand side.

Following the notations and parameterization introduced by \cite{Colpi+2000}, 
we write:
%
%
\begin{equation}
\label{eq:B-deriv}
\frac{dB_{\rm dip}}{dt} = - A B_{\rm dip}^{1+\alpha} = - \frac{B_{\rm dip}}
{\tau_{\rm d}(B_{\rm dip})}\ , 
\end{equation}
where we define the field decay timescale $\tau_{\rm d} (B_{\rm dip})
\equiv (A B_{\rm dip}^{\alpha})^{-1}$.  The solution of Eq.~(\ref{eq:B-deriv})
is:
\begin{equation}
\label{eq:B-vs-time}
B_{\rm dip}(t)  = B_{\rm dip,i} \left\{\matrix{
(1+\alpha\,t/\tau_{\rm d,i})^{-1/\alpha} \quad &(\alpha\neq 0)&\ , \cr
\exp(-t/\tau_{\rm d,i})\quad &(\alpha=0)&\ ,}\right.
\end{equation}
where $B_{\rm dip,i} = B_{\rm dip}(t=0)$ is the initial dipole field
strength, and $\tau_{\rm d,i} = \tau_{\rm d}(B_{\rm dip,i}) =
1/AB_{\rm dip,i}^\alpha$ is the initial field decay time. Note that
$\alpha = 0$, which corresponds to an exponential field decay on the
timescale $\tau_{\rm d,exp} = 1/A$, is the only value of $\alpha$ for
which $\tau_{\rm d}$ remains constant. We do not consider $\alpha<0$
(for which the field vanishes within a finite time, $t = -\tau_{\rm
d,i}/\alpha$). Substituting Eq.~(\ref{eq:B-vs-time}) into
Eq.~(\ref{eq:P-evol}) gives:
\begin{equation}
\label{eq:p-t_general}
P^2(t) = P_i^2+ f \frac{8\pi^2 R^6_*}{I c^3} B^2_{\rm dip,i}
\tau_{\rm d,i}\left\{\matrix{
\frac{1}{2-\alpha}\left[1-(1+\alpha t/\tau_{\rm d,i})^{(\alpha-2)/\alpha}\right] 
\quad &(\alpha\neq 0,\,2)\ ,\cr
\frac{1}{2}\left[1-\exp(-2t/\tau_{\rm d,i})\right] \quad &(\alpha=0)\ ,\cr
\frac{1}{2}\ln(1+2t/\tau_{\rm d,i})\quad &(\alpha = 2)\ ,}\right.
\end{equation}
where $P_i = P(t=0)$ is the initial spin period, at the birth of the
NS.  For $0\leq\alpha < 2$ the spindown essentially freezes out at
late times (or at very late times for $\alpha$ close to 2) and the
rotation period $P$ appraoches a constant value,
\begin{equation}\label{eq:P_infty}
P_\infty^2 = P_i^2 + f \frac{8\pi^2 R^6}{I c^3} 
B^2_{\rm dip,i}\frac{\tau_{\rm d,i}}{(2-\alpha)}\ .
\end{equation}
The field decay thus proceeds at a nearly constant spin at such late
times.  Usually $P_\infty\gg P_i$, and in this case:
\begin{equation}
\label{eq:def-pinfty}
P_{\infty} \simeq \sqrt{\frac{f\,8\pi^2R^6}{I c^3(2-\alpha)}}
\,B_{\rm dip,i} \tau^{1/2}_{\rm d,i}\approx \frac{7.65\;{\rm s}}{\sqrt{2-\alpha}} 
f^{1/2}R^3_{6} I^{-1/2}_{45} B_{\rm i,15} \left(\frac{\tau_{\rm d,i}}
{10^3~{\rm yrs}}\right)^{1/2} \propto B_{\rm dip,i}^{\frac{2-\alpha}2}\ ,
\end{equation}
where hereafter we use the notation\footnote{To avoid too many
subscript, the initial value of the dipole field is indicated simply
as $B_{\rm i,n}$, when it is normalized to the n-th power of 10.}
$Q_n = 10^n \times Q$ in c.g.s units. 

It is convenient to rewrite Eq.~(\ref{eq:B-deriv}) in terms of the
spindown time, $\tau_c = P/(2\dot{P})$, using the relation $dB_{\rm
dip} / d\tau_c = \dot{B}_{\rm dip} {\rm d} \tau_c / {\rm
dt}$: 
%
\begin{equation}
\label{decay-vs-td}
\frac{d B_{\rm dip}}{d \tau_{c}} = - \frac{B_{\rm dip}}{\tau_d} \frac{1}{1+ 2 (\tau_c/\tau_d)}\ .
\end{equation}
For the initial conditions $\tau_{c} = \tau_{\rm c,i}$, $B_{\rm dip} =
B_{\rm dip,i}$ and $\tau_{\rm d} = \tau_{\rm d,i}$ at $t=0$, the
solution to this equation is:
\begin{equation}\label{eq:tau_c(B)}
\tau_{c}(B_{\rm dip}) = \left\{\matrix{
\frac{\tau_{\rm d,i}}{2-\alpha} \left\{ \left[1+(2-\alpha) \frac{\tau_{\rm
      c,i}}{\tau_{\rm d,i}} \right] \left(\frac{B_{\rm dip,i}}{B_{\rm
    dip}}\right)^2 - \left(\frac{B_{\rm dip,i}}{B_{\rm dip}}\right)^{\alpha}\right\}
\quad &(\alpha\neq 2)\ ,\cr\cr
\fracb{B_{\rm dip,i}}{B_{\rm dip}}^2\left[\tau_{\rm c,i}+
\tau_{\rm d,i}\ln\fracb{B_{\rm dip,i}}{B_{\rm dip}}\right] 
\quad &(\alpha=2)\ .\cr}\right.
\end{equation}
\vspace{0.3cm}
Substitution of Eq.~(\ref{eq:B-vs-time}) into this solution yields:
\vspace{0.3cm}
\begin{equation}\label{eq:tau_c(t)}
\tau_{c}(t) = \left\{\matrix{
\frac{\tau_{\rm d,i}}{2-\alpha} \left\{ \left[1+(2-\alpha) \frac{\tau_{\rm
      c,i}}{\tau_{\rm d,i}} \right] \left(1+\frac{\alpha t}{\tau_{\rm d,i}}\right)^{2/\alpha} 
   - \left(1+\frac{\alpha t}{\tau_{\rm d,i}}\right)\right\}
\quad &(\alpha\neq 0,\,2)\ ,\cr\cr
\frac{\tau_{\rm d,i}}{2} \left\{ \left[1+\frac{2\tau_{\rm c,i}}{\tau_{\rm d,i}} 
\right] \exp\left(\frac{2t}{\tau_{\rm d,i}}\right) - 1\right\}
\quad &(\alpha=0)\ ,\cr\cr
\left(1+\frac{2t}{\tau_{\rm d,i}}\right)\left[\tau_{\rm c,i} +
\frac{\tau_{\rm d,i}}{2}\ln\left(1+\frac{2t}{\tau_{\rm d,i}}\right)\right]
\quad &(\alpha=2)\ .\cr}\right.
\end{equation}
\vspace{0.3cm}
For the specific case of $\alpha = 0$ the expression for $\tau_c(t)$
in Eq.~(\ref{eq:tau_c(t)}) can be easily inverted to obtain:
\begin{equation}\label{eq:t_tau_c}
t(\tau_c) = \frac{\tau_{\rm d,i}}{2}\ln\fracb{1+
\frac{2\tau_c}{\tau_{\rm d,i}}}{1+\frac{2\tau_{\rm c,i}}{\tau_{\rm d,i}}}\ ,
\end{equation}
but this is not possible for a general value of
$\alpha$. Nevertheless, at early times we can write:
\begin{equation}
\tau_c(t\ll\tau_{\rm d,i}/\alpha) \approx \tau_{\rm c,i}+ t\ .
\end{equation}
Equations (\ref{inferred-dipole}) and (\ref{eq:P_infty}) imply that:
\begin{equation}
(2-\alpha) \frac{\tau_{\rm
      c,i}}{\tau_{\rm d,i}} = \left[\fracb{P_\infty}{P_i}^2 - 1\right]^{-1} 
\approx \fracb{P_i}{P_\infty}^2\ .
\end{equation}
Since one expects $P_i\ll P_\infty$, then the term in
Eqs.~(\ref{eq:tau_c(B)})--(\ref{eq:t_tau_c}) involving the initial
conditions (i.e. that with $\tau_{\rm c,i}$) can be neglected at
$t\gg\tau_{\rm c,i}$, where $\tau_c(\tau_{\rm c,i}\ll t\ll \tau_{\rm
d,i}/\alpha) \approx t$.

The limits of the expressions above at late times, $t\gg\tau_{\rm
d,i}/\alpha$, are:
\begin{equation}
\label{eq:B-t_late}
B_{\rm dip}(t\gg\tau_{\rm d,i}/\alpha) \approx B_{\rm dip,i} \left\{\matrix{
(\alpha\,t/\tau_{\rm d,i})^{-1/\alpha} \quad &(\alpha > 0)&\ , \cr
\exp(-t/\tau_{\rm d,i})\quad &(\alpha=0)&\ }\right.
\end{equation}
and 
\begin{equation}
\label{eq:p-t_late}
P(t\gg\tau_{\rm d,i}/\alpha) \approx \left\{\matrix{
P_\infty(\alpha)\left[1-\frac{1}{2}\exp(-2t/\tau_{\rm d,i})\right]
\approx P_\infty(\alpha) \quad &(\alpha=0)\ ,\cr
P_\infty(\alpha)\left[1-\frac{1}{2}(\alpha t/\tau_{\rm d,i})^{(\alpha-2)/\alpha}\right]
\sim P_\infty(\alpha) \quad &(0<\alpha<2)\ ,\cr
P_\infty(\alpha=0)\sqrt{\ln(2t/\tau_{\rm d,i})}\quad &(\alpha = 2)\ ,\cr
P_\infty(4-\alpha)\,(\alpha t/\tau_{\rm d,i})^{(\alpha-2)/2\alpha} \quad &(\alpha>2)\ .}\right.
\end{equation}
For $\alpha > 2$ the period grows at late times as $P\propto
t^{(\alpha-1)/2\alpha}$. This reproduces the familiar $\dot{P}\propto
P^{-1}$ evolution for spin-down at a constant magnetic field ($\alpha
\to \infty $). For $0 \leq \alpha < 2$, $P$ approaches $P_\infty$ at
late times. We note, however, that for $0 < 2-\alpha \ll 1$ the asymptotic spin
period $P_\infty$ is approached extremely slowly, where $P(t)/P_\infty
\approx g < 1$ at $t \approx (\tau_{\rm
d,i}/\alpha)(1-g)^{-\alpha/(2-\alpha)} \to (\tau_{\rm
d,i}/2)(1-g)^{-2/(2-\alpha)}$, e.g. $P$ reaches half of its asymptotic
value at $t \approx (\tau_{\rm
d,i}/\alpha)2^{\alpha/(2-\alpha)}\to\tau_{\rm
d,i}2^{2/(2-\alpha)}$. When $P_\infty$ is approached, then
Eq.~(\ref{inferred-dipole}) (or equivalently, Eq.~[\ref{eq:tau_c(B)}])
implies that:
%
%
\begin{equation}
\label{eq:Bdipvstauc}
B_{\rm dip} = \sqrt{\frac{I c^3}{f\,8 \pi^2 R^6}}\,
\frac{P_{\infty}}{\tau^{1/2}_c} = 1.07\times 10^{15}f^{-1/2}
I_{45}^{1/2}R_6^{-3}\left(\frac{P_{\infty}}{10~\rm s}\right)
\left(\frac{\tau_c}{1\;{\rm kyr}}\right)^{-1/2}\;{\rm G}\ ,
\end{equation}
Note that, although this expression is independent on $\alpha$, it holds only 
for $\alpha <2$. This simple scaling  will indeed hold for any scenario 
in which the spin period freezes, provided the appropriate value of $P_{\infty}$
is used. In fact, it corresponds to the general relation
\ref{inferred-dipole}, with the spin period kept constant to the asymptotic 
value.

From Eqs.~(\ref{eq:tau_c(B)}) and (\ref{eq:tau_c(t)}) we obtain the
late time behaviour of $\tau_c$, at $B_{\rm dip}\ll B_{\rm dip,i}$ or
$t\gg\tau_{\rm d,i}$:
\begin{equation}\label{eq:tau_c(B)_late}
\tau_{c}\approx \left\{\matrix{
\frac{\tau_{\rm d,i}}{2}\left(\frac{B_{\rm dip,i}}{B_{\rm dip}}\right)^2 
&\approx& \frac{\tau_{\rm d,i}}{2} \exp\fracb{2t}{\tau_{\rm d,i}} 
\quad &(\alpha = 0)\ ,\cr\cr
\frac{\tau_{\rm d,i}}{2-\alpha}\left(\frac{B_{\rm dip,i}}{B_{\rm dip}}\right)^2 
&\approx& \frac{\tau_{\rm d,i}}{2-\alpha}\fracb{\alpha t}{\tau_{\rm d,i}}^{2/\alpha} 
\quad &(0<\alpha < 2)\ ,\cr\cr
\tau_{\rm d,i}\fracb{B_{\rm dip,i}}{B_{\rm dip}}^2\ln\fracb{B_{\rm dip,i}}{B_{\rm dip}}
&\approx& t\ln\fracb{2t}{\tau_{\rm d,i}}
\quad &(\alpha=2)\ ,\cr\cr
\frac{\tau_{\rm d,i}}{\alpha-2}\left(\frac{B_{\rm dip,i}}{B_{\rm dip}}\right)^{\alpha}
&\approx& \frac{\alpha}{\alpha-2}t
\quad &(\alpha > 2)\ ,
}\right.
\end{equation}
\vspace{0.3cm}

\begin{figure*}
  \begin{center}
\leavevmode\includegraphics[width=8.cm, height=6.2cm]{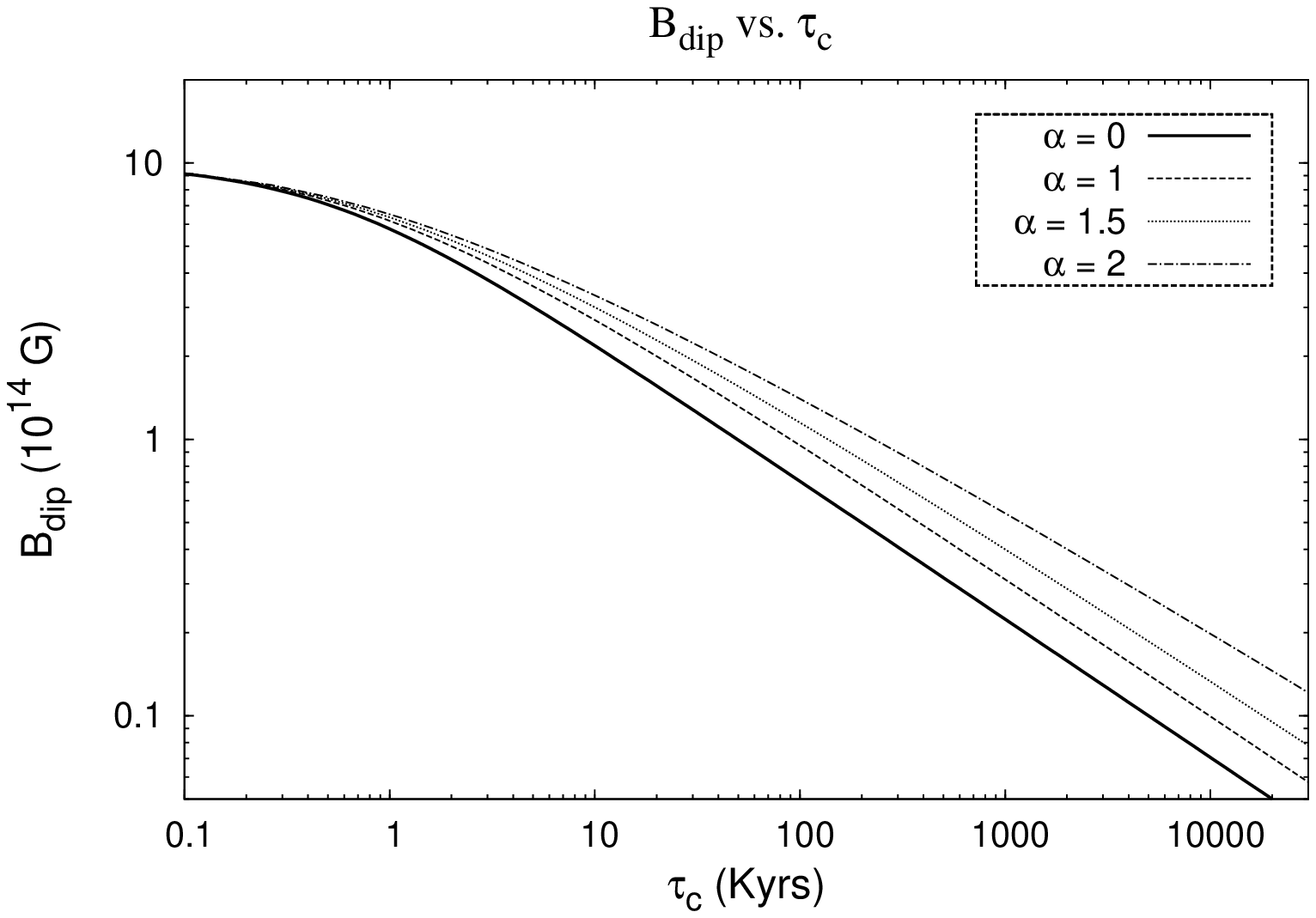}  
\hspace{3mm}
\leavevmode\includegraphics[width=8.cm, height=6.2cm]{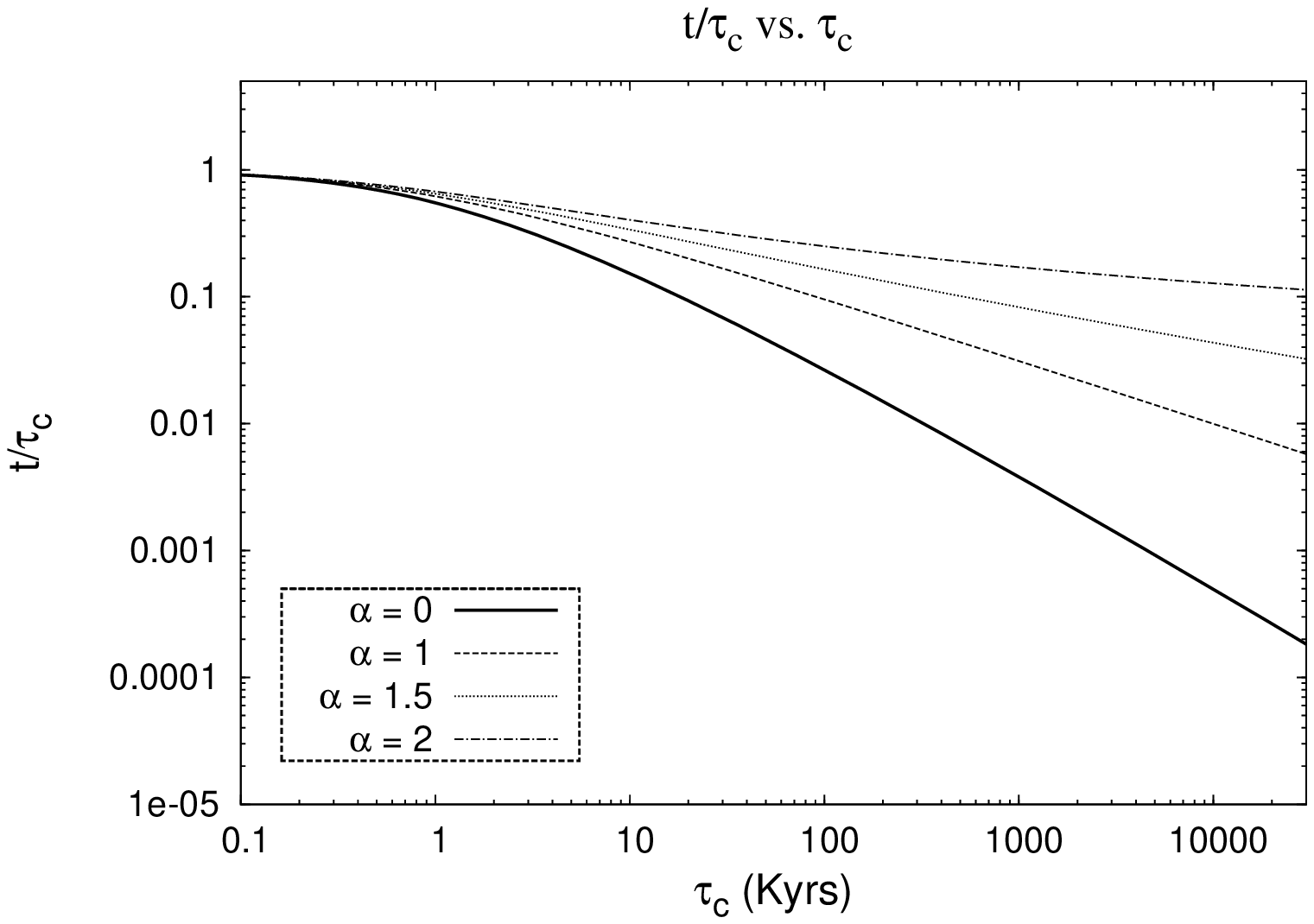}
  \caption{Different cases of field decay with asymptotic spin
  period. \textit{Left Panel}: Dipole magnetic field, $B_{\rm dip}$,
  as a function of spindown age, $\tau_{\rm c}$, for 4 selected values
  of the decay index, $\alpha \leq 2$. All curves have the same
  initial $B_{\rm dip,i} = 10^{15}\;$G. Smaller values of the
  asymptotic period $P_{\infty}$ correspond to lower asymptotic curves
  where $B_{\rm dip}\propto P_\infty/\sqrt{\tau_{\rm
  c}}$. \textit{Right Panel}: The ratio $t/\tau_{\rm c}$ between real
  age and spindown age as a function of the spindown age, for the same
  models. The $y$-axis gives the correction factor that must be
  applied to the measured spindown age, $\tau_{\rm c}$, to derive the
  real age of an object. Lower values of $\alpha$ correspond to much
  younger objects, at a given $\tau_{\rm c}$, because they produce a
  faster decay of the field and, accordingly, of $\dot{P}$.}
  \label{fig:summary}
\end{center}
\end{figure*}

The  are summarized in
Fig.~\ref{fig:summary} summarizes the main results of this section. It shows the expected relation
 $B_{\rm dip}$ vs. $\tau_c$ and the evolution of the ratio $(t/\tau_{\rm c})$ 
with  $\tau_c$, for 4 different values of $\alpha \leq 2$. All plots are 
obtained for the same initial value $B_{\rm dip,i} =10^{15}$ G and assume a 
normalization of the decay time, $\tau_{\rm d,i} = 10^3 {\rm yrs} /
B^{\alpha}_{\rm i,15}$ yrs. 
%
 
\section{Summary of physical mechanisms causing field decay}
\label{summary-fieldecay}
General modes of fied decay in non-superfluid NS interiors were studied by 
\citealt{GR92} (GR92 hereafter), who identified three avenues
for field evolution: Ohmic decay, ambipolar diffusion and Hall drift.
While the first two mechanisms are intrinsically dissipative, leading directly 
to a decrease in field energy, the third one is not. GR92 proved its potential 
relevance, however, for the transport and dissipation of magnetic energy
within NS crusts (cfr. \citealt{Jones+1988}), by speeding up ohmic dissipation of  the field. 

The analysis by GR92 showed that ambipolar diffusion is conveniently split in two different
components, according to their effect on the stable stratification of NS core
matter. The solenoidal component does not perturb chemical equilibrium of
particle species, thus its evolution is opposed only by particle collisions. 
The irrotational mode does perturb chemical equilibrium, so it also activates
$\beta$-reactions in the NS core. At high temperatures ($T > 7\times
10^8$ K) 
the two modes are degenerate, as $\beta$-reactions are very efficient
and particle collisions  represent the only effective force against which
ambipolar diffusion works (GR92, TD96) 
The distinction between the two components becomes essential at lower temperature, as perturbations
of the chemical equilibrium are not erased quickly, thus
significantly slowing down the irrotational mode (GR92, TD96).
%
Following GR92, the relevant timescales for the two modes of ambipolar
diffusion can be written as:
%
\begin{eqnarray}
\tau^{(\rm s)}_{\mbox{\tiny{ambi}}} & \simeq & 
3 \times 10^3 \frac{L^2_5 T^2_8}{B^{2}_{15}}~\mbox{~yr} \nonumber \\
\tau^{(\rm ir)}_{\mbox{\tiny{ambi}}} & \simeq & 
\frac{5 \times 10^9} {T^6_8 B^{2}_{15}}~\mbox{~yr} +
\tau^{(s)}_{\mbox{\tiny{ambi}}}~~.
\end{eqnarray}
Note the strikingly different
dependence on temperature of the two modes, which turns out to be a key factor.
Since field dissipation releases heat in the otherwise cooling core,
a balance between field-decay heating and neutrino cooling (through
  modified URCA reactions) is expected to be reached. This determines 
an equilibrium relation between core temperature and strength of the core
magnetic field. 
The two following relations, for either mode, are obtained (TD96; \citealt{DSS09}, hereafter DSS09):
\begin{eqnarray}
\label{eq:TvsB-relation}
T^{(\rm s)}_{8, {\rm eq}} & \simeq & 
2.7~B^{2/5}_{15} \left(\frac{\rho_{15}}{0.7}\right)^{-2/3}\ , 
\nonumber \\
T^{(\rm ir)}_{8,{\rm eq}} & \simeq & 
2.4 \left(\frac{B}{10^2~B_{\rm QED}}\right)^2 \left(\frac{\rho_{15}}{0.7}\right)^{-1}\ ,
\end{eqnarray}
where $\rho_{15}$ is the density of matter normalized to $10^{15}$ g
cm$^{-3}$. 
Using these relations we eventually obtain expressions for the relavant timescales as
a function of $B$ alone:
\begin{eqnarray}
\tau^{(s)}_{\rm ambi} & \simeq & \frac{1.5 \times
  10^5}{B^{6/5}_{15}} \left(\frac{\rho_{15}}{0.7}\right)^{-2/5}~\mbox{~yrs}
~~~~~~(\alpha=6/5)\ , \nonumber \\
\tau^{(ir)}_{\rm ambi} & \simeq & 3.7 \times
  10^5 \left(\frac{\rho_{15}}{0.7}\right)^{22/3} \left(\frac{B}{10^2 
B_{\rm QED}}\right)^{-14}~\mbox{~yrs}~~~~~~(\alpha=14)\ .
\end{eqnarray}

Hall-driven evolution of the magnetic field is characterized by 
the timescale $\tau_{\rm Hall} = 4\pi e n^2_e L^2/B$. Here 
$n_e$ is the electron number density and $L$ is a characteristic length on
which significant gradients $n_e$ and $B$ develop.
GR92 argued that the main effect of the Hall term would be that 
of driving a cascade of magnetic energy 
from the large scale structure of the field to increasingly smaller scales.
The ohmic dissipation timescale is $\tau_{\rm ohm} = 4 \pi \sigma_o
L^2/c^2\propto L^2$ (GR92, \citealt{Cumming+2004}), where $\sigma_o$ is 
the electrical conductivity of NS matter. Thus, very efficient dissipation of 
sufficiently small-scale structures would eventually be reached. Such a 
"turbulent" field evolution would be of particular relevance in NS crusts, 
where  \textit{(1)} matter density is lower than in the core and the Hall 
timescale is then short enough
\textit{(2)} ions are locked in the crystalline lattice and B-field evolution 
is thus coupled to the electron flow only. The importance of the Hall
 term relative to ohmic dissipation is typically quantified by the Hall
  parameter, $\omega_{\rm B} \tau = eB \tau /(m_* c) = \tau_{\rm ohm}/\tau_{\rm Hall}$.
  Here $m_* = E_{\rm F}/c^2 \gg m_e$ is the electron effective mass, $E_{\rm
    F}$ being its Fermi energy, and $\tau$ is a typical electron collision
  time. \cite{Cumming+2004} have shown that the Hall parameter in a NS crust 
is always much larger than unity, if the magnetic field is $> 10^{14}$ G, apart
from, possibly, the lowest density regions of the outer crust (cfr. their
Fig. 4). For weaker fields, on the other hand, the ohmic term might largely
dominate at temperatures lower than a few $\times 10^8$ K. Hence, a prominent
role of the Hall term is expected for magnetar-strength fields.

%
%
\cite{Cumming+2004}
highlighted the prominent role played by a realistic electron density profile
in determining the Hall evolution of the crustal field. As Hall modes first 
get excited at the base of the crust, where $\rho 
\approx 10^{14}$ g cm$^{-3}$ and the Hall time is longer, they can propagate 
to the surface with their wave-vector $k$ progressively decreasing, because of 
the decreasing electron density. The ohmic dissipation time decreases
accordingly and
the overall decay rate of the field is set by the longest timescale at the
base of crust. Using the local pressure scale height $P/(\rho g)$ as the
natural lengthscale $L$, \cite{Cumming+2004} derive the expression:
\begin{equation}
\label{eq:cummingetal}
\tau_{\rm Hall} \simeq 1.2 \times 10^4 \frac{\rho^{7/3}_{14}}{B_{\rm
      dip,15}}~{\rm yrs}
\end{equation}
for the Hall decay timescale in NS crusts. Note the different
normalization of the density compared to ambipolar diffusion timescales,
reflecting the different locations of the two processes.

Several authors \citep{Vainshtein+2000, Rhein+2002, Pons+2010} further
investigated this scenario numerically, generally confirming
the tendency of crustal fields to develop shorter scale structures on the Hall
timescale. However, numerical calculations may not fully support the 
idea that the Hall timescale actually characterizes the decay timescale of 
magnetic modes (see \citealt{ShalUrp97, Holl02, Pons+2007, Koji12}). 
The situation in this case is likely more complex than the basic picture given above.

Alternatively, ohmic dissipation can proceed at a fast rate if the
  electric currents and the field are initially rooted in relatively outer
  layers of the crust, where the electrical conductivity $\sigma_{\rm o}$ is rather
  low (cfr. \citealt{Peth95}).
Due to the subsequent diffusion of electrical currents into deeper
  crustal layers, $\sigma_{\rm o}$ progressively increases and field decay 
slows down accordingly. The decay of the dipole field can effectively be described as a
  sequence of exponentials with a growing characteristic time
  $\tau_{\rm d,exp}(t)$. In this framework it can be shown \citep{Urpetal+94,
    Urpetal+97, Urpiko+08} that, although the ohmic
  dissipation rate is field-independent, the resulting field decay proceeds as
  a power-law in time, after a short ``plateau'' which is set  by the initial distribution of currents.
 The decay index of the power-law is determined by the rate at which $\sigma_{\rm o}$ changes
with time. This depends on both the depth reached by currents, $h(t)$, and the
temperature of the crust, $T_{\rm c}(t)$. It is found \citep{Urpetal+94} that a power-law decay with index $=3/2$ results when $\sigma_{\rm o}$
is independent on temperature, as is the case when impurity scattering
dominates the conductivity. An index $=11/6 \approx 1.83$ is obtained, instead,
when $\sigma_{\rm o}$ is dominated by electron-phonon scattering and, thus, 
scales with $T_{\rm c}^{-2}$ (cfr. Fig. 2 and 3 of \citet{Urpetal+97} and
the discussion of ohmic decay in \citealt{Cumming+2004}).

Note that an asymptotic spin period $P_{\infty} \propto \sqrt{\tau_{\rm
      d,i}}~B_{\rm dip,i}$ exists also in this model. Since the decay time $\tau_{\rm d,i}$ is field independent, the
  expected scaling between $P_{\infty}$ and ${\rm B}_{\rm dip,i}$ is
  linear, like in the exponential case ($\alpha=0$) previously discussed.

\section{Observational evidence for field decay in neutron stars with strong magnetic fields}
\label{sec:two}
%


\begin{figure}
\begin{center}
\includegraphics[width=8.8 cm, height=7.5cm]{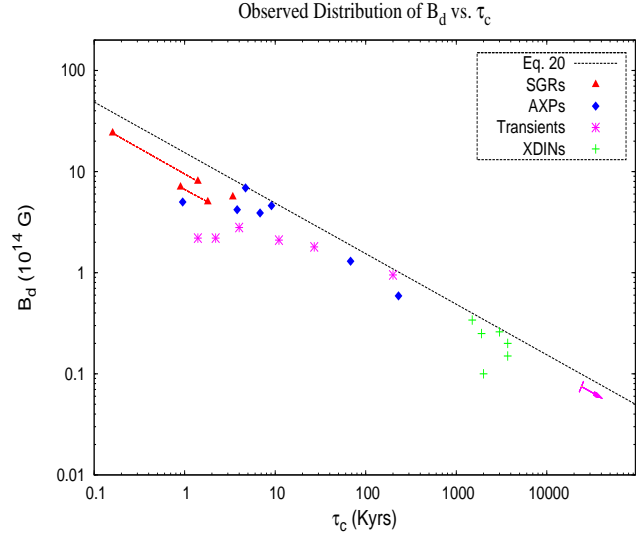}
\caption{Inferred Dipole Field ($B_{\rm dip} = 
3.2\times 10^{19}\sqrt{P \dot{P}}\;$G) vs. characteristic spindown age
($\tau_{\rm c} = (1/2) P/\dot{P}$) for SGRs, AXPs, transients and XDINs.
The dashed line in Fig.~\ref{fig:B-tau} represents the scaling
of Eq.~(\ref{eq:Bdipvstauc}), expected from field decay with $\alpha
<2$, using $P_{\infty} = 11.77\;$s that corresponds to the longest
measured spin period (for AXP~1841-045). The magenta arrow indicates the
current upper limit on the position of SGR 0418+5829. The red triangles joined by dotted 
lines represent the whole range of values spanned by SGR 1806-20 and
 SGR 1900+14, as explained in the text (cfr. Tab. \ref{table}).}
\label{fig:B-tau}
\end{center}
\end{figure}

We begin  by assessing whether the distribution of dipole magnetic fields
of magnetar candidates, as inferred from timing observations, can provide indications for
field decay. Fig.~\ref{fig:B-tau} shows the inferred dipole fields, $B_{\rm dip}$, of magnetar candidates 
plotted versus their spindown age, $\tau_c$. This represents an alternative projection of the usual $P-\dot{P}$ diagram.

Two points stem most clearly from Fig.~\ref{fig:B-tau}.
First, the absence of old objects (with  relatively large spin-down ages) 
with strong dipole fields and the tendency for objects with increasingly
stronger fields to be found only at increasingly younger spin-down
ages. The dashed line in Fig.~\ref{fig:B-tau} represents the scaling
of Eq.~(\ref{eq:Bdipvstauc}), expected from field decay with $\alpha
<2$, using $P_{\infty} = 11.77\;$s that corresponds to the longest
measured spin period (for AXP~1841-045). The apparently prohibited
region in parameter space is  equivalent to
the existence of a limiting spin period for the objects considered. 
As discussed in previous sections, precisely such an asymptotic
spin period is expected if the dipole field decays  and its decay is
governed by a physical mechanism with $\alpha < 2 $. 

Second, all ``old'' sources (spin-down age $\tau_c\gtrsim 10\;$kyr)
lie in a narrow strip corresponding to $3.5\;{\rm s}\lesssim
P_\infty\lesssim 11\;$s, i.e. within a factor of $\approx 3$ in their
spin period\footnote{Note that, with the notable exception of the XDIN RX 
J0420, all other sources would lie between 7s and 11 s, a range $\simeq 1.6$ in 
spin period.}. In the context of magnetic field decay models, the fact
that this strip is so narrow can be interpreted in
one of two ways. Either (1) all ``old'' AXPs, XDINs and SGR0418+5729
came from a narrow distribution of initial dipole fields $B_{\rm
dip,i}$, in which case a relatively large range of $\alpha < 2$ values 
would be compatible with observations, or (2) $\alpha$ is sufficiently
close to 2 that field decay has largely washed out the spread in
$B_{\rm dip,i}$ values, which could have been significantly larger in this case.
 
\begin{figure*}
  \begin{center}
\leavevmode\includegraphics[width=8.cm, height=6.2cm]{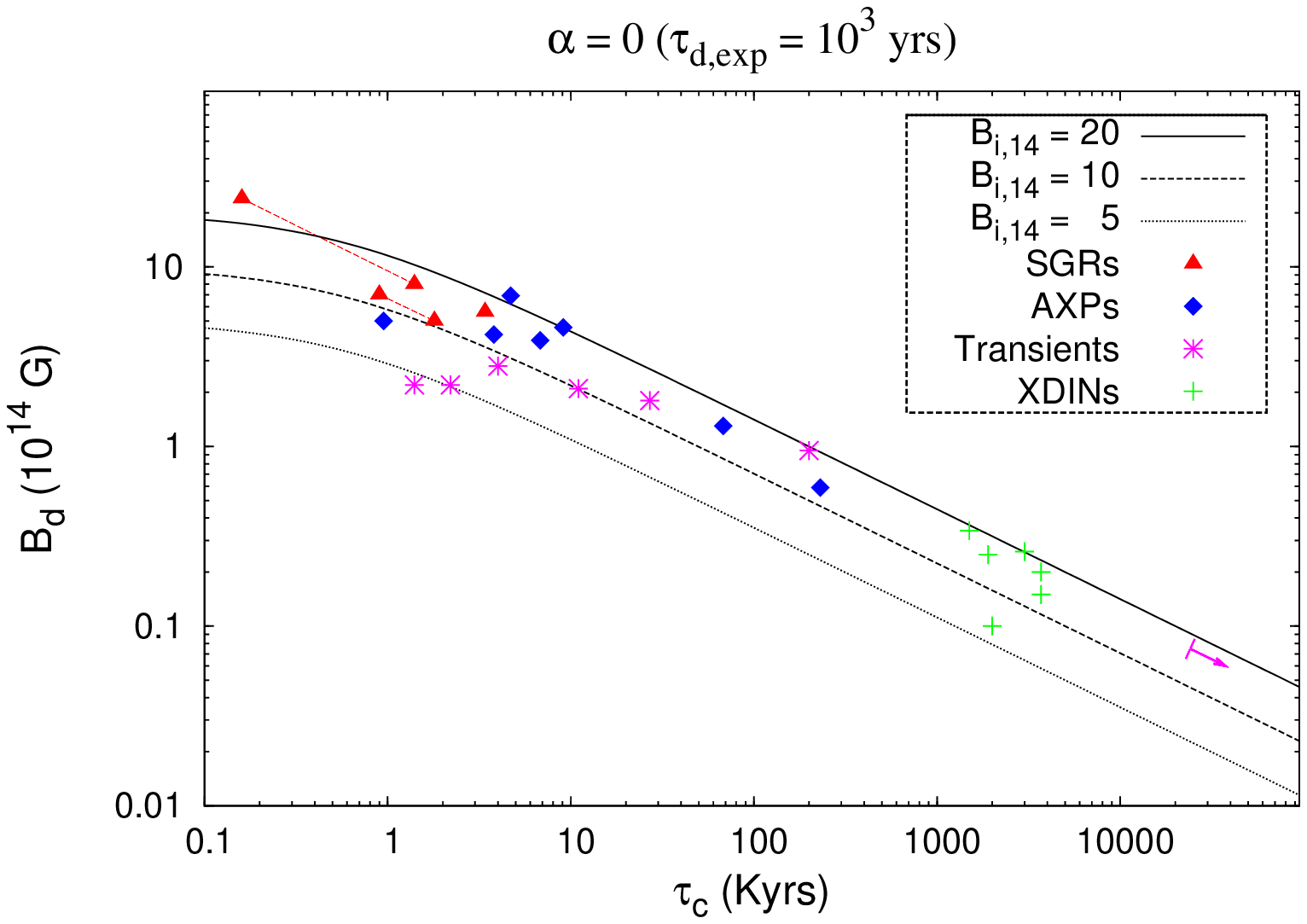}                \hspace{3mm}
\leavevmode\includegraphics[width=8.cm, height=6.2cm]{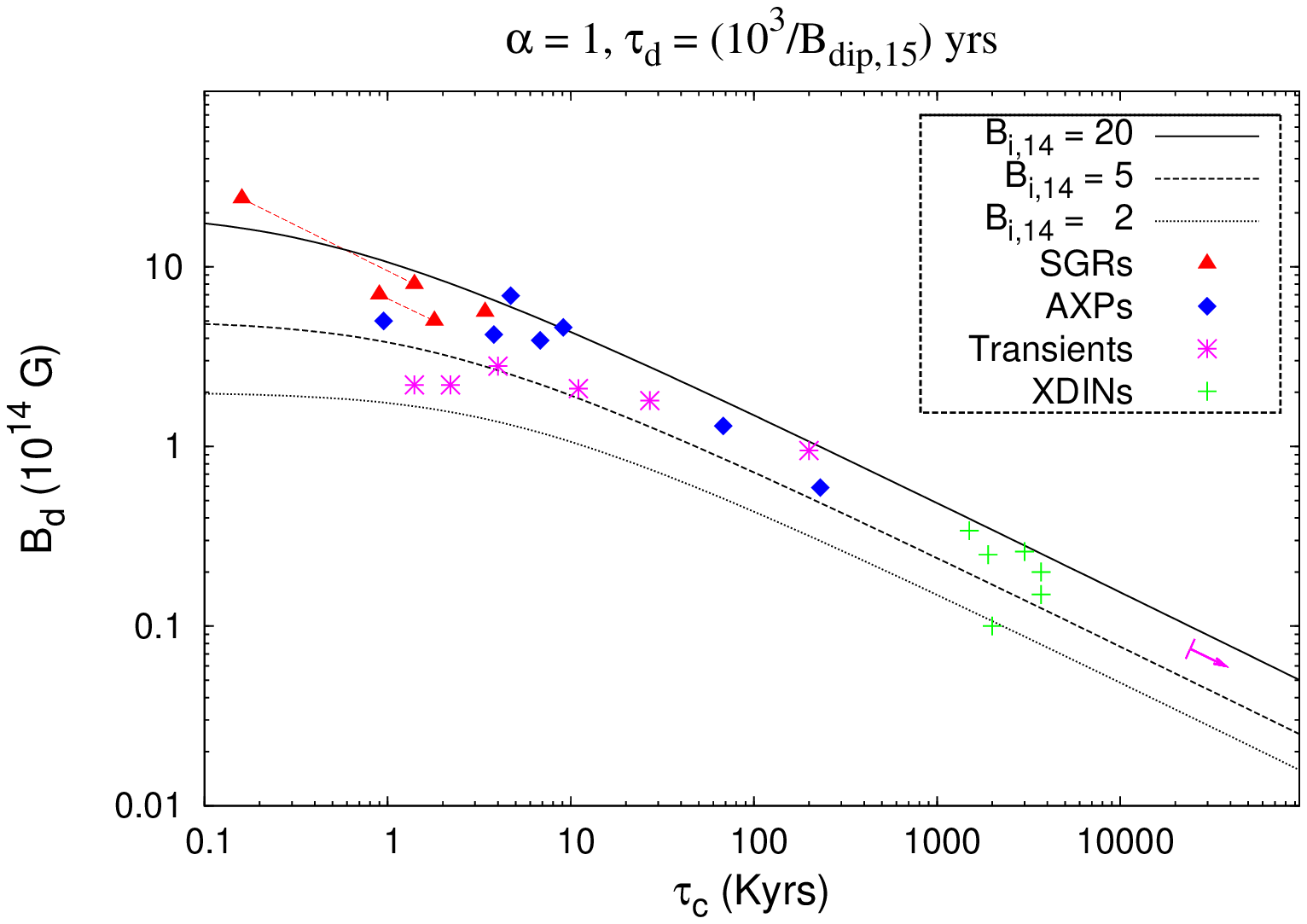}
\leavevmode\includegraphics[width=8.cm, height=6.2cm]{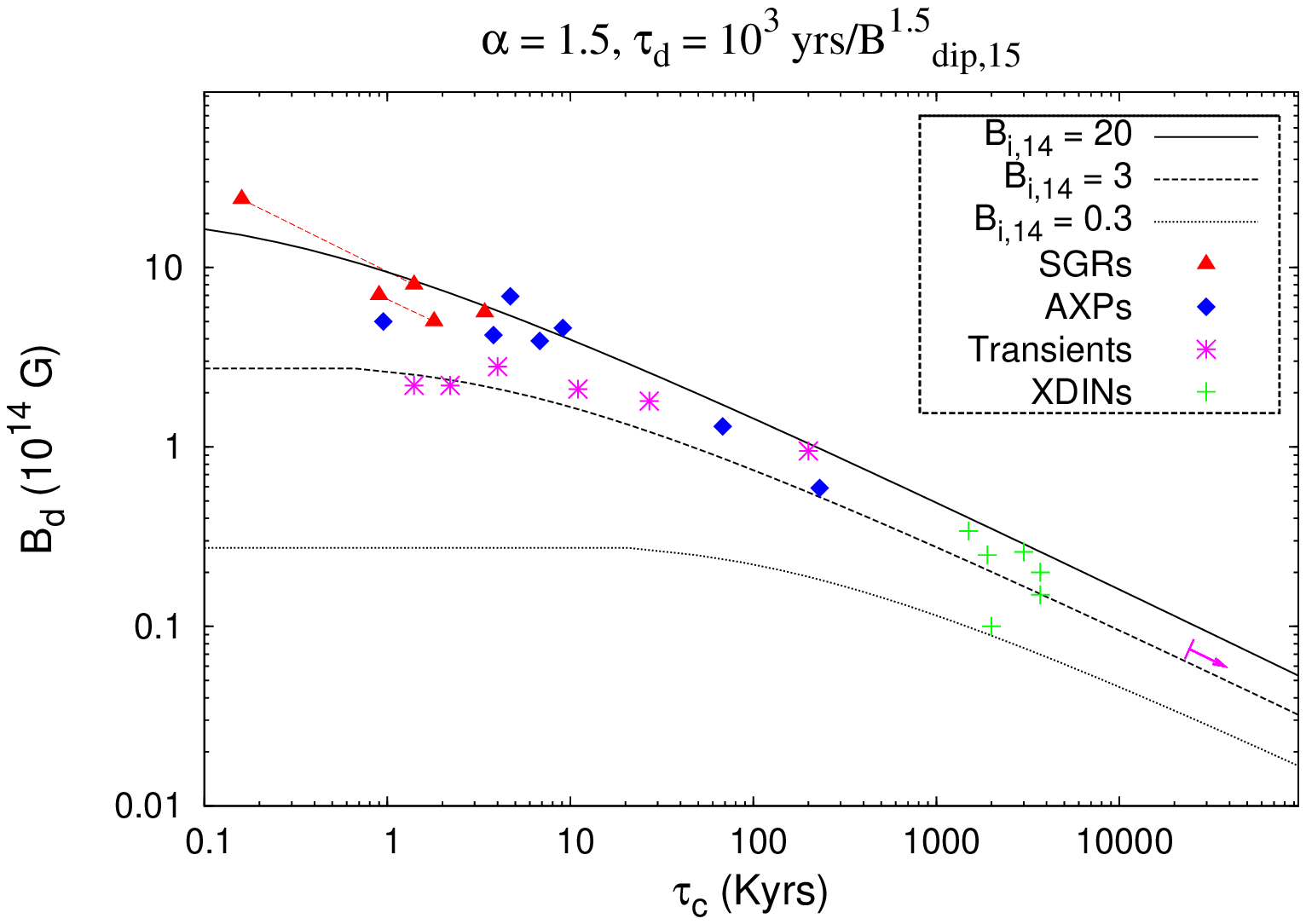}
\leavevmode\includegraphics[width=8.cm, height=6.2cm]{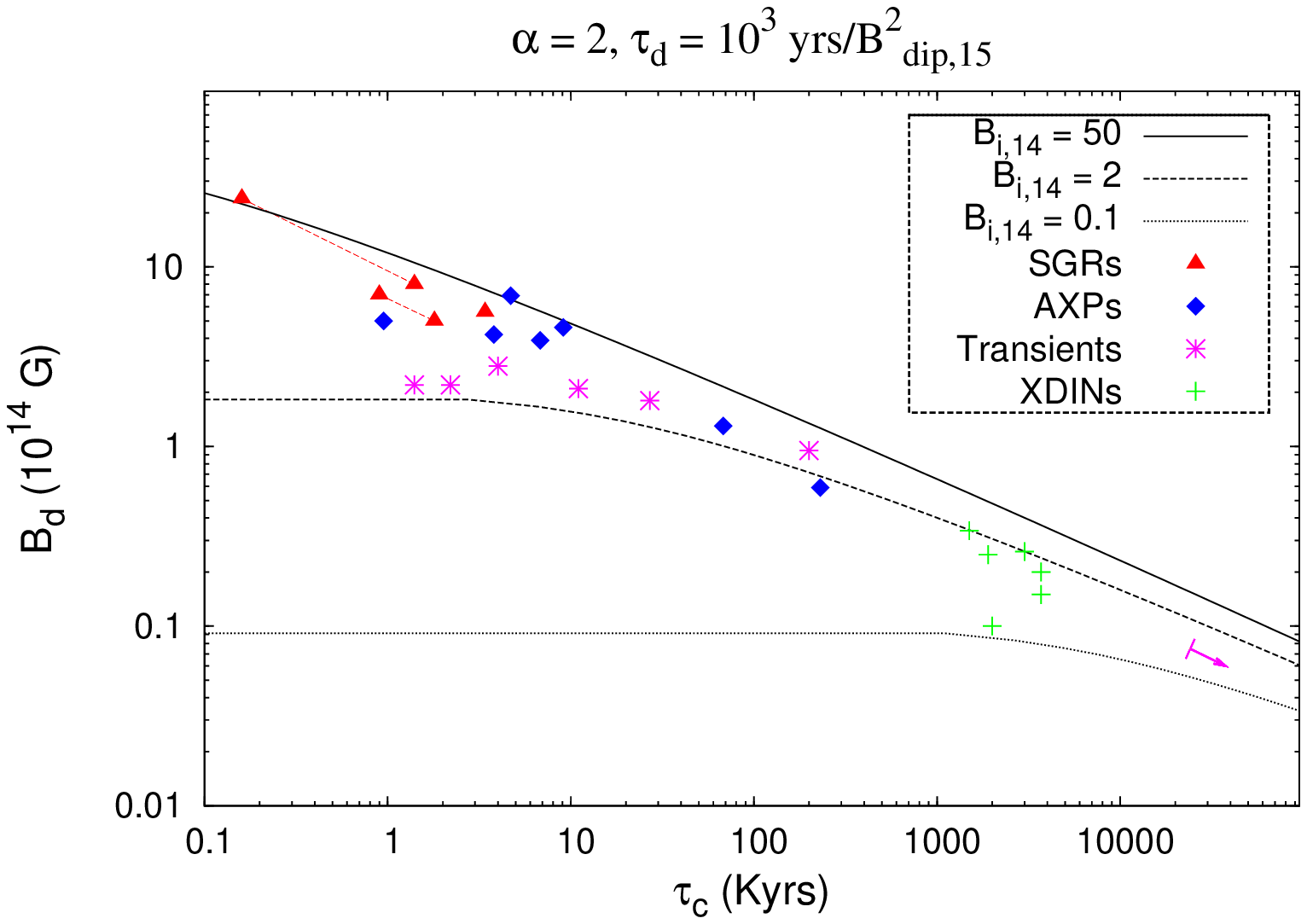}
  \caption{Different models of field decay with asymptotic spin period, 
($\alpha < 2$) 
    compared with the inferred dipole values of magnetar candidates from timing
 measurements. The magenta arrow represents the current upper limit to the
 position of
 SGR 0418+5729. The plotted curves don't represent fits to data but the normalization of
    the decay timescale ($A$) was chosen, for each value of $\alpha$,   
to match the position of sources.}
  \label{fig:B-tau-fieldecay-models}
\end{center}
\end{figure*}

\begin{figure*}
  \begin{center}
\leavevmode\includegraphics[width=8.0cm, height=6.2cm]{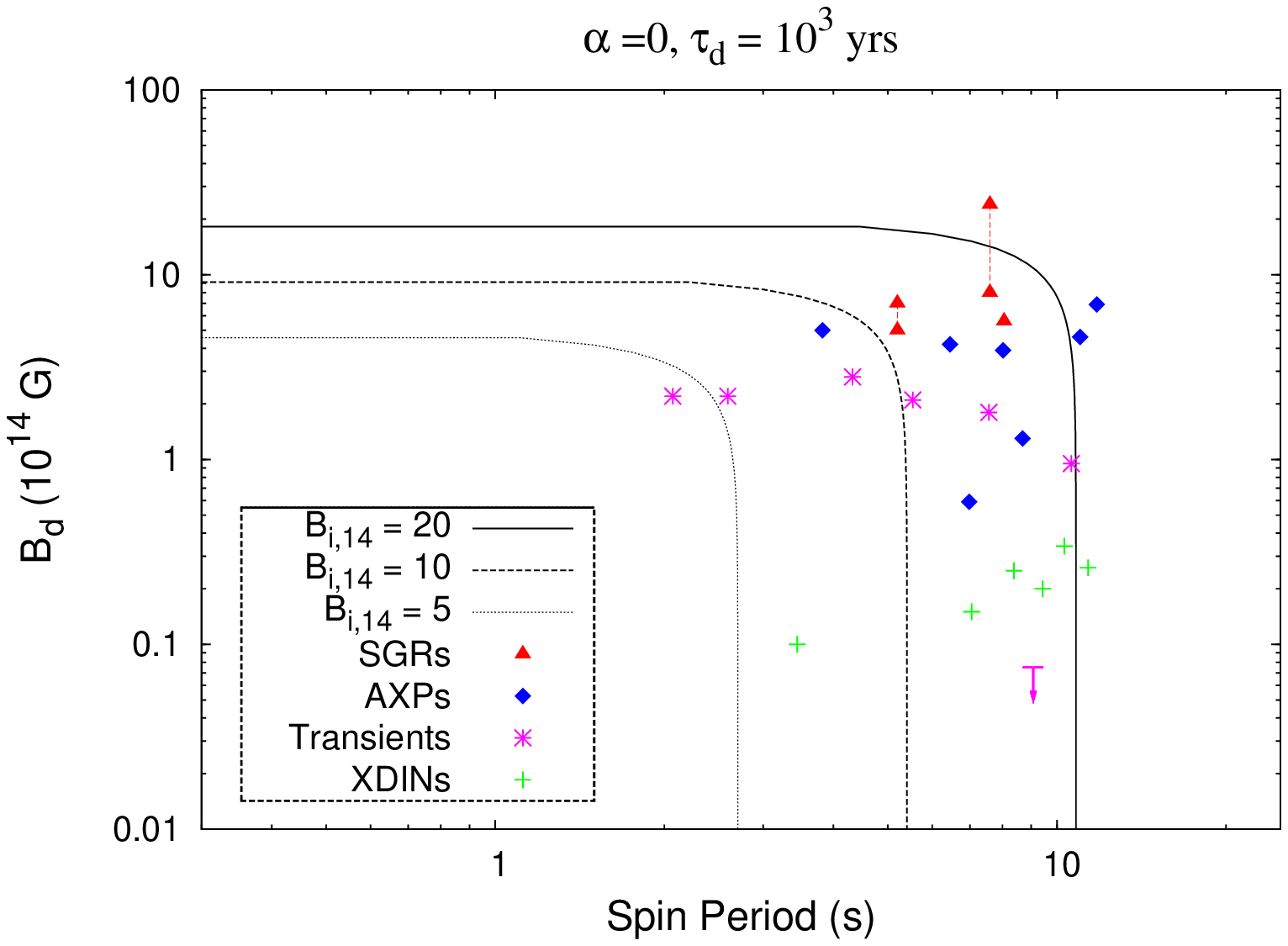}              
  \hspace{3mm}
\leavevmode\includegraphics[width=8.cm, height=6.2cm]{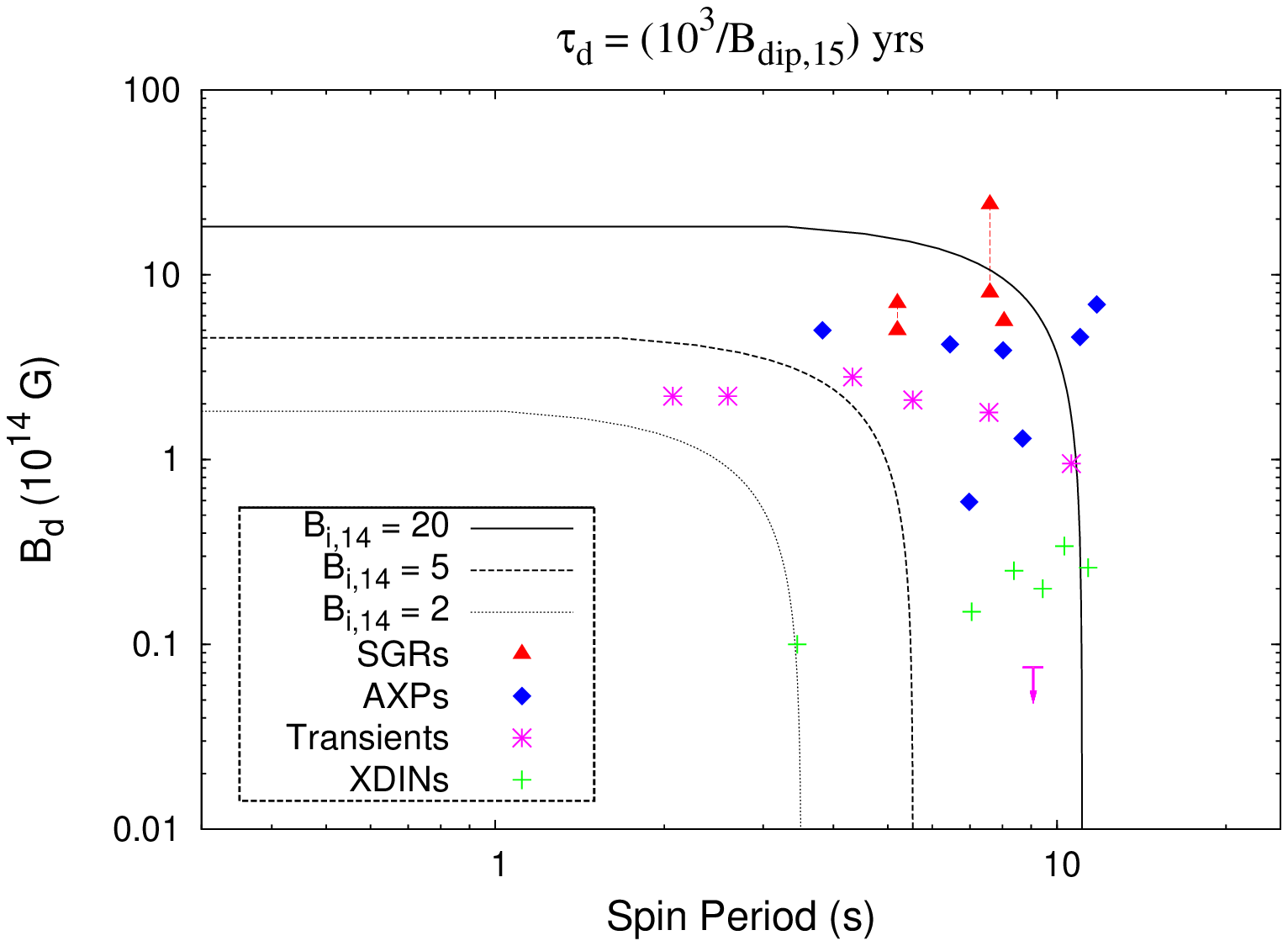}
\hspace{3mm}
\leavevmode\includegraphics[width=8.cm, height=6.2cm]{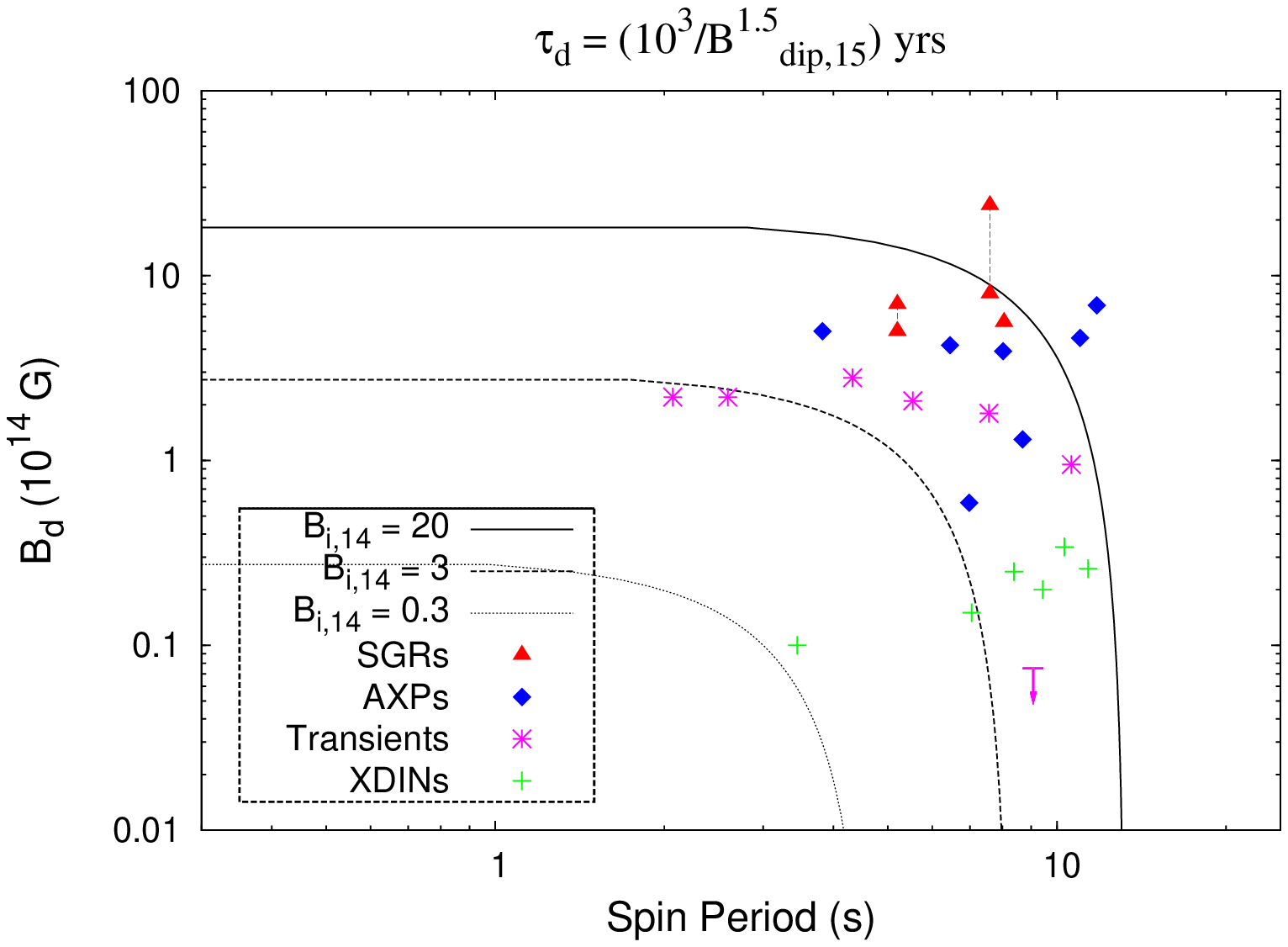}
\hspace{3mm}
\leavevmode\includegraphics[width=8.cm, height=6.2cm]{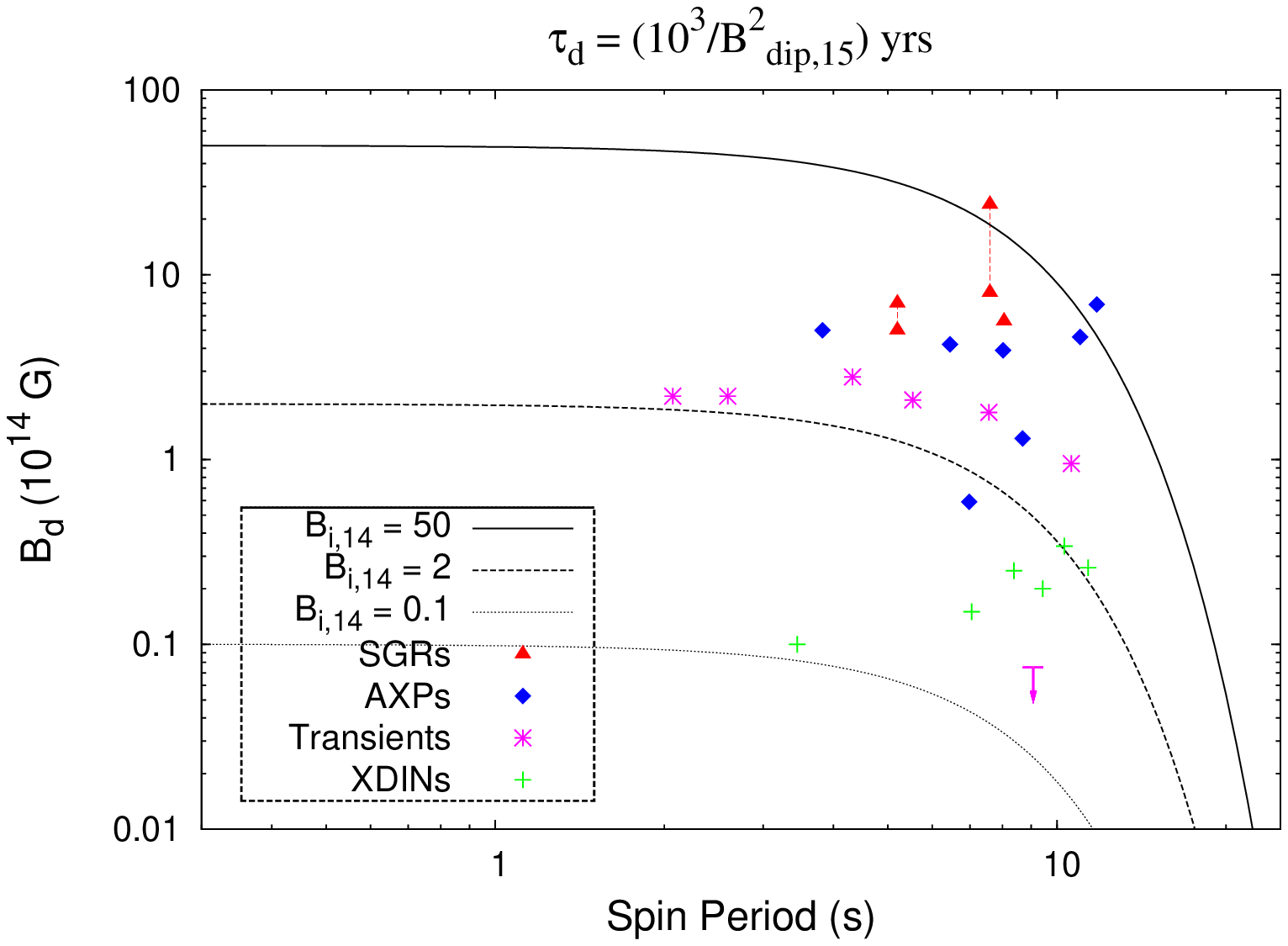}
  \caption{Different models of field decay with asymptotic spin period  ($\alpha < 2$) in the $B_{\rm dip}$ vs. P plane.
   The magenta arrow indicates the current upper limit to the position of SGR
   0418+5729. The  curves correspond to the same models of the previous figure. }
  \label{fig:B-P-fieldecay-models}
\end{center}
\end{figure*}

The distribution of sources shown in Fig.~\ref{fig:B-tau} can be understood
as follows.  For young objects, $t \ll \tau_{\rm
d,i}$, the magnetic field does not have time to decay and it 
is almost constant. This  implies a constant $\dot{B}$
 and a constant  $\dot{E}_{B_{\rm dip}}\propto B_{\rm
dip}^{2+\alpha}$. At this stage the period $P$ and the
spin-down age grow while the dipole field $B_{\rm dip}$ doesn't vary
 (corresponding to a horizontal trajectory in the $B_{\rm
dip}\,$--$\,\tau_c$ plane). Moreover, the object has a nearly constant
power output  due to magnetic field decay  up to $\tau_{\rm c} \lesssim \tau_{\rm d,i}$, at which point 
the power drops  following the decay of the 
magnetic field.
This evolution scenario implies that objects with age $t\leq \tau_{\rm
d,i}$ are most likely to be detected close to $\tau_{\rm d,i}$ since, for a
constant power output, they spend more time at $\tau_{\rm c} \sim \tau_{\rm d,i}$. 

This is particularly expected for SGRs, since they are detected
predominantly through their bursting activity, which is thought to be
directly powered by the decay of their magnetic field\footnote{This assumes
that their bursting activity is powered by their dipole field, rather
than by their internal field. If the latter decays on a longer
timescale this might account for SGRs further along the dashed line in
Fig.~\ref{fig:B-tau}.}, rather than through their quiescent emission
(which might have a non-negligible contribution from the NS residual heat). 
Therefore, the fact that we do not detect any SGRs with magnetic
fields of several $\times 10^{14}\;$G but larger spin periods ($P\gg
10\;$s), strongly supports a decay of the dipole field on a timescale
of $\tau_{\rm d,i}\approx 10^3(B_{\rm dip,i}/10^{15}\,{\rm
G})^{-2}\;$yr, in these objects. Note that, if the SGR's magnetic
dipole field did not decay on such a timescale, a large population
of SGRs 
with spin periods of tens or even hundreds of seconds would exist.
Such large period SGRs should be easy to detect if they maintained similar
magnetic fields and thus similar bursting activity as the observed SGR
sample. Even objects with much weaker or no bursting activity but larger 
spin-down ages are detected (e.g., AXPs, transients, XDINs) which, if there 
was no magnetic field decay, would imply similarly larger ages.
Therefore, we find it highly unlikely that long period
SGRs, with $P\gg 10\;$s, exist in much larger numbers than the
observed SGR sample but are not detected for some unknown reason.

Objects with ages $t\ll \tau_{\rm d,i}$, which populate the lower-left
region of the $B_{\rm d}\,$--$\,\tau_c$ plane, are detected with
proportionally smaller probability, unless their bursting/flaring activity
is stronger at younger ages, resulting in a greater detectability that
could compensate for the short time  spend in this part of the
diagram.
Relatively old objects with an age $t\gg\tau_{\rm d,i}$ 
will have reached their asymptotic period $P_\infty$ (for $\alpha <
2$) and will thus be found on the asymptotic line, $B_{\rm
d}\propto\tau^{-1/2}_c$. These will have a much lower luminosity than
their younger brethren and will
be detected only if, e.g., they are sufficiently close to us or if
they maintain some level of bursting activity.

Finally, we note that 5 out of 7 transient SGRs/AXPs lie, in the $B_{\rm dip}$
vs. $\tau_c$ diagram, below the 
asymptotic line. This suggests that $\tau_{\rm c}$ is not much longer than
their $\tau_{\rm d,i}$. 
CXO J164710.2-455216 and, of course, of SGR 0418+5729 represent two notable
exceptions, as both appear to lie well on the asymptotic line (hence, $\tau_{\rm
  c} > \tau_{\rm d,i}$). In particular, the former object has both $\tau_{\rm
  c}$ and $B_{\rm dip}$ comparable to the persistent AXP 1E2259+586.

\begin{figure*}
  \begin{center}
\leavevmode\includegraphics[width=8.cm, height=6.2cm]{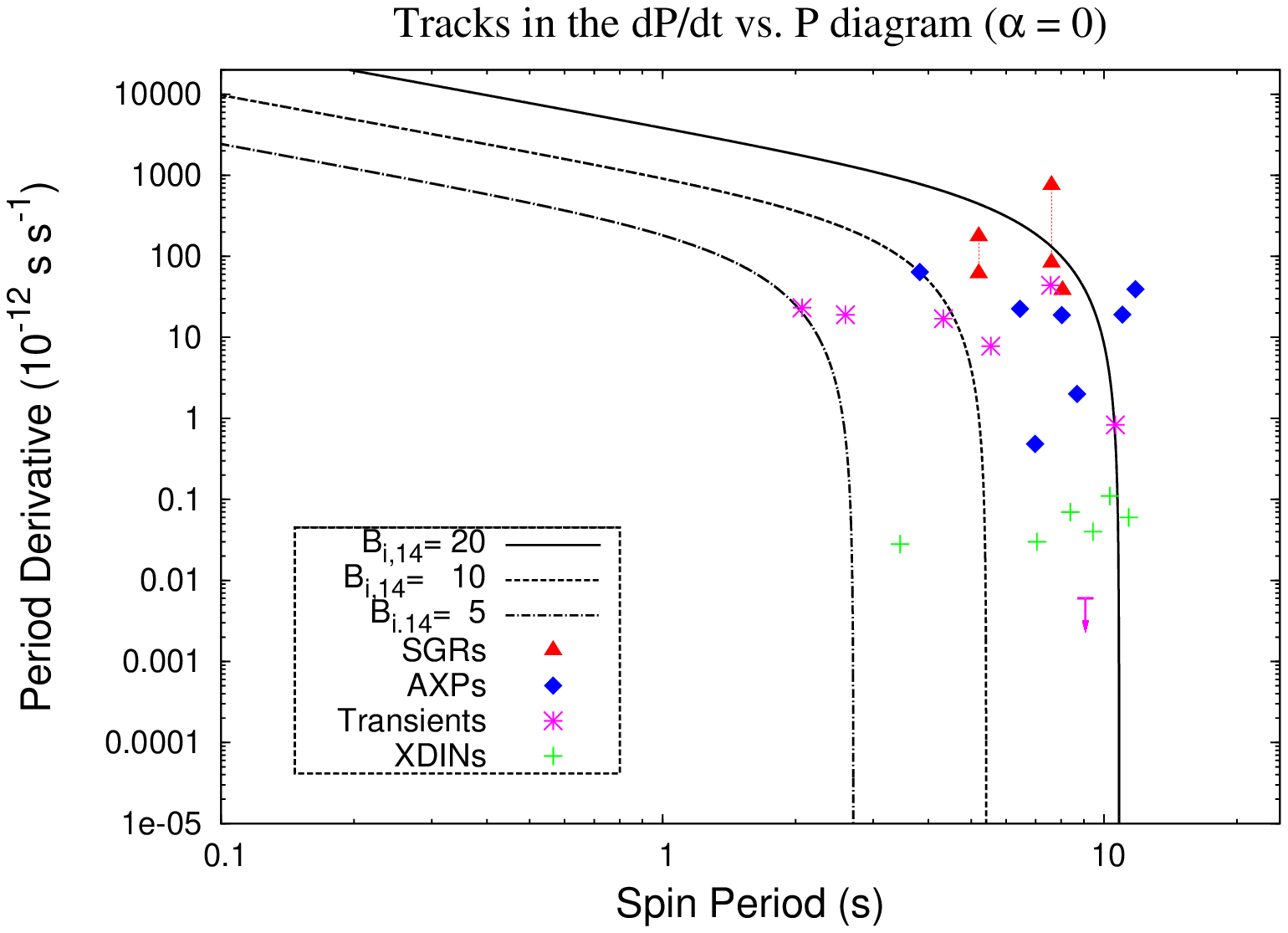}                \hspace{3mm}
\leavevmode\includegraphics[width=8.cm, height=6.2cm]{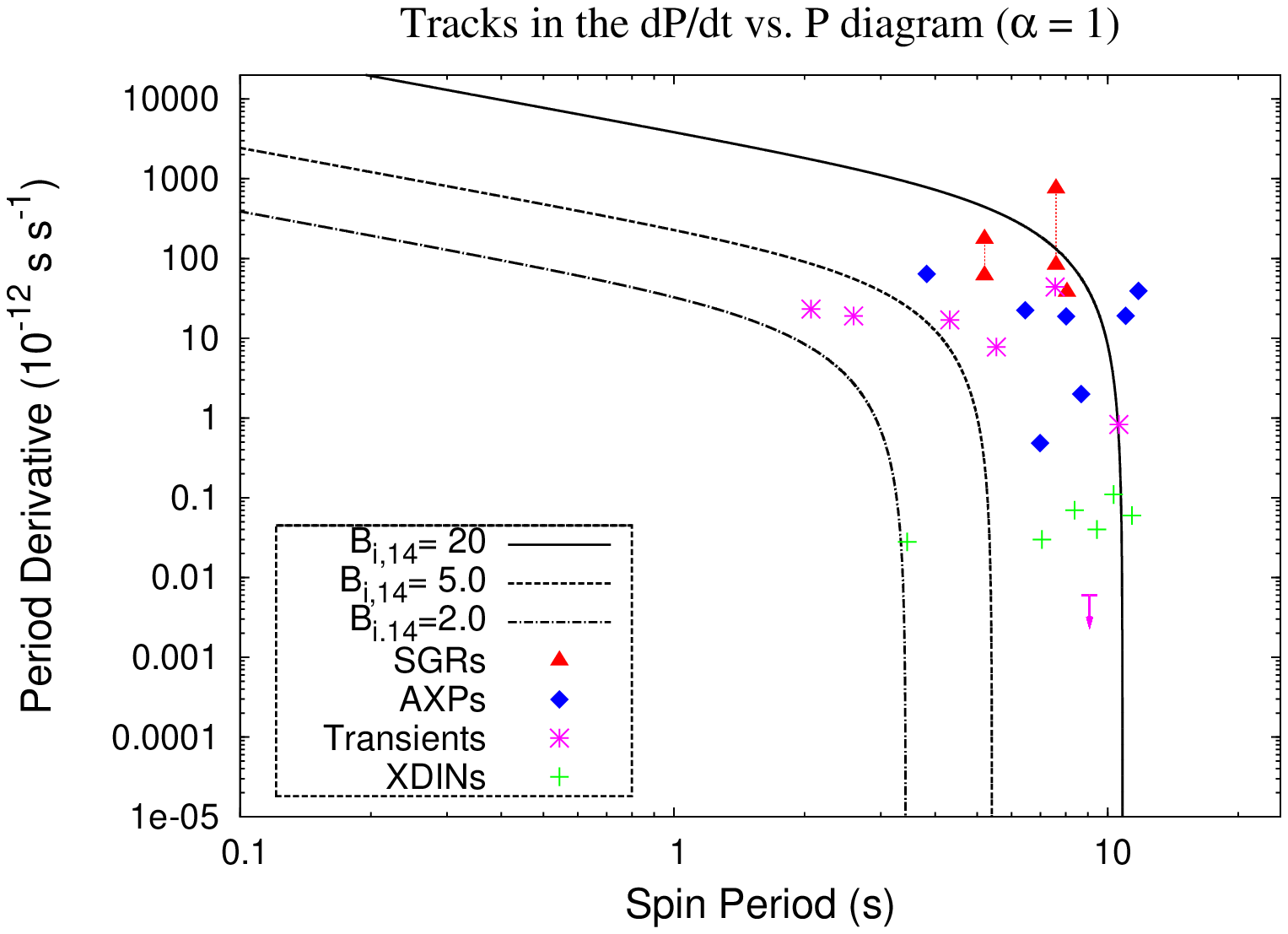}
\leavevmode\includegraphics[width=8.cm, height=6.2cm]{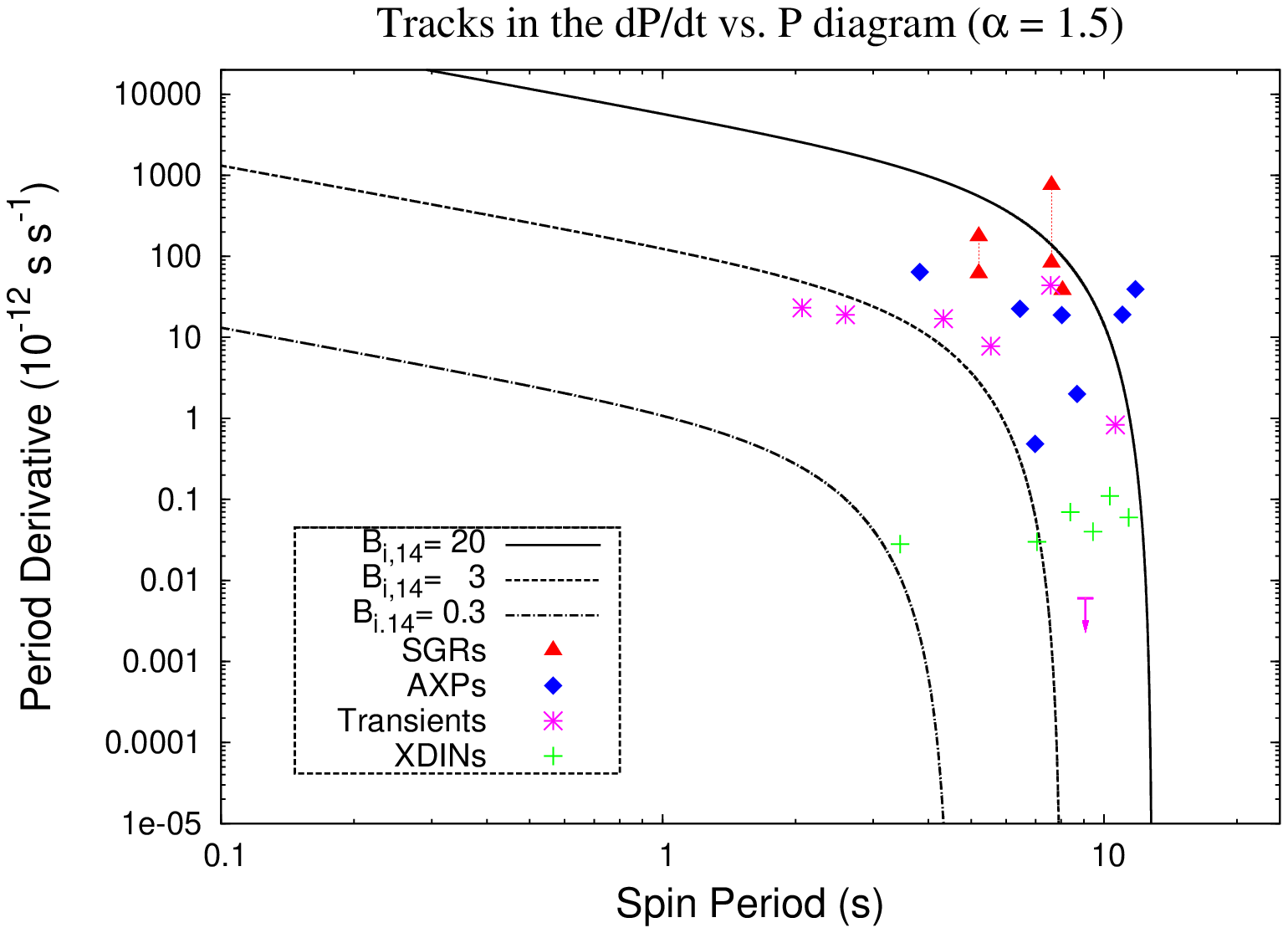}
\leavevmode\includegraphics[width=8.cm, height=6.2cm]{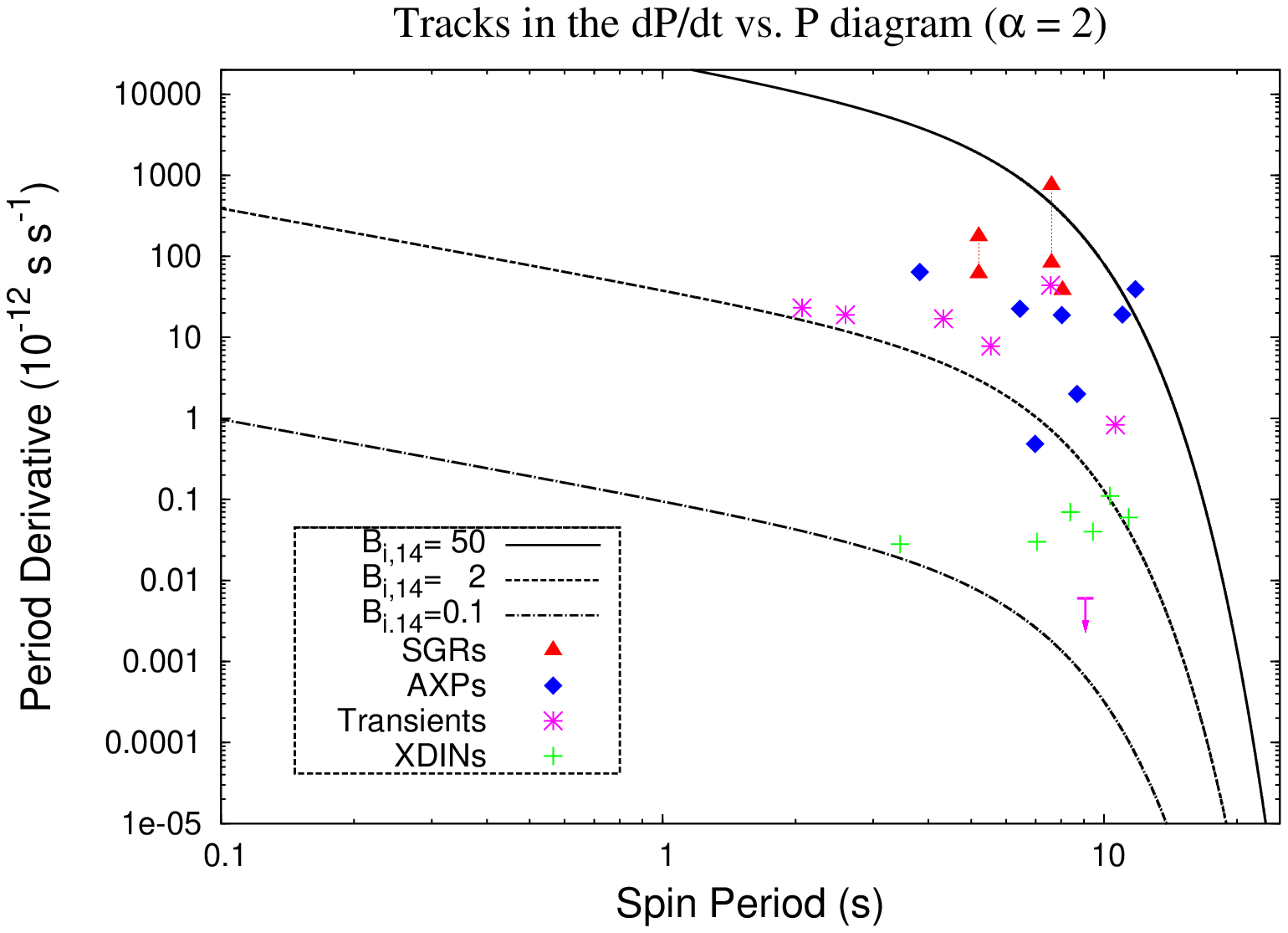}
  \caption{Tracks in the $\dot{P}$ vs. $P$ plane expected for the different 
 cases  of field decay previously discussed, compared to measured source
positions. Plotted curves do not represent fits to data. Different symbols and
lines are as in previous figures.}
  \label{fig:p-pdot-tracks}
\end{center}
\end{figure*}

We summarize the effect of field decay in
Figs. \ref{fig:B-tau-fieldecay-models} and \ref{fig:B-P-fieldecay-models}. We
also show in Fig. \ref{fig:p-pdot-tracks} the corresponding tracks in the
usual $P$-$\dot{P}$  diagram. 
As already stated, previous plots just represent different projections 
of these fundamental measured quantities. 

We can estimate a minimal spread in $B_{\rm dip,i}$ from Fig. \ref{fig:B-tau}
by restricting attention to objects lying well below the asymptotic line, 
so that $\tau_{\rm c} < \tau_{\rm d,i}$. The strongest value of the field is 
$8\times 10^{14}$ G in SGR 1806-20, while the weakest is $1.8 \times 10^{14}$ 
G in SGR 1833-0832. 
These numbers imply
a minimal spread $\approx 4.5$ for the initial field distribution of the whole
population.

\begin{figure*}
  \begin{center}
\leavevmode\includegraphics[width=8.cm, height=6.2cm]{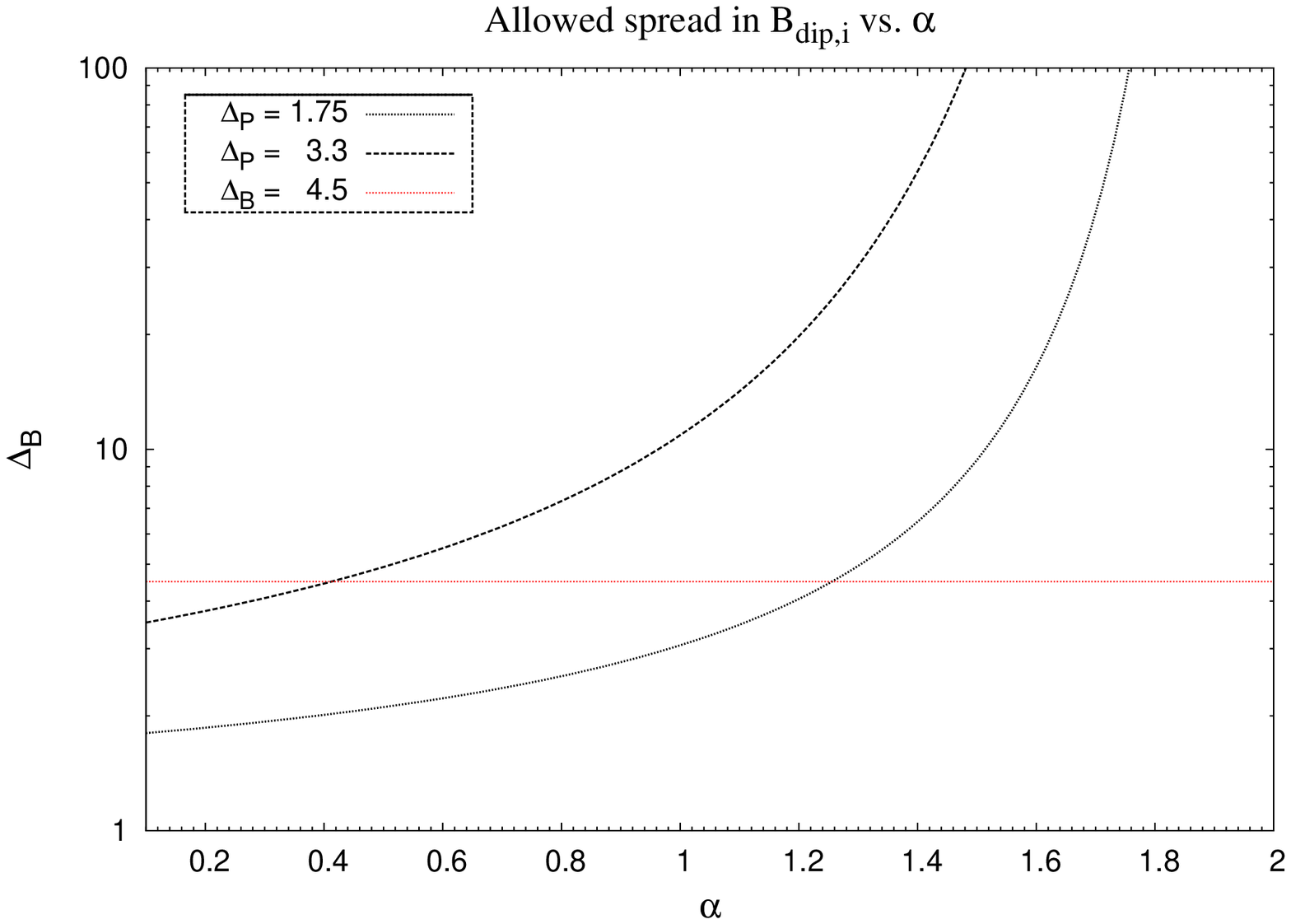}            
\hspace{3mm}
\leavevmode\includegraphics[width=8.cm, height=6.2cm]{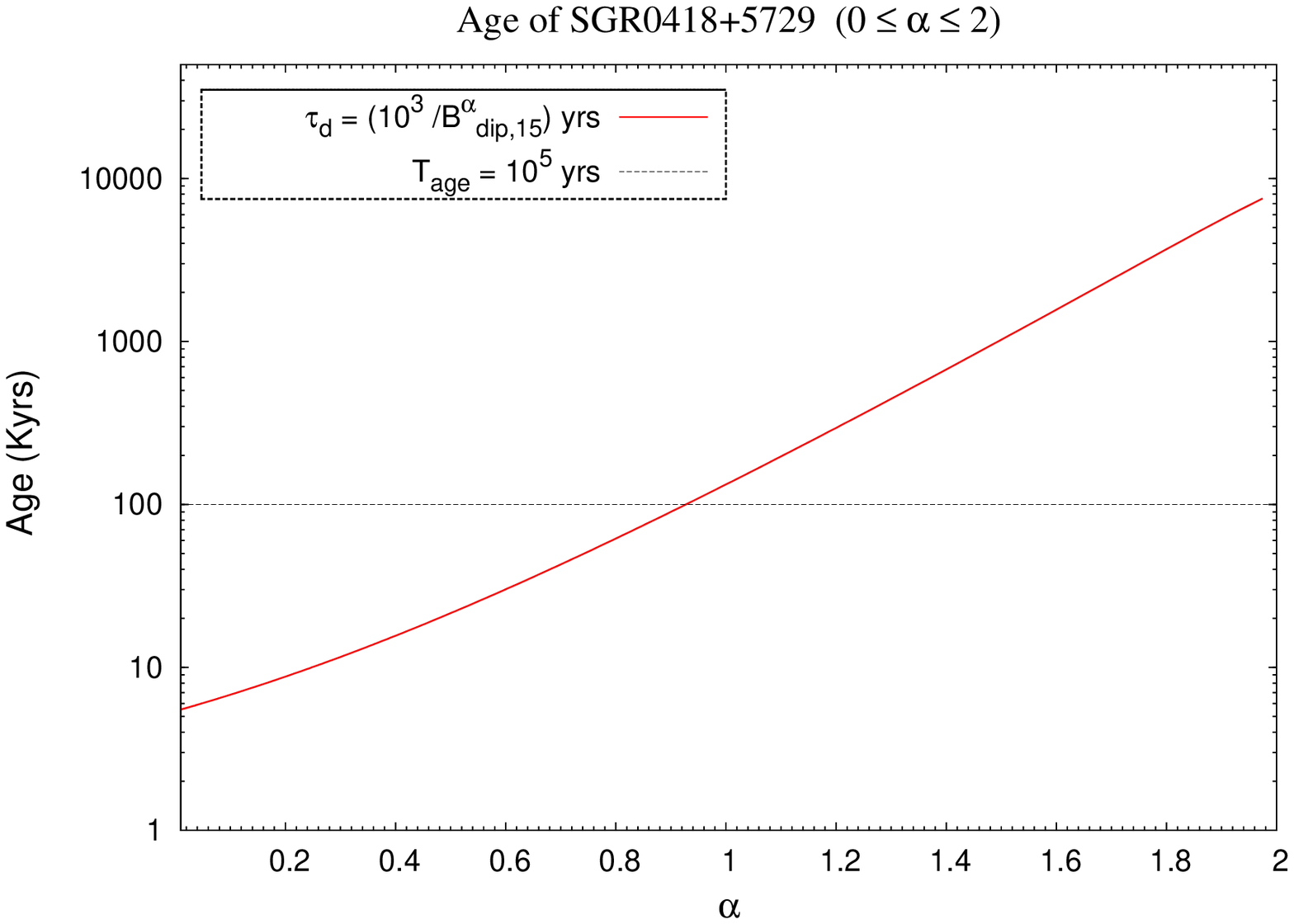}
\hspace{3mm}
  \caption{Constraints on the value of $\alpha$. \textit{Left Panel} : The
    allowed spread in the initial magnetic field distribution as a function of
    $\alpha$, for 3 different values of the spread in asymtpotic spin
    period. \textit{Right Panel} : The inferred age of SRG 0418+5729, as a
    function of the chosen value of $\alpha$. The likely lower limit $\sim
    10^5$ yrs to the age of this source is drawn as a horizontal line.}
  \label{fig:sgrold}
\end{center}
\end{figure*}

For a given value of $\alpha < 2$ (and a fixed normalization $A$),
Eq.~(\ref{eq:def-pinfty}) can be used to derive a general relation
between the observed spread in $P_{\infty}$ at late times and the
initial spread in $B_{\rm dip,i}$ that produced it. For a distribution
of initial dipoles in the range
$B_{\rm i,min} < B_{\rm dip,i} < B_{\rm i,max}$ (or $1 < B_{\rm
dip,i}/B_{\rm i,min} < \Delta_B \equiv B_{\rm i,max}/B_{\rm i,min}$)
a spread $P_{\infty,{\rm min}} < P_\infty < P_{\infty,{\rm max}}$ (or $1 <
P_\infty/P_{\infty,{\rm min}} < \Delta_P \equiv P_{\infty,{\rm
max}}/P_{\infty,{\rm min}}$) is expected. Therefore 
\begin{equation} 
\label{eq:relation-spread}
\Delta_B = \Delta_P^{2/(2-\alpha)} \quad\Longleftrightarrow\quad
\log\Delta_P\equiv\log\fracb{P_{\infty,{\rm max}}}{P_{\infty,{\rm min}}} 
= \frac{2-\alpha}{2}\log\fracb{B_{\rm i,max}}{B_{\rm i,min}} 
\equiv \frac{2-\alpha}{2}\log\Delta_B\ .
\end{equation}
In order to quantify $\Delta_P$ from the data, we evaluate the average
spin period, $\langle P \rangle $, and standard deviation,
$\sigma_{\rm P}$, of those sources which are old enough to be
considered close to the asymptotic line, e.g.  having $\tau_{\rm c}
\geq 10^4$ yrs. The resulting subsample contains 13 sources, with
$\langle P \rangle \approx 8.42$ s and $\sigma_{\rm P} \approx 2.2$ s.
The parent population
would thus be characterized by a $2\sigma$ spread of $\simeq 3.3$, corresponding to the ratio between the
slowest and fastest spinning objects of the subsample.  We plot
relation (\ref{eq:relation-spread}) in the left panel of
Fig. \ref{fig:sgrold}, where the two curves correspond to the
1$\sigma$ and $2\sigma$ values of $\Delta_P$, as derived above.

A totally independent constraint on $\alpha$ can be derived considering
the very weak, thermal, X-ray emission of SGR 0418+5729. In relative
quiescence (on 2010 July 23rd, \citet{Rea+2010}) this is dominated by 
a black-body component with a temperature of $kT = 0.67\pm
0.11\;$keV, and corresponding 0.5$\,$--$\,10\;$keV luminosity, 
$L_{\rm X}\sim 6.2\times 10^{31}\;{\rm erg\;s^{-1}}$ 
In the earlier, more active period (June to November 2009), the temperature gradually
decreased from $kT\approx 1.0\;$keV to $\approx 0.8\;$keV and $L_X$
was between one and two orders of magnitude higher \citep{Esposito+2010}, 
  while the corresponding emitting areas gradually decreased. This suggests 
that even this weak X-ray emission is more likely to originate from a
localized heating event associated to the recent bursting activity, rather
than being powered by the secular cooling of the NS. Hence, the luminosity of 
$6 \times 10^{31}\;{\rm erg\;s^{-1}}$ can be considered as a solid upper 
limit to the quiescent X-ray emission of this object. 
This value compares well with the X-ray luminosity of $\sim 10^5$ yr-old,
passively cooling objects, like B0656+14 \citep{Yakovlev+2004}, while younger $10^3\,$--$\,10^4\;$yr old isolated, 
passively cooling NSs are at least one order of magnitude brighter than the
upper limit for SGR 4018+5729. Since field decay is likely to provide
additional heat in the latter object, the age of B0656+14 represent as a
robust lower limit to the true age of SGR 0418+5729. 

In the right panel of Fig. \ref{fig:sgrold} we plot the estimated age of 
SGR~0418+5729 as a function of $\alpha$, for the whole range of values that
give an asymptotic spin period ($0\leq\alpha<2$). The line corresponding to 
the above lower limit, $10^5\;$yr, is drawn for clarity. Note that as 
$\alpha$ approaches 2 from below the true age of SGR~0418+5729 becomes 
progressively closer to its spindown age. 

 Combining the 
constraints from these two figures, we can  rule out, or at least to consider very unlikely,   values of $\alpha 
\lesssim 1$.  Combined with the earlier constrains this implies 
$1\lesssim \alpha < 2$.

The points discussed above are illustrated quantitatively in 
Fig.~\ref{fig:B-tau-fieldecay-models}, where trajectories in the
$B_{\rm dip}\,$--$\,\tau_c$ plane are plotted for different values of
$\alpha$ and for a given initial distribution of dipole fields,
$B_{\rm dip,i}$, comparing them to the objects in our
sample. Fig.~\ref{fig:B-P-fieldecay-models} shows trajectories in the
$B_{\rm dip}\,$--$\,P$ plane for the same sources, same field decay models
and $B_{\rm dip,i}$ distributions as Fig.~\ref{fig:B-tau-fieldecay-models}. 
This represents an alternative way of illustrating the same argument.
Below we draw some conclusions from these plots regarding the
viability of different $\alpha$ values.

\underline{\textbf{Exponential decay}   ($\alpha = 0$)} can explain 
the observed distribution of sources in the $B_{\rm
dip}\,$--$\,\tau_c$ or $B_{\rm dip}\,$--$\,P$ planes if the field
decay timescale is particularly short, $\tau_{\rm d} = \tau_{\rm d,i}
\approx 1\;$kyr. Longer timescales would fail to halt the spindown
at sufficiently short spin period, so this is quite a strict
requirement.  This model could work if all the sources considered
here came from a very narrow distribution of initial dipole
fields (as can be seen in Figs.~\ref{fig:B-tau-fieldecay-models} and
\ref{fig:B-P-fieldecay-models}). In particular, 
Fig.~\ref{fig:B-P-fieldecay-models} shows that even a factor 4 spread
in $B_{\rm dip,i}$ would lead to a significantly wider distribution of
spin periods than observed (i.e a factor of 4 spread in $P_{\infty}$ compared 
to the observed factor of $\lesssim 3$). Indeed, Eq.~(\ref{eq:def-pinfty})
shows that the value of $P_{\infty}$ is linear in the initial field, $B_{\rm
  d,i}$, for $\alpha=0$.

Given the short timescale required by the exponential decay, the implied
age of SGR0418+5729 would have to be younger than, at most, 
$5\;$kyr, as can be derived from Eq.~(\ref{eq:t_tau_c}). This is in sharp 
contrast with the minimal age that we derived for SGR 0418+5729 based on its 
weak thermal emission. Additionally, the age of SGR 0418+5729 
would not be much larger than that of other transients with much smaller 
$\tau_{\rm c}$, while its X-ray emission would be a factor 10-50 lower. 
Explaining such a fast decrease in $L_{\rm X}$ at these young ages would
also represent a major challenge. Overall, exponential decay 
of the field appears ruled out based on the very weak emission of
SGR 0418+5729.

\underline{\textbf{The case $\alpha = 1$}}  provides a reasonable 
description of the distribution of sources in
Fig.~\ref{fig:B-tau-fieldecay-models} and
Fig.~\ref{fig:B-P-fieldecay-models}. 
Although this case suggests a straightforward relation to the basic
  Hall decay mode (cfr. sec. \ref{summary-fieldecay}), note that the curves in Fig. 
\ref{fig:B-tau-fieldecay-models} and \ref{fig:B-P-fieldecay-models} were drawn
by adjusting the normalization of the decay timescale, $A$, to be 10 times
smaller than the value provided by \cite{Cumming+2004} and GR92. Namely, 
we assumed
\begin{equation}
\label{eq:enhance-ohmic}
\tau_{\rm d} \simeq \frac{10^3} {B_{\rm dip,15}} {\rm yrs}
\end{equation}
The decay for sources whose initial field was larger than $\simeq 2 \times
10^{14}$ G would be to slow with the Hall decay time of  Eq. \ref{eq:cummingetal} 
and in this case most  AXPs/SGRs would evolve to significantly longer spin 
periods at later times. These older counterparts would 
occupy a region right above XDINS and SGR0418+5729 in the $(\tau_c,B)$ plane (Fig. \ref{fig:B-tau})  where no object is actually found.
Note, however, that the timescale of Eq. \ref{eq:cummingetal} could match our
empirical scaling if the decay of Hall modes were regulated by processes 
occurring at somewhat lower crustal densities, $\lesssim 3 \times 10^{13}$ g 
cm$^{-3}$, well above the crust/core interface.

This model for field decay implies that all sources come from a  distribution
of initial dipoles in the range $(0.2 - 2) \times 10^{15}$ G, although
most of them (20 out of 23) were between $5 \times 10^{14}$ G and $2 \times
10^{15}$ G. The XDIN RX J0420 would represent a notable exception, having
reached a remarkably short asymptotic spin period of $\approx 3.45$ s because 
its initial dipole was weaker. The two transients 1E 1547-5408 and 
SGR 1627-41, which are slightly below the range of the other 20 objects, 
would also populate the weak-field tail of the distribution of 
$B_{\rm dip,i}$.

 Overall, this scenario appears to account well for the observed distribution 
of sources, although a full self-consistent   population synthesis model is needed
to verify this quantitatively.

\underline{\textbf{Phenomenological decay law with intermediate value of $1<\alpha<2$}
  (e.g.  $\alpha \approx 1.5$)}. 
Although there are no existing models for field decay predicting $1< \alpha<2$, 
we choose this particular value of $\alpha$ as representative of cases in 
which a wider spread in
$B_{\rm dip,i}$ ($\Delta_B \gg 1$) is allowed, despite the observed 
modest spread in $P_\infty$ ($\Delta_P \approx 3$). 

Fig.~\ref{fig:B-tau-fieldecay-models} shows $B_{\rm dip}$ vs. $\tau_c$
trajectories with  this choice of $\alpha$. One can see that these  
converge to a narrow strip despite an initial wide
distribution of $B_{\rm dip,i}$. This scenario can well reproduce also the 
distribution of sources in the $B_{\rm dip}\,$--$\,P$ plane, including the 
apparent absence of spin periods longer than $\simeq 12$ s. 
Note that the normalization for the decay timescale, $\tau_{\rm
d,i}= 10^3\;$yr  for $B_{\rm dip,i} = 10^{15}\;{\rm G}$,  was chosen
to match the scaling of Eq. \ref{eq:Bdipvstauc}, represented by
the dashed line in Fig. \ref{fig:B-tau}. 

An effective power-law decay with index $\approx 1.5-1.8$
  is expected in the ohmic decay model by \cite{Urpetal+94}. This is not completely equivalent to our phenomenological decay law, though,
  since in that model the ratio $B_{\rm dip}(t)/B_{\rm dip,i}$ is still a universal
  function, determined by the field-independent parameter $\tau_{\rm d,i}$. In
  particular, this implies a linear relation between $P_{\infty}$ and 
 $B_{\rm dip,i}$. Thus, in order for the asymptotic spins of our sources to all
  fall in the observed narrow range, a correspondingly narrow range in ${\rm B}_{\rm
    dip,i}$ is required. This is the same problem as in the
  $\alpha=0$ case. 
To circumvent it, one has to assume that $\tau_{\rm d,i}$, the initial
decay time, is itself  a function of $B_{\rm dip,i}$. This extra assumption would make the
crustal decay model totally equivalent to our phenomenological model. Since
$\tau_{\rm d,i}$ is determined by the initial location of the electric currents
that sustain the field, a strict (anti)correlation between the initial
field strength and the initial depth at which currents flow is implied. This is far from
trivial and an account of its implications is left for future study.
 
\underline{\textbf{The limiting case, $\alpha =2$}} is shown for the 
sake of completeness.  As the previous case, it provides a narrow range of
asymptotic spin periods, even starting from a  very wide distribution of
initial magnetic field values. However, as already discussed, it does not have an actual
asymptotic spin period, which implies that spin periods of order 20$\;$s would be 
expected in older objects. Thus, this case is disfavored by the data compared 
to $\alpha \approx 1.5$.
Also in this case, the normalization for the decay timescale was chosen 
arbitrarily as above.

It is also possible that two different mechanisms for field decay operate, 
respectively  above or below  
some threshold value of the dipole field, $B_*$. 
We refer to these as the early and late mechanism for field decay, according 
to the time when they dominate, and denote their decay index  as $\alpha_e$ 
or $\alpha$, respectively. At each value of $B_{\rm dip}$ the mechanism with 
the shorter decay time $\tau_{\rm d}$ determines the overall field decay and, 
of course, the two times are equal at $B_*$, with  the value $\tau_*$.

In such a scenario, 
a wide distribution of $B_{\rm dip,i}$ at birth can also result in a narrow 
distribution of $P_\infty$. 
As such, it is only meaningful for $\alpha$ much below 2 (say 
$\alpha =1$) and $\alpha_e$ significantly larger than\footnote{If $\alpha_e 
< 2$, it will cause sources to reach the asymptotic spin period before the 
late mechanism becomes operative, thus correspondingly to only one
effective mechanism.} 2. 
Sources with $B_{\rm dip} > B_*$ would initially evolve along a shallow trajectory, $\propto 
(t/\tau_{\rm  d,i})^{-1/\alpha_e}$ in the $B_{\rm dip}\,$--$\,\tau_{\rm c}$
plane. Because $\alpha_e >2$, trajectories for 
different initial fields would converge to a narrow bundle of curves while
$\tau_{\rm c} < \tau_*$. Upon reaching the point where $\tau_{\rm c} =\tau_*$
they would then meet the line 
corresponding to the late mechanism ($\alpha <2$) and follow that beyond 
$\tau_*$. 

Applying this reasoning to  Fig. \ref{fig:B-tau} we derive the values of $B_* \sim 6.9
\times 10^{14}\;$G and $\tau_*\sim 4.5\;$kyr, corresponding to 
the position of the AXP 1E~1841-045. However, only SGR~1806-20 is
above this threshold, implying that the early mechanism could still be dominant only in it.
Stated differently, in all the observed sources - but one at
most - invoking a second, early decay mechanism doesn't help to reduce
the initial spread in $B_{\rm dip,i}$.

\section{The persistent X-ray luminosity}
\label{sec:explaining}

Having assessed the decay of the dipole field in magnetar candidates, we now
turn to test whether such a decay can account for their
observed X-ray luminosity, $L_X$, which typically exceeds their
spin-down power, $|\dot{E}_{\rm rot}|$.

We define the magnetic luminosity of the dipole
field, $L_{\rm B,dip} \equiv 2 E_{\rm dip}/\tau_{\rm d}$, as the available
power of a dipole field decaying on the timescale $\tau_{\rm
d}$. Here, the total energy of the field is $E_{\rm dip} =
(4\pi R_*^3/3) B_{\rm dip}^2/(8\pi) = R_*^3B_{\rm dip}^2/6 \simeq
1.7\times 10^{47} B^2_{\rm dip,15}R^3_{*,6}\;$erg and, following the definition of Eq.~(\ref{eq:B-deriv}):
\begin{equation}
\label{eq:def-lb}
L_{\rm B,dip} = -\frac{dE_{\rm dip}}{dt} =
-\frac{R^3_*}{6}\frac{d(B^2_{\rm dip})}{dt} = \frac{R^3_*B^2_{\rm dip}}{3\tau_{\rm d}} =
\frac{2E_{\rm dip}}{\tau_{\rm d}} \simeq 5.3\times 10^{36}B^2_{\rm
dip,15}R^3_{*,6}\tau_{\rm d,kyr}^{-1}\;{\rm erg\;s^{-1}}\ .
\end{equation}

\begin{figure*}
  \begin{center}
\leavevmode\includegraphics[width=8.cm, height=6.2cm]{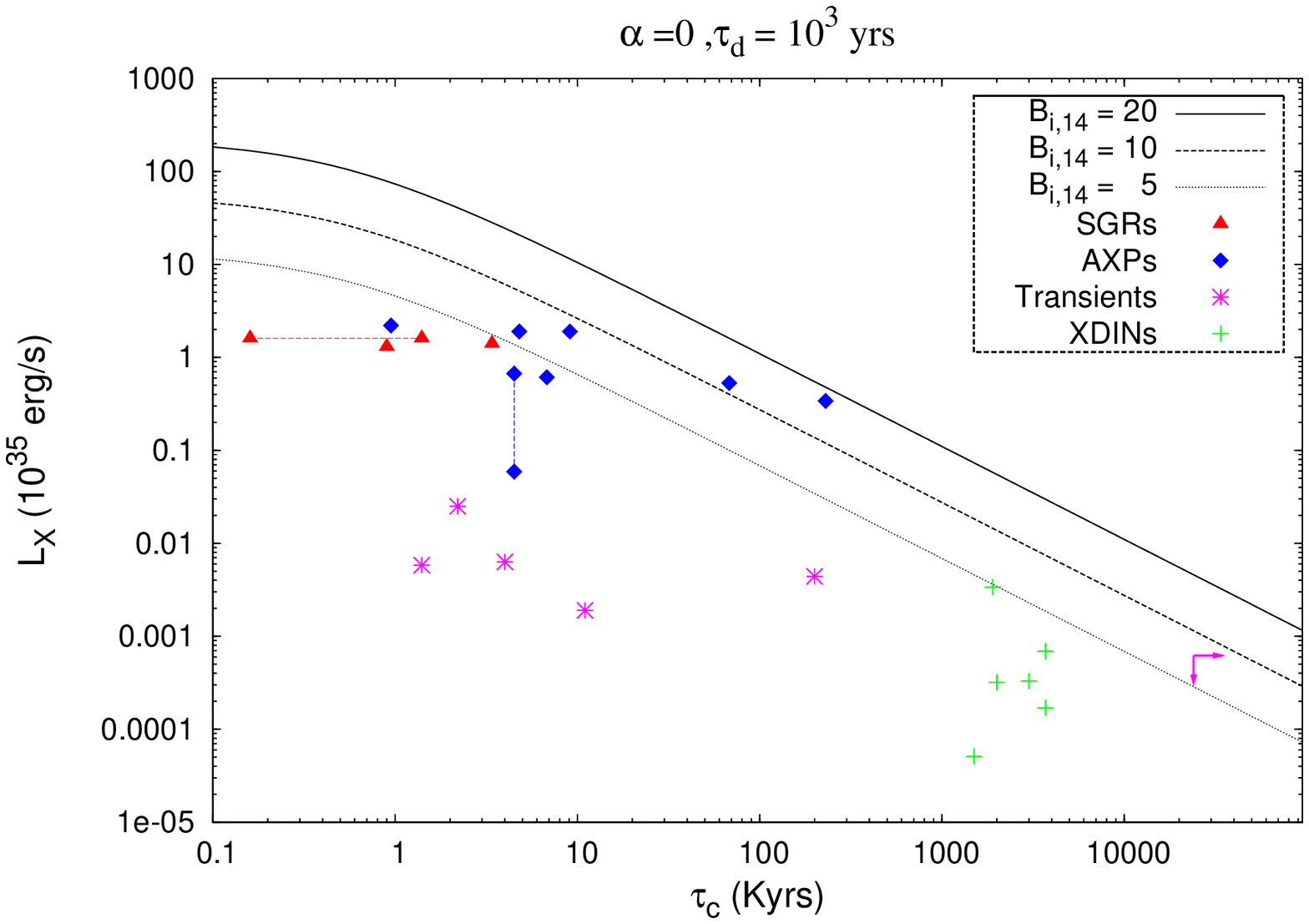}  
\hspace{3mm}
\leavevmode\includegraphics[width=8.cm, height=6.2cm]{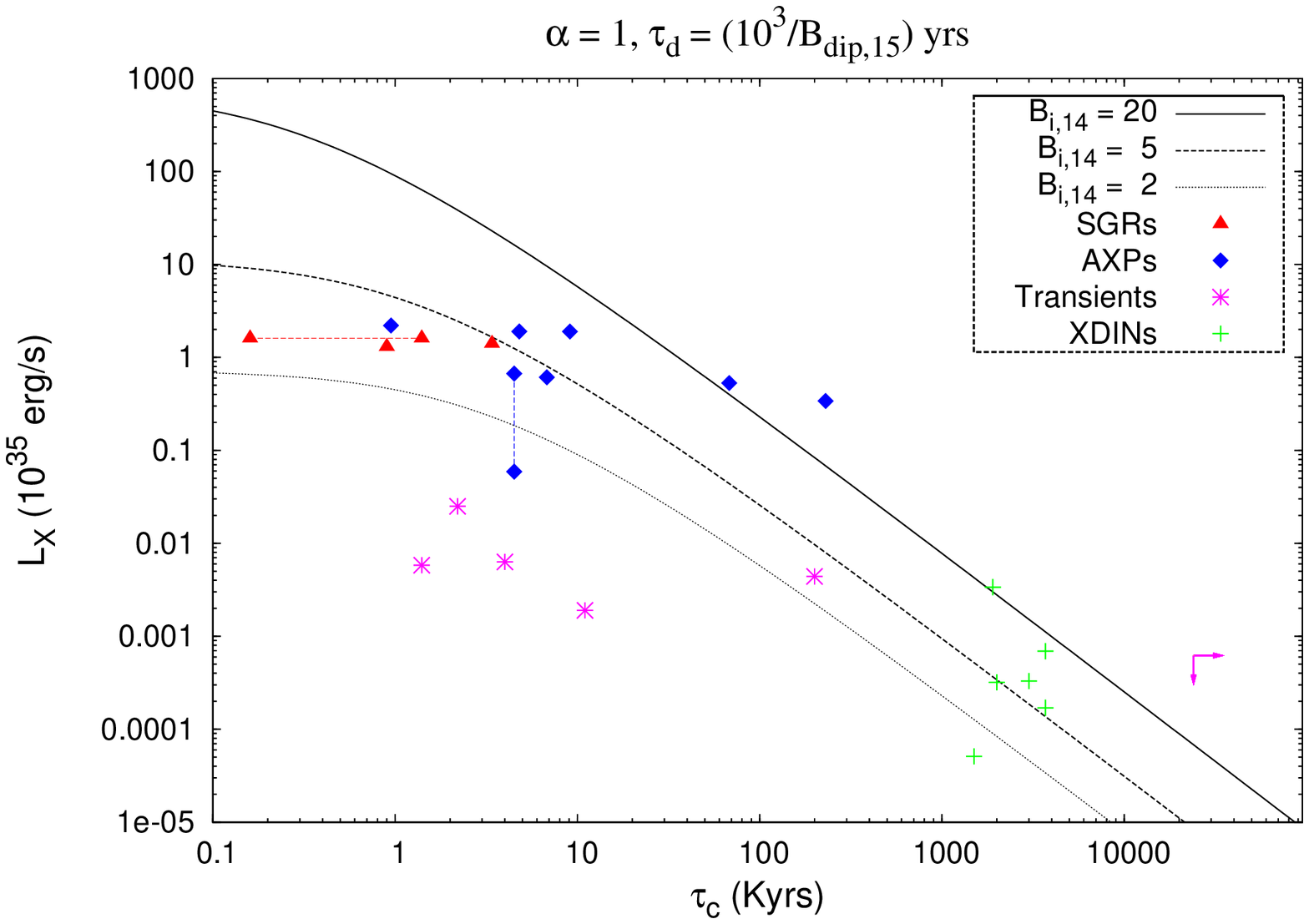}
\hspace{3mm}
\leavevmode\includegraphics[width=8.cm, height=6.2cm]{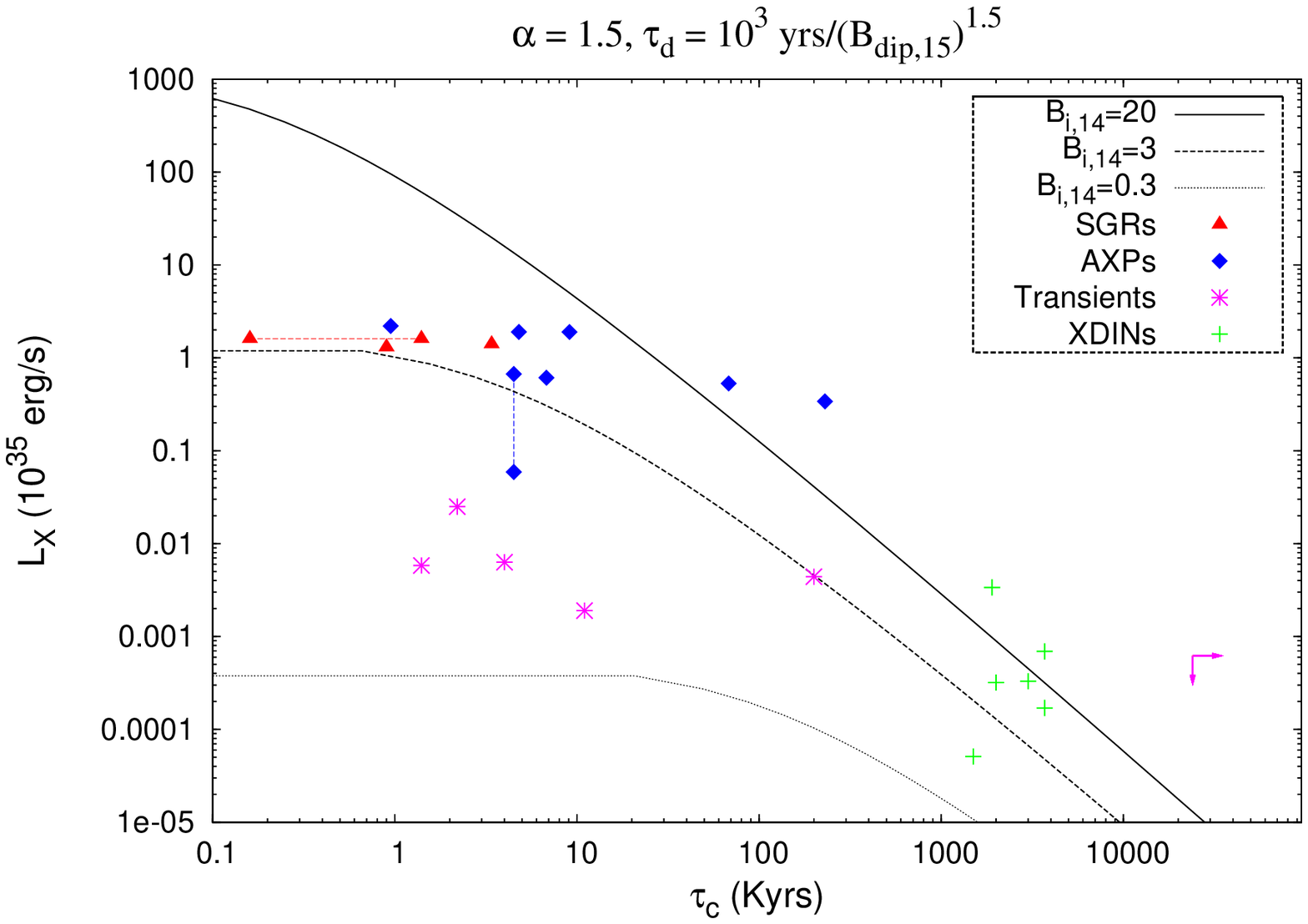}
\hspace{3mm}
\leavevmode\includegraphics[width=8.cm, height=6.2cm]{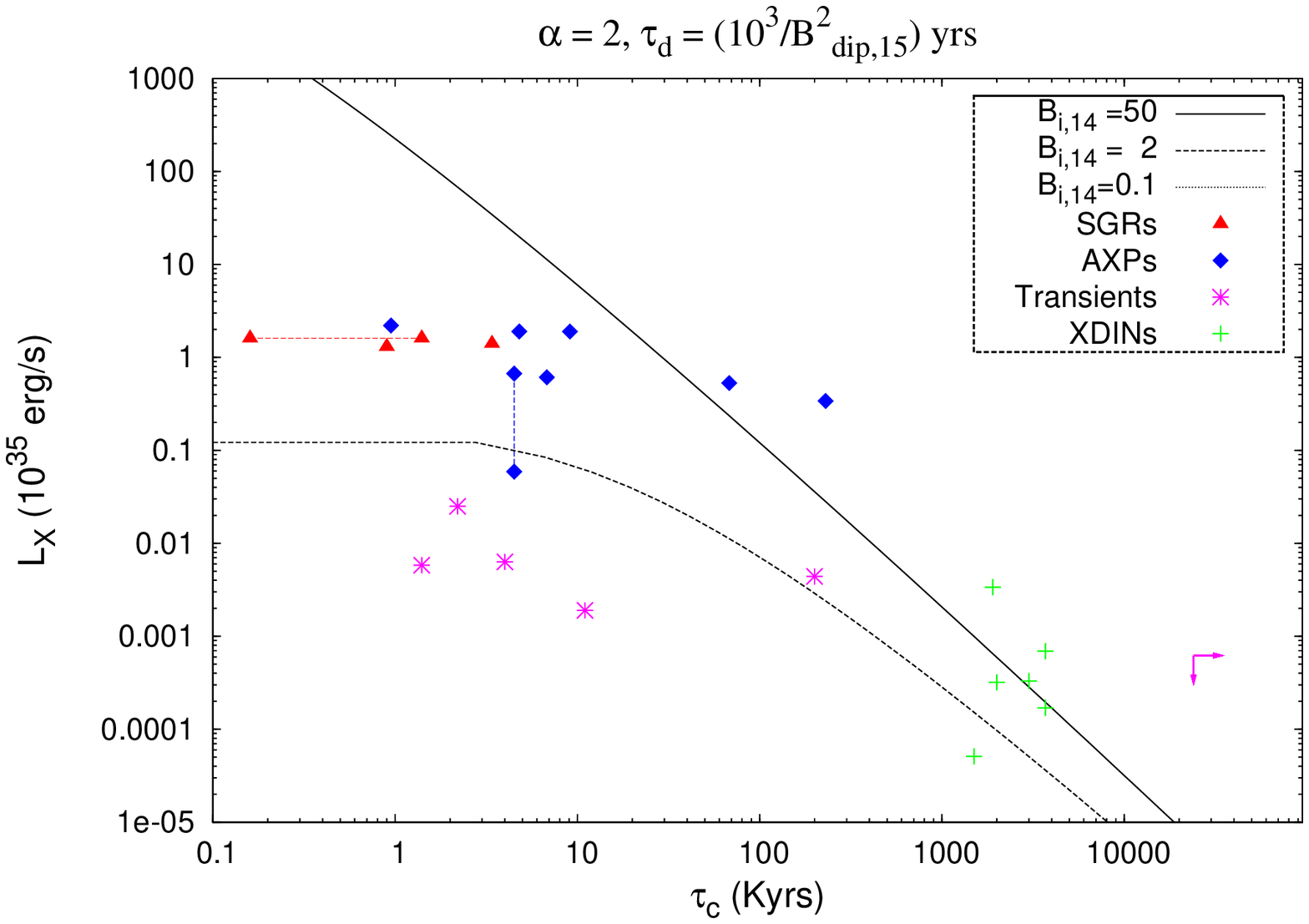}
  \caption{A comparison of the maximal bolometric luminosity as a function of $\tau_{\rm c}$
    expected from the decay of the dipole field, according to the models of
    section 5, with the observed X-ray luminosity of different
    classes of magnetars. The magenta arrows indicate current upper limits to the
    spindown age and $L_{\rm X}$ of SGR 0418+5729.}
  \label{fig:Lx-tau-fieldecay-models}
\end{center}
\end{figure*}

Fig. \ref{fig:Lx-tau-fieldecay-models} depicts a comparision of the 
 total available magnetic power for the
same four field decay models examined in sec. \ref{sec:two}
with the observed X-ray emission of SGRs, AXPs, transients and XDINs.
The three main groups of sources (SGRs/AXPs, transients and XDINs)
populate three separate regions in parameter space and in the following we comment on
them separately. We  focus, first, on the persistent AXPs/SGRs, for
which results and conclusions are much clearer, and comment later on 
the more problematic XDINs and transients.

\subsection{Persistent SGRs, AXPs and SGR 0418+5729}
\label{sec:persistents}
Overall, the evolution of $L_{\rm B,dip}$ with $\tau_{\rm c}$
as derived from the above models for field decay does not match well the distribution
of sources in parameter space. In particular, relatively old sources ($\tau_{\rm c} \geq
10^5$ yrs) tend to be more luminous, in the 2$\,$--$\,$10$\;$keV energy range, 
than the maximum available power from decay of the dipole field {\it namely, } $L_X > L_{\rm
B,dip}$. These old sources thus provide the strongest indication that
even the decay of their dipole field is not 
able to explain the X-ray luminosity.

An exponential decay with  $\alpha = 0$ and with a short $\tau_d$ appears
consistent with all observations.
However, as previously discussed, an exponential field decay on a $\sim 1\;$kyr timescale 
would imply an implausibly young age ($\leq 5\times 10^3$ yrs) for SGR
0418+5729 and would require a very narrow initial magnetic field
  distribution. On these grounds, exponential decay of the field is discarded anyway.

Decay with  $\alpha =1$ falls short of the observed emission of the 
two AXPs with $\tau_{\rm c} \gtrsim 10^5\;$yr. Although the mismatch is apparently 
by a small factor, we stress that curves represent the \textit{total} 
available magnetic power, while data points include only the observed emission
in the 2$\,$--$\,$10$\;$keV energy range. The bolometric 
luminosity is a few times larger. Gravitational redshift also
introduces a non-negligible correction to the observed luminosity (see
sec. \ref{sec:coredecay}). 
Finally, the radiative efficiency may well be less than unity since field decay 
is also expected to dissipate energy in other channels (such as neutrino
emission or bursting activity). The presence of two AXPs beyond the
upper curve must therefore be considered as a significant failure
of this model. 

For $\alpha=1$, the observed X- ray luminosity of SGR 0418+5729 is 30 times
larger than the available power from dipole field
decay, according to Eq. \ref{eq:def-lb}.
Although, as discussed in section \S \ref{sec:source-classes}, this  may overestimate 
the quiescent power, we don't expect the  average power
output to be lower by more than a factor of 10 than that and,  even with this
correction, the dipole energy is still insufficient.

For values of $\alpha >1$ the mismatch between curves and observations is 
striking. 
The distribution of persistent sources in the $L_{\rm X}\,$--$\,\tau_{\rm c}$ plane is at best
marginally consistent with being powered by the decay of the dipole field, if $\alpha =1$, while
it is totally inconsistent for larger values of $\alpha$. In general, the X-ray
luminosity appears to decay on a longer timescale, and to be associated with a 
larger energy reservoir, than can be provided by the decaying dipole fields.
Sources are indeed found at a fairly constant level of X-ray luminosity up 
to $\tau_{\rm c} \sim 10\;$kyr. The few older objects known clearly show a 
decline of $L_{\rm X}$ with (spindown) age which, however, is 
flatter than expected from the decay of the dipole field.

In the framework of the magnetar model, the most natural explanation
for these findings is that the persistent X-ray emission of these
sources is powered by the decay of an even larger field component,
stored in their interior
In the next section we try to assess the location of such a (presumably magnetic) energy reservoir and its
most likely decay modes.

\subsection{X-ray Dim Isolated Neutron Stars}
\label{sec:XDINs}
The thermal X-ray luminosity of XDINs is much dimmer than that of
persistent magnetar candidates. It is even possible, in principle, to
explain it within the so-called ``minimal cooling'' scenario for
passively cooling NSs. However, it is difficult to reconcile the
relatively bright emission of RXJ 0720 with its apparent old age
($\sim 2\times 10^6\;$yr) within that scenario. This suggests a
possible role of strong internal heating in this object \citep{page+2004}. On the other hand, the possibility of mild internal heating in most XDINs
has also been recently suggested, based on a tendency for their effective
temperatures to be higher than that of normal pulsars with the same age
\citep{Kaplan+2011}.

From Fig. \ref{fig:B-tau} we conclude that XDINs should be interpreted
as objects whose dipole field has decayed substantially, hence
their true ages must be younger than their spin-down ages, which are
narrowly clustered around $\tau_{\rm c} \sim {~\rm a~few~} \times
10^6\;$yr.
This favors passive cooling models in accounting for the effective
temperatures and X-ray luminosities of XDINs, possibly removing the need for 
internal heating in all these sources. On the other hand, for $\alpha =1$ the energy
 released by the dipole field decay is
not negligible compared with their X-ray luminosities (see  Fig. \ref{fig:Lx-tau-fieldecay-models}). 
Whether this can significantly affect their temperatures
as compared to normal pulsars can only be checked through detailed cooling modeling,
which are beyond the scope of this work. 

If, however, detailed modeling will reveal that these luminosities are 
too large for passive cooling then we have to consider  heating sources. 
 RXJ 0720 stands here as a unique object. Even with $\alpha=1$, 
 Eq. \ref{eq:def-lb} yields $L_{\rm B,dip} \approx  8 \times 10^{31} ~{\rm  erg~s}^{-1}$ , which 
is a factor $\sim 5$ lower than its measured $L_X$.   This  could
not be explained by $L_{\rm B,dip}$ and would thus hint to the presence of an 
additional energy reservoir, siimilar to magnetar candidates. In this case, some level of bursting activity would accordingly 
be expected from this source.
For $\alpha >1$, the available $L_{\rm B,dip}$ is insufficient even to account for 
the X-ray luminosity of RXJ 2143 and  it 
is marginal compared  with other XDINs. 

We conclude that the decay of their dipole field implies that XDINs are younger than
their spindown age, $\tau_{\rm c}$, which could improve the match between cooling models 
and their X-ray emission properties. On the other hand, there is no compelling 
evidence for a dominant contribution to their X-ray emission from the 
decay of the dipole field. The case $\alpha=1$ is, at most, marginally consistent 
with this hypothesis while, as $\alpha$ becomes $>1$, 
it is increasingly hard for $L_{\rm B,dip}$ to match the observed $L_{\rm X}$.

We can use the observed emission properties of XDINs to further constrain
the possible values of $\alpha$. It was recently shown that the spindown ages 
of XDINs are systematically longer, by a factor $\sim$ 10, than the ages of NSs 
with similar effective temperatures \citep{Kaplan+2011}. The simplest 
intepretation of this fact is that, due to dipole field deacy, 
their $\tau_{\rm c}$ overestimate their true ages by approximately one order
of magnitude. We can compare quantitatively this statement with models for
decay of the dipole field. For a given value of $\alpha$, the
formulae of \S~\ref{sec:two} allow to derive 
the value of $B_{\rm dip,i}$ corresponding to each individual XDIN
 from its $\tau_{\rm c}$ and $B_{\rm dip}$ (or equivalently, $P$
and $\dot{P}$) and, from its $\tau_{\rm c}$, obtain its real age, $t_{\rm
age}$. These estimates, for the cases $\alpha=1$ and
$\alpha =1.5$, are summarized in  Table~\ref{tab:one}. For the sake of completeness, we report
in the last column of this table the effective temperature, $kT_{\rm  eff}$, for each
source, as derived by the spectral fits. 
\newcommand{\rb}[1]{\raisebox{1.5ex}[0pt]{#1}}
\begin{table}
\begin{center}
{\small
\begin{tabular}{|c|c|c|c|c}
\hline 
          & $\tau_{\rm c}$ [kyr]  & $t_{\rm age}$ [kyr]  & $t_{\rm age}$  [kyr]  & $ kT_{\rm eff}$  [eV] \\
\rb{Name} & & ($\alpha = 1$) & ($\alpha=1.5$) &  \\
\hline
\hline
RXJ 1308  &  1500 &  30  &  100  &  102 \\ \hline
RXJ 0806  &  3000 &  40  &  160  &  96  \\ \hline
RXJ 0720  &  1900 &  42  &  160  &  90  \\ \hline
RXJ 2143  &  3700 &  50  &  200  &  100  \\ \hline
RXJ 1856  &  3700 &  66  &  350  &  63  \\ \hline
RXJ 0420  &  2000 &  100 &  580  &  44  \\
\hline
\end{tabular}}
\end{center}
\caption{Spindown age, true age (for two different values of the decay
  index, $\alpha$) and effetive temperature of the six XDINs considered in
  this work.}
\label{tab:one}
\end{table}
Note that the derived age distributions match fairly well the
distribution of effective temperatures, as opposed to the $\tau_{\rm
c}$ distribution.  It is clear, however, that low $\alpha$
values give  corrections to the spindown ages of
XDINs by a factor $\lesssim$ a hundred, which it too large.
The case $\alpha =1.5$, on the other hand, gives just the right correction factors
of order ten.

Finally we note the striking clustering of XDINs at $\tau_{\rm c} \in [1.5
  - 3.7]$ Myr.  Together with their comparatively 
wider spread in X-ray luminosities, $[\lesssim 10^{31} - 3 \times 10^{32}]$ 
erg s$^{-1}$, it suggests that these sources may have reached some threshold
age, at which cooling becomes very efficient,  the luminosity drops  sharply and the  sources become  
undetectable. This can happen, for example ,if  the sources enter the photon-cooling dominated regime 
(\cite{page+2004} and references therein). In this framework, the actual age
of XDINs should be just slightly larger than the time at which the transition 
to photon-cooling occurs. 

From Tab. \ref{tab:one} we see that, for $\alpha=1$, all XDINs would have 
ages $\leq 10^5$ yr, most of them being significantly younger than that. In this case, 
a transition to photon-cooling at $t_{\rm  age} < 3\times 10^4$ yrs would have
to be invoked, however, in most ``standard'' cooling models this transition
occurs at $\gtrsim 10^5\;$yr
\citep{Yakovlev+2004, page+2004}. Only the fastest cooling models have this transition
at $\sim 3\times 10^{4}\;$yr \citep{page+2004, page+2011} and are
thus just marginally consistent. Even in this case, however, the weak X-ray 
emission of XDINs at such young ages would represent by itself a major
problem. No cooling model predicts X-ray luminosities $< 10^{32}$ erg s$^{-1}$
at ages $t_{\rm age} < 10^5$ yrs. As a conclusion, the case $\alpha =1 $ is 
clearly very problematic from this point of view.

For $\alpha =1.5$, on the other hand, all ages are $\sim [1 - {\rm a~few}] \times 10^5$
yrs. This  allows an easier interpretation of the XDINs as cooling NSs having 
just passed the transition to photon-cooling, the older ones being
progressively cooler and dimmer.

The observed properties of XDINs thus strongly argue against a value
of $\alpha =1$ and are much more consistent with $1.5\lesssim
\alpha < 2$.
In particular, the value $\alpha \approx 1.5$ provides an overall good
agreement with the main observed properties of  these sources.
Altogether, taking into account the requirement of $\alpha$
sufficiently below $2$ to account for lack of periods well above 10~s,
we conclude that $1.5\lesssim\alpha\lesssim 1.8$ is strongly favored
by the data.

\subsection{Transients}
\label{sec:transients}
Transient AXPs are the most enigmatic objects. Most of them appear to
be young sources whose field has not significantly decayed yet and their
weak quiescent emission testifies of an extremely low efficiency in converting 
magnetic energy into X-rays. This could be related to magnetic dissipation 
(heat release) occurring only deep in the NS core, involving only a fraction 
of the whole volume and
resulting in large neutrino energy losses and a low efficiency for the
X-ray emission. As opposed to this, magnetic dissipation would be more 
distributed, or would occur closer to the NS surface, in persistent sources, 
thus reducing neutrino losses and leading to a higher efficiency of the X-ray emission.
However, during outbursts transients become temporarily very similar to 
the persistent sources and get much closer to them in the $L_{\rm
  X}\,$--$\,\tau_{\rm c}$ plane. The origin of this behaviour is still
unclear and we will not discuss it further.
%


\section{Decay of the \textit{internal} magnetic field}
\label{sec:internal-field}

The presence of internal fields larger than the dipole components in
magnetars has been considered as a likely theoretical possibility
since the suggestion made by TD96.   In the previous sections we
have provided, for the first time, evidence based on the observed
properties of persistent SGRs/AXPs that such a component must exist,
if magnetic energy is indeed powering their X-ray emission.

The decay of the internal field is even harder to constrain than the
dipole, as the only observational guidance is the  evolution of $L_{\rm X}$
and  the level of bursting activity (the latter being more qualitative in
nature as it is harder to accurately quantify it). The X-ray
  luminosity depends on several physical details of the NS structure other
than the properties of field decay. To keep the focus on the salient effect,
we  adopt a different approach here and, instead of dealing with general 
phenomenological decay models, we will calculate the expected evolution of 
$L_{\rm X}$ with $\tau_{\rm c}$ adopting two simple and general prescriptions 
from selected, physically-motivated models of field decay in NS interiors.

There are two main possible locations for the decay of the internal field. 
It could either take place in the liquid core of the NS, at $ 10^{14} \lesssim
  \rho < 10^{15}$ g cm$^{-3}$
  (TD96, \citealt{Heyl+1998, TD01, Colpi+2000, Arras+2004, DSS09}), or in the 
rigid lattice
  of the inner crust, at $\rho \lesssim 10^{14}$ g cm$^{-3}$ \citep{Vainshtein+2000,
  Konenk+2001, Arras+2004, Pons+2007, Pons+2009}. In either case, heat 
released locally
  by field dissipation is subsequently conducted to the surface,  thereby
 powering the  enhanced X-ray emisson. Note that energy release may also take
 place at the NS surface, or just below it ($\rho \lesssim 10^{10}$ g cm$^{-3}$, \citealt{Kamink+2009}, \textit{e.g.} due to a gradual dissipation of electrical currents
  in a global/localized magnetospheric twist \citep{TLK+2002, Belobod+2007, Belobod+2009}. This would
likely allow a larger radiative efficiency. However, the twist and 
associated radiation would still draw their energy from that of an
evolving, strongly twisted internal field, either in the deep crust or 
 core.

As far as field decay in the liquid core is concerned, the state of matter
there plays an important role. If it is normal $npe$ matter, as opposed to
superfluid, then ambipolar diffusion is expected to be the dominant channel 
through which the magnetic field decays. If, instead, either protons or
neutrons (or both species) were in a condensed state, particle interactions
may be significantly affected, reducing or possibly quenching 
the mode (GR92, \citealt{TD96, Jones+2006, Glampe+2011}). 
We do not discuss the role of core condensation. Here we note that, even if ambipolar diffusion were completely 
quenched by core condensation, field decay in the crust, driven by the Hall
effect, would still continue unaffected by the changing conditions in the core
(\cite{Arras+2004} give a quantitative account of this). 

Therefore, with the aim of illustrating just the salient effects on the X-ray luminosity
of the decay of a strong internal field, we will consider here only the two 
limiting cases: ambipolar diffusion in a normal core of $npe$ matter, or
Hall-driven field decay in the inner crust, with a magnetically inactive
core. The latter could be considered as a \textit{``minimal heating scenario''} for magnetars. 
A study of realistic models, which include the contribution to field dissipation
of hydromagnetic instabilities \citep{Arras+2004} and magnetospheric currents 
\citep{TLK+2002, Belobod+2007}, along with 
the effects of strong crustal magnetic fields and light-element
envelopes/atmospheres on radiative transfer, is clearly beyond the scope of
this work.

\subsection{Field decay in the NS core}
\label{sec:coredecay}
We consider evolution of the internal field ($B_{\rm int}$) through the 
solenoidal mode only\footnote{The irrotational mode evolves too slow to be of
  interest in the low-T regime and is very likely quenched by core condensation.}.
 In addition to previous treatments
\citep{Heyl+1998, Colpi+2000, Arras+2004}, we allow
explicitly for a different decay law for the dipole field, according to our 
conclusions of sec. \ref{sec:two} and sec. \ref{sec:explaining}. 
In our picture the decay of $B_{\rm int}$ heats the core and 
powers the surface X-ray emission, which will then decline following the 
decrease of the internal field. The decay of the dipole field will, on the 
other hand, determine the relation between real time, $t$, and  spindown age, 
$\tau_{\rm c}$ (see Fig. \ref{fig:summary}). We restrict attention to the two more realistic 
cases, $\alpha = 1$ or $1.5$, as emerged in previous sections. 

For a core of normal $npe$ matter the decay time for the solenoidal mode of 
ambipolar diffusion is $\tau_{\rm d,int} \approx 10^4/B^{6/5}_{\rm int,16}$ 
yrs. The total available magnetic luminosity is then, according to Eq. \ref{eq:def-lb},  
$L_{\rm B,int} = (R^3_*/3) B^2_{\rm int}/
\tau_{\rm d,int} \approx 10^{38} B^{16/5}_{\rm int,16} R^3_{*,6} 
{\rm erg\ s}^{-1}$.

As discussed in sec. \ref{summary-fieldecay}, the equilibrium between heating and
neutrino cooling determines the temperature as a function of\footnote{We consider only
modified URCA processes in the NS core. Note also that the equilibrium temperature
in this case also depends on density, strictly speaking. However, we will carry out calculations at a
fixed $\rho = 7 \times 10^{14}$ g cm$^{-3}$, the average density of a 1.4
  $M_{\odot}$ NS with 10 Km radius, and focus only on the B-dependence.}
$B_{\rm int}$, when ambipolar diffusion is active, as expressed by Eq. \ref{eq:TvsB-relation}.
The evolution of the core temperature, $T_{\rm c}$, will thus track directly
that of $B_{\rm int}$.

Finally, an appropriate relation between the
core and surface temperatures is needed in order to calculate the expected 
surface X-ray emission.
This is the so-called $T_{\rm b} - T_{\rm s}$ relation, where $T_{\rm b}$
is the temperature at the base of the crust, the core being isothermal due to 
its large density and heat conduction. We adopt the
minimal scaling for an unmagnetized, Fe envelope (Potekhin \& Yakovlev 2001)
\begin{equation}
\label{eq:tb-ts-simple}
 T_{\rm s} \simeq 1.17 \times 10^6 ~{\rm K} ~\left[(7 \zeta)^{9/4} +
   \left(\frac{\zeta}{3}\right)^{5/4}\right]^{1/4}
\end{equation}
where
$\zeta \equiv T_{\rm c,9}-0.001~g^{1/4}_{14} (7~T_{\rm c,9})^{1/2}$ and $g_{14}
\approx 1.87$ is the surface gravity.

The above relation is known to be sensitive to the strength and topology of
crustal magnetic fields \citep{page+2007,Kamink+2009, Pons+2009} and also to 
the chemical composition of the outer layers of the crust. 
We comment later on these issues and their possible relevance for our calculations.

The surface luminosity,  $L_{\rm X} = 4\pi R^2_* \sigma_{\rm sb} T^4_{\rm s}$,
is eventually obtained from Eq. \ref{eq:tb-ts-simple}. However, 
due to general relativistic corrections, an observer at infinity will measure 
the luminosity (Page et al. 2007 and references therein)
\begin{equation}
\label{eq:lumix-Bint}
L_{\rm X,\infty} = \left(1 -\frac{2G M_*}{R_* c^2}\right) L_{\rm X}
\approx 10^{33}~{\rm erg ~s}^{-1}~
\left[(7 \zeta)^{9/4} +
   \left(\frac{\zeta}{3}\right)^{5/4}\right]
\end{equation}
which is the quantity we will compare with observations. 
%
The term in square parenthesis  contains the dependence
on $B_{\rm int}$ through the $T_{\rm c}$-dependence of $\zeta$. 

Eq. \ref{eq:lumix-Bint} implies that fields in the $10^{16}$ G range are strictly
required 
to approach or even exceed $\sim 10^{35}$ erg s$^{-1}$, as observed in the 
youngest ($\tau_{\rm c}\leq 10^4 $ yrs) sources. Note, however, that the total available power, $L_{\rm
  B,int}$, would be 2-3 orders of magnitude larger during this early stage. Indeed, 
when magnetic energy is released in the core, most of it is carried away by
neutrinos resulting in a very low efficiency of X-ray radiation. This can be 
estimated to be
%
$\epsilon_{\rm X} \equiv L_{\rm X}/L_{\rm B,int} \approx 8.3 \times
10^{-4} B^{-19/9}_{16}$

As the field decays, the equilibrium temperature in the core drops
accordingly, the efficiency of neutrino emission decreases and $\epsilon_{\rm
  X}$ grows. When it becomes close to unity, the NS thermal evolution 
becomes dominated by photon cooling and the equilibrium condition leading to 
Eq. \ref{eq:TvsB-relation} does not hold anymore.
To treat this transition self-consistently we follow the evolution of $B_{\rm
  int}$ (and thus $T_{\rm c}$ and $L_{\rm X}$) with time $t$ from our chosen initial 
conditions. For a given choice of $\alpha$ and $B_{\rm dip,i}$, this is
also known as a function of $\tau_{\rm c}$. The temperature at
which photon cooling becomes dominant, $T_*$, is defined by requiring that 
$\epsilon_{\rm X} \geq 1/2$, which is of course 
equivalent to finding where $L_{\rm  X}$ first equals the
neutrino luminosity. We denote by $L_{\rm X,*}$, $\tau_{\rm d,*}$, $B_{\rm
  int,*}$ and $t_*$ the quantities at this transition point. From here on, 
the heat released by the decay of the internal field will have to balance
the energy lost to photon emission from the NS surface. That this equilibrium 
can be maintained is implied by the scaling $L_{\rm X} \propto T^{20/9}_{\rm
  c}$, while the total magnetic power is $L_{\rm B,int} \propto T^2_{\rm
  c}$. Hence, the latter decays (slightly) slower than the former as heat is lost and
$T_{\rm c}$ decreases.
 
The new equilibrium relation beyond $T_*$  thus reads
%
\begin{equation}
\label{eq:photon-cooling-equlibrium}
T_{\rm c} \approx 2 \left(\frac{B_{\rm int,15}}{0.5}\right)^{18/19}
  \left(\frac{\rho_{15}}{0.7}\right)^{3/19}~~,
\end{equation}
with which we can eventually express the field decay time as a function
of $B_{\rm int}$ 
\begin{equation}
\label{eq:decaytime-late}
\tau^{{\rm(ph)}}_{\rm d} \approx 
3.27 \times 10^5 ~{\rm yrs} ~
B^{-2/19}_{\rm int,15}
\left(\frac{\rho_{15}}{0.7}\right)^{20/57}
\end{equation}
showing that field decay becomes much faster now, with an effective $\alpha_{\rm int,ph} = 2/19$.

Solutions for $B_{\rm int}$, $T_{\rm c}$ and $L_{\rm X}$ as a function of real
time are written straightforwardly 
\begin{eqnarray}
\label{eq:solutions-photon-cooling}
B_{\rm int} (t - t_*) & = & \frac{B_{\rm int,*}}{\left(1+ \frac{2}{19}
  \frac{t-t_*}{\tau_{\rm d,*}}\right)^{19/2}} \nonumber \\
L_{\rm X,\infty} (t-t_*) & = & \frac{L_{\rm *,\infty}}{\left(1+\frac{2}{19}
  \frac{t-t_*}{\tau_{\rm d,*}}\right)^{20}}
\end{eqnarray}
and, for a given choice of $\alpha$ and $B_{\rm dip,i}$, can be plotted versus 
the corresponding values of $\tau_{\rm c}$. 
%
\begin{figure*}
  \begin{center}
\leavevmode\includegraphics[width=8.cm, height=6.2cm]{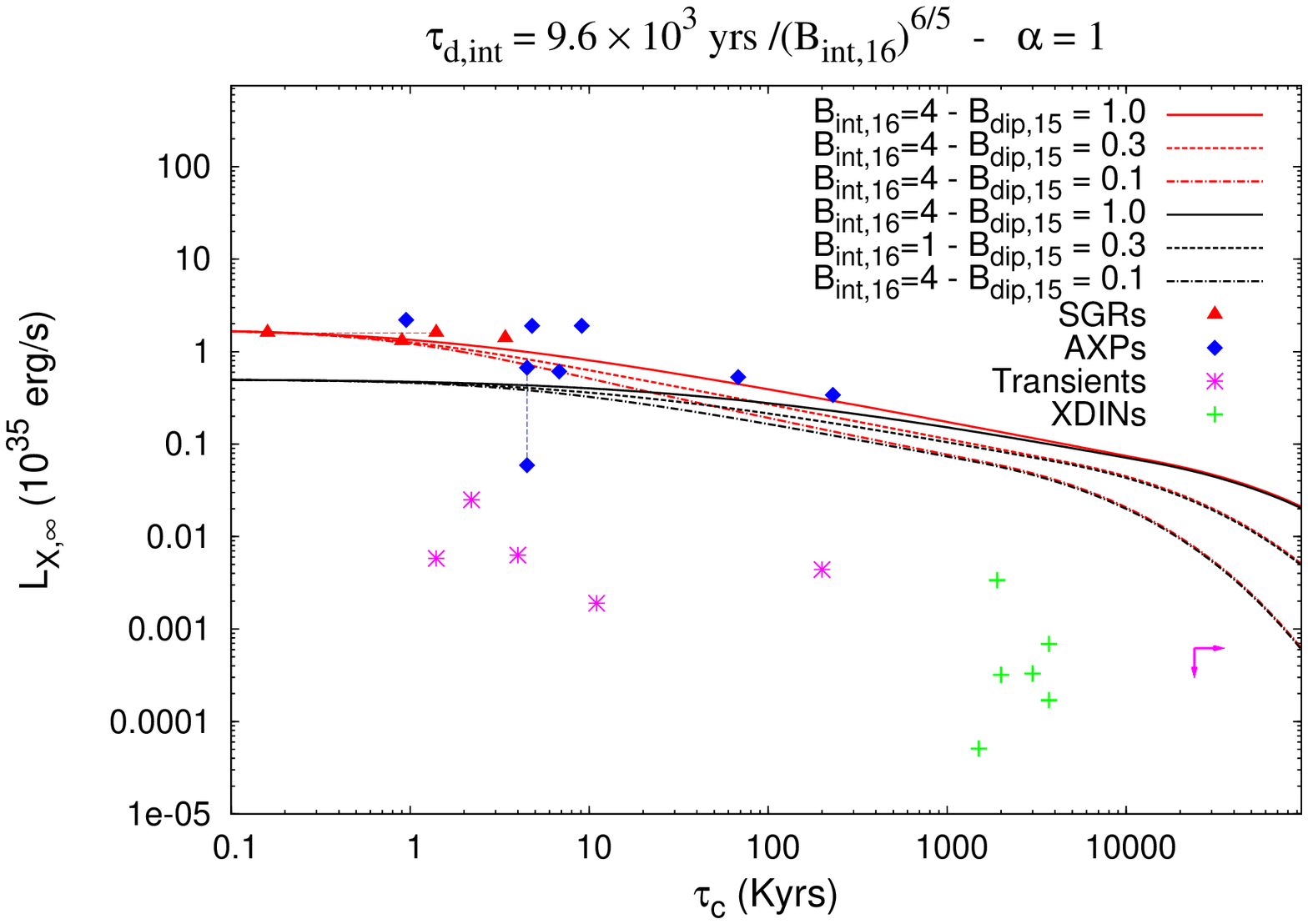}
\hspace{3mm}
\leavevmode\includegraphics[width=8.cm, height=6.2cm]{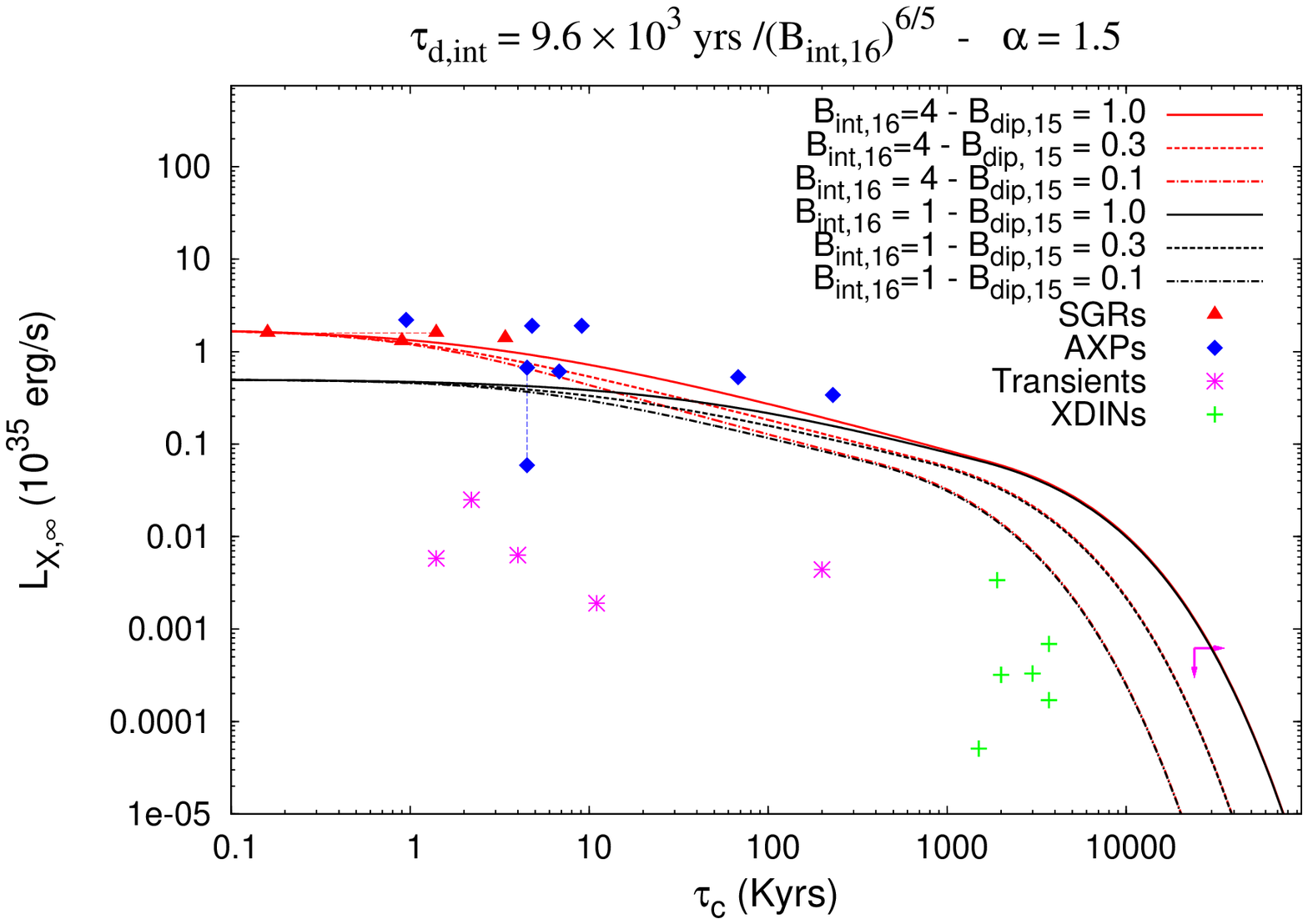}
\hspace{3mm}
  \caption{Maximum surface luminosity as a function of $\tau_c$ of a NS with
    an internal magnetic field decaying through the solenoidal mode of
    ambipolar diffusion, with $\tau_{\rm d}\approx 9.6\times 10^3  
    B^{-6/5}_{\rm int,16} ~{\rm yrs}$. The magenta arrows indicate the current
    upper limits on the spindown age and $L_{\rm X}$ of SGR 0418+5729. Two initial values for the internal field are chosen, 
   $B_{\rm int,i} = 4\times10^{16}$ G corresponds to red curves and 
   $10^{16}$ G corresponds to black curves. For each value of the internal 
   field, three values of the initial dipole, $B_{\rm dip,i} = 
   10^{15}, 3\times 10^{14} $ and $10^{14}$ G were chosen, corresponding to the 
   continuous, dashed and dot-dashed lines, respectively. \textit{Left Panel}: 
   The dipole field decays according to the $\alpha=1$ scaling, $\tau_{\rm
     d} = 10^3~{\rm yrs}~B^{-1}_{\rm dip,15}$;  \textit{Right Panel} : The
   dipole field decays according to the $\alpha=1.5$ scaling, $\tau_{\rm
     d} = 10^3 ~{\rm yrs}~ B^{-1.5}_{\rm dip,15}$.}
  \label{fig:Lx-vs-tau-solenoidal}
\end{center}
\end{figure*}
%
%

We stress that, in this regime, the former scaling $L_{\rm
    B,int} \propto B^{16/5}_{\rm int}$ does not hold
  anymore. We obtained $\alpha_{\rm int} = 2/19 $ which implies
$L_{\rm B,int} \propto B^{40/9}_{\rm int}$, matching the evolution of $L_{\rm X}$ as
  it must, given the equilibrium condition we imposed in the photon-cooling phase.

In Fig. \ref{fig:Lx-vs-tau-solenoidal} we show $L_{\rm X,\infty}$
vs. $\tau_{\rm c}$ curves for two different values of the initial internal field
and three different values of the initial dipole. Two different choices for
the decay index, $\alpha = 1$ or 1.5, are shown in the left
and right panel of that figure, respectively. Note that curves for different 
values of $B_{\rm int,i}$ and the same value of $B_{\rm dip,i}$ become 
coincident, at late times. This happens because 
the surface luminosity tracks the instantaneous value of $B_{\rm int}$ and 
all models reach the same value of the internal field, at late times. On the 
other hand, curves with different $B_{\rm dip,i}$ and the same $B_{\rm int,i}$
maintain memory of the initial conditions although dipole fields also reach 
to the same value at late times. This happens because the relation between 
$\tau_{\rm c}$ and $t$ depends explictly on the initial dipole (cfr. Eq. 
\ref{eq:tau_c(t)}). At a given $\tau_{\rm c}$, objects that had a different
$B_{\rm dip,i}$ are not co-eval. Those who had a weaker initial dipole are
older, have a weaker $B_{\rm int}$ and, thus, have a lower luminosity too.

The relatively flat evolution of $L_{\rm X}$ for sources with $\tau_{\rm c} <
10^4$ yrs can be explained quite naturally, in
this scenario. Larger initial fields produce larger luminosities
(Eq. \ref{eq:lumix-Bint}) but, at the same time, they have shorter decay times 
($\tau_{\rm d,i} \propto B^{-6/5}_{\rm int,i}$). Hence, $L_{\rm X}$ is larger
and bends earlier downwards for increasingly strong initial fields, with the
result that a flat, narrow strip of sources is produced. Its spread in 
$L_{\rm X}$ is smaller than the spread in $B_{\rm int,i}$ and its maximum
extension in $\tau_{\rm c}$ roughly corresponds to the decay time, $\tau_{\rm  d,i}$
of the minimum field. We estimate  
$B^{{\rm (min)}}_{\rm int,i} \gtrsim 10^{16}$ G.

Two further properties of the above plots are worth noticing.
First, the model with a faster decay of the dipole, $\alpha =1$, matches
well the luminosities of persistent sources up to $\tau_{\rm c} \gtrsim 10^5$
yrs, but largely fails to account for the position of SGR 0418+5729. 
A direct link between the latter object and the persistent sources would then 
be impossible. Although
the possibility that SGR 0418+5729 is linked to the transient sources is an 
interesting alternative, it has a serious drawback, if $\alpha=1$. A population 
of older ($\tau_{\rm c} \geq 10^6$ yrs), relatively bright magnetars would
be expected, which is clearly not seen. All known sources at that $\tau_{\rm
  c}$ (or beyond) are much dimmer and, accordingly, this possibility seems
ruled out.

The model with $\alpha = 1.5$, on the other hand, provides a viable option to 
intepret SGR 0418+5729 as an older relative ($t_{\rm age} \sim 10^6$ yrs) of 
the persistent sources whose dipole, as well as internal, fields have strongly 
decayed. In particular, the two lower curves in the right panel of Fig.
\ref{fig:Lx-vs-tau-solenoidal} give an internal field $B^{\rm {(0418)}}_{\rm int,14}
\approx 1.6 (1.1)$, a dipole field $B^{\rm{(0418)}}_{\rm
  dip,12} \approx 6 (4)$ and an X-ray luminosity $L^{\rm {(0418)}}_{\rm X,30}
\approx 30 (0.4)$, respectively, from top to bottom.  
The predicted range of quiescent luminosities is quite wide, as opposed to the 
relatively narrow range for both $B_{\rm int}$  and $B_{\rm dip}$. This
reflects the fact that SGR 0418+5729 is already in the photon-cooling dominated 
regime, where a steep drop in $L_{\rm X}$ occurs.
A careful assessment of its actual quiescent emission would thus in
principle put additonal constraints on the pysical parameters of this
object.

\subsection{Hall decay of the field in the inner crust}
\label{sec:Halldecay}
The calculation of the previous section rules out small values of $\alpha$ and
 points to $1.5\lesssim\alpha\lesssim 1.8$
as a viable option. Transition to core
superfluidity ($T_{\rm c} < 5\times 10^8$ K, Page et al. 2011) is expected to occur
at an age $\gtrsim {\rm a~few} \times 10^4$ yrs, if $B_{\rm int,i} > 10^{16}$ G,
which for the viable values of $\alpha$ corresponds to $\tau_{\rm c}
\sim 10^5$ yrs. As stated before, we currently don't have a clear
understanding of what will happen at this transition. The most
conservative option is assuming that evolution of the core field will
suddenly freez, its influence on the NS temperature soon becoming
negligible.

Even in this case, however, an internal field that threaded the NS crust would
still be actively decaying, due to the Hall term in the indcution equation. We 
consider here this ``minimal heating scenario'' for magnetars, focusing on the 
effect of field decay in the NS crust.  

The timescale of Hall-driven decay of the magnetic field in NS crusts has a
dependence on the actual field geometry, as was shown by Cumming et al. (2004).
The field component that we are considering could either be a twisted
(toroidal) field threading both core and crust (case $I$) or an 
azimuthal/multipolar field anchored only in the NS crust (case $II$). We 
consider the two cases separately.
\subsubsection{Case $I$ Hall decay}
\label{sec:caseI}
If the decaying crustal field were threading both the NS core and crust, its 
decay timescale would be sensitive to the global structure and is expected to be 
longer than Eq. \ref{eq:cummingetal}  by a factor $\sim R_*/h$, where $h$ is 
crust thickness \cite{Cumming+2004}. A more accurate expression was provided 
by \cite{Arras+2004},
\begin{equation}
\label{eq:hall-internal-arras}
\tau^{(I)}_{\rm H} = \frac{2.4 \times 10^4~\rho^{5/3}_{14}}{B_{16}} ~{\rm yrs}~~,
\end{equation}
who also integrated the field induction equation through the crustal volume 
adopting this formula and a realistic density profile, to estimate the 
associated power output. The resulting expression
\begin{equation}
\label{eq:Lb-Hall}
L^{(I)}_{\rm B,int} \approx 2 \times 10^{36} B^3_{16} R_{*,6}~~{\rm erg\ s}
^{-1}~~,
\end{equation}
exceeds the observed X-ray emission of younger sources if $B_{\rm
int,i} \gtrsim 5\times 10^{15}\;$G, which accordingly represents a
strict lower limit to the required crustal field. We evaluate the
radiative efficiency of this model by considering, in a crude way, the
impact of neutrino emitting processes within the NS crust. Neutrino
bremsstrahlung and plasmon decay, in particular, become quickly very
efficient in carrying away heat as the temperature rises, effectively
limiting the maximum temperature that can be reached at the surface
(cfr. TD96).
An approximate, analytical expression for the implied maximum surface temperature, 
$T^{({\rm max})}_{\rm s}$, which also includes the effects of the magnetic
field, was recently provided by \citep{Pons+2009}
\begin{equation}
\label{eq:Tmax-surf}
T^{({\rm max})}_{\rm s} \simeq \frac{3.6 \times 10^6~{\rm K}}{1+ 0.002~{\rm
    Log B_{12}}} \Rightarrow  L^{({\rm max})}_{X,\infty} \approx 9\times 10^{34} R^2_{*,6}~{\rm
  erg s}^{-1}
\end{equation}
where we neglect the very weak dependence on magnetic field in the last
step. The effect on $T^{({\rm max})}_{\rm s}$ would be much more pronounced
if the field were predominantly tangential to the surface, thus strongly
inhibiting heat conduction in the radial direction (Yakovlev et al. 2007;
Page et al. 2007; Pons \& Geppert 2009). However, the 
X-ray emission in the sample of \textit{persistent} sources is very close to 
the limit of Eq. \ref{eq:Tmax-surf} and much higher than the limit 
derived for a strong tangential field. The effects of such a component do not 
appear to be relevant, at least for these sources, as was already pointed
out by Kaminker et al. (2009). On the other hand, the effects of a strong
tangential field would be qualitatively consistent with the weak quiescent 
emission of transient sources. A proper account of this issue is beyond our
scope here and is postponed to future work.

With the above formulae we can build approximate luminosity curves. As
long as $L_{\rm B,int}$ calculated through Eq. \ref{eq:Lb-Hall} is
larger than $L^{({\rm max})}_{\rm s}$ (Eq. \ref{eq:Tmax-surf}), our
curves are limited by the latter value. Once $L_{\rm B,int}$ becomes
comparable to $L^{({\rm max})}_{\rm X}$ the radiative efficiency is
close to one and our curves track the evolution of $L_{\rm B,int}$
from here on. The latter thus represents the \textit{bolometric}
luminosity of sources at later times. As in the previous section, we
include the effect of gravitational redshift on the resulting $L_{\rm
X}$. Note that the Hall timescale is independent of temperature,
so the evolution of $L_{\rm B,int}$ extrapolates from the previous,
$\nu$-limited regime, into the photon-cooling regime.

\begin{figure*}
  \begin{center}
\leavevmode\includegraphics[width=8.cm, height=6.2cm]{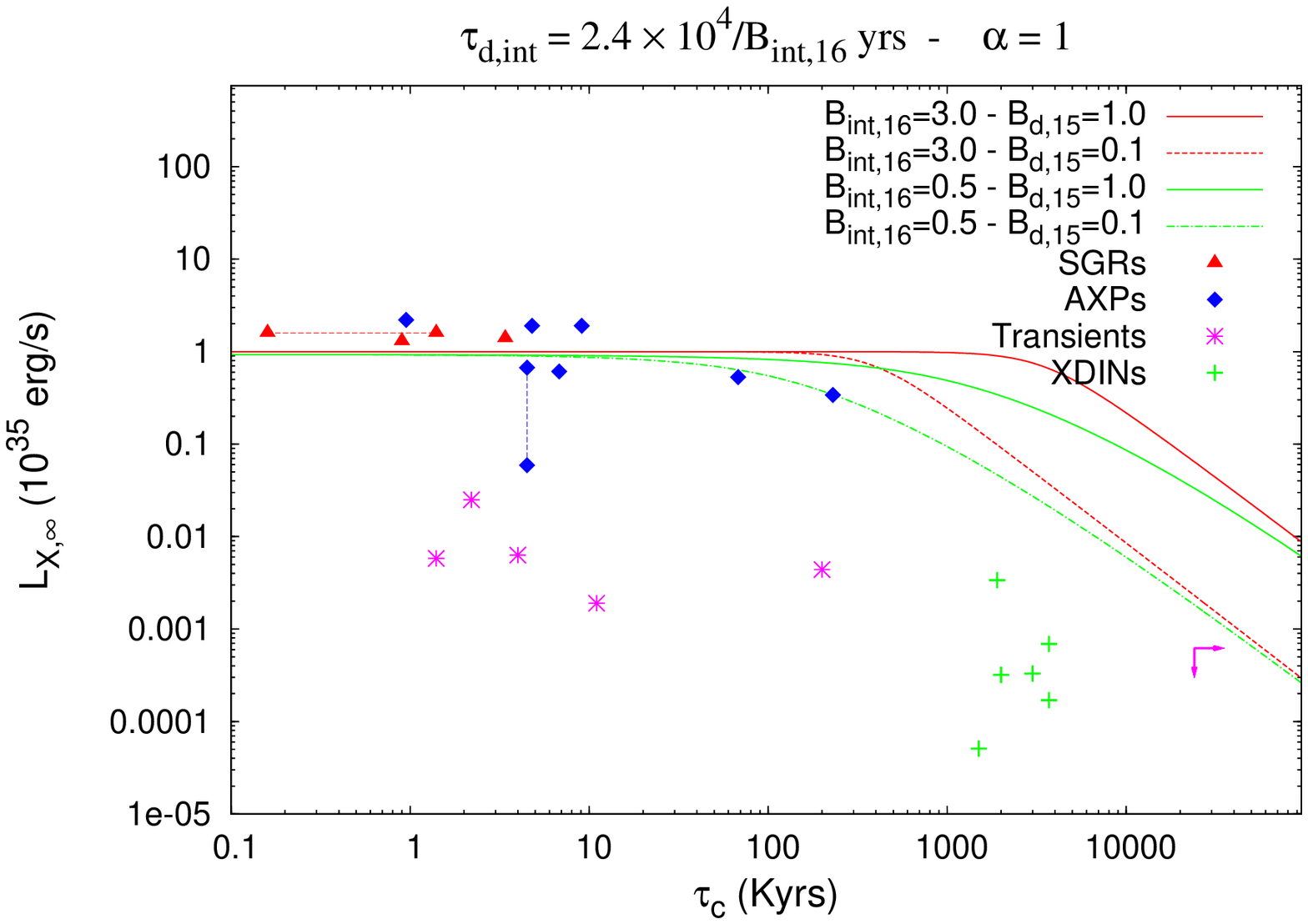}              
\hspace{3mm}
\leavevmode\includegraphics[width=8.cm, height=6.2cm]{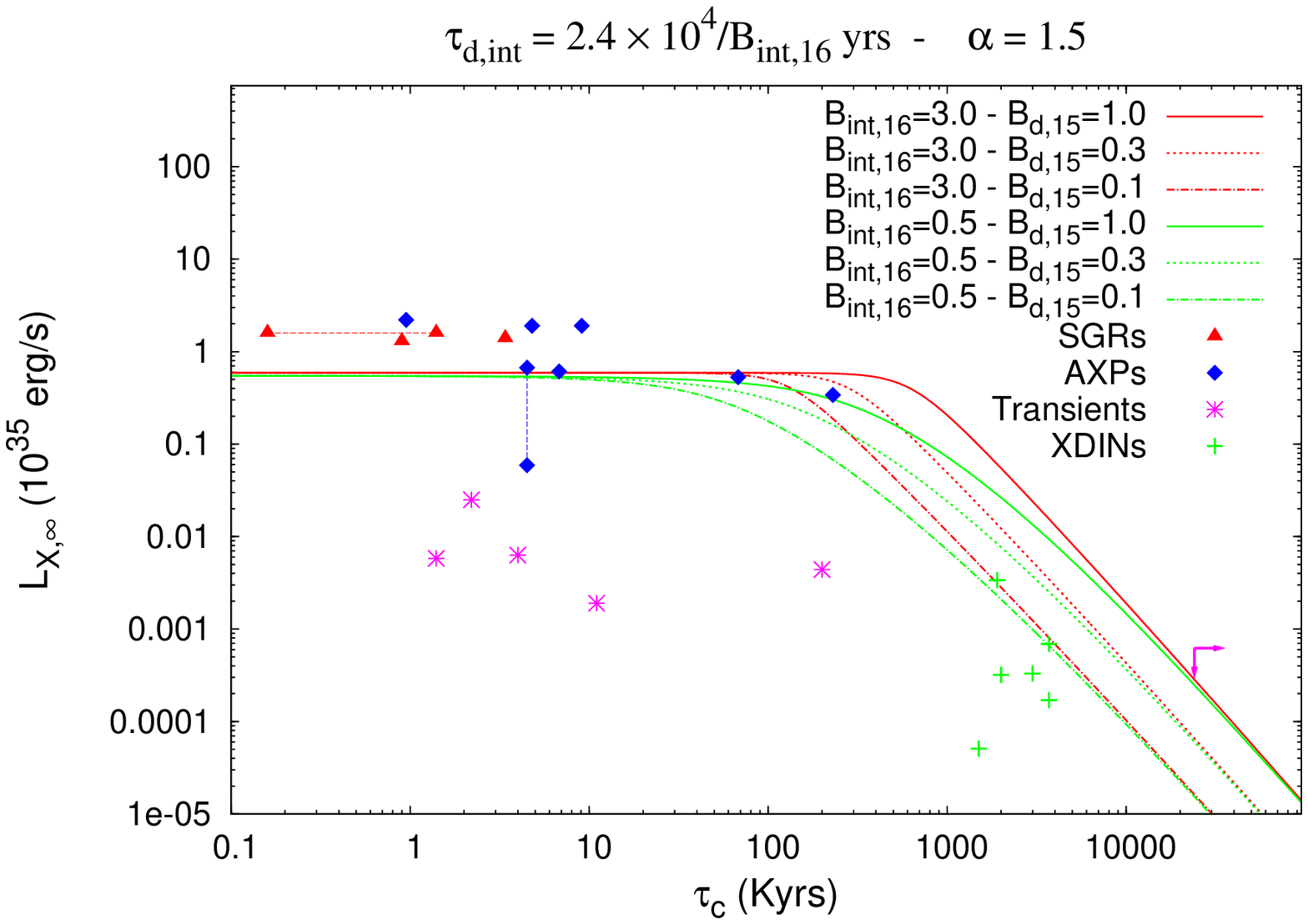}
  \caption{Curves show the evolution of the NS surface luminosity, as a function of $\tau_{\rm c}$, for Hall-driven 
   decay of a field threading core and crust. The flat part of the curves
   corresponds to the maximum surface luminosity of Eq. \ref{eq:Tmax-surf} 
   and lasts as long as $L_{\rm B,int}$ is larger than this limit. Once it 
   drops below that, curves switch to $L_{\rm B,int}$ and, thus, represent 
   the maximum available power from field decay (\textit{i.e.} an upper limit 
   to $L_{\rm X,\infty}$ in the (2-10) keV range). Points mark the measured 
   luminosities in the $(2-10)$  keV range of magnetar candidates. The magenta
   arrows indicate the current upper limits on the spindown age and $L_{\rm
     X}$ of SGR 0418+5729. Different
   symbols are explained in the figure. \textit{Left Panel:} the dipole field 
   decays according to the case $\alpha=1$ of sec. \ref{sec:two}, with the 
   scaling $\tau_{\rm d} = 10^3 B^{-1}_{15}$ yrs.
   Two values for the initial internal field are chosen, $3\times 10^{16}$ G and 
   $5\times 10^{15}$ G. For each value of $B_{\rm int,i}$, two curves are 
   shown, corresponding to two different choices for the initial dipole 
   field, $10^{15}$ G (upper curves) and $10^{14}$ G (lower curves). 
   \textit{Right Panel:} same as left panel, but the dipole field is assumed 
   to follow the $\alpha = 1.5$ case of sec. \ref{sec:two}. For each value of
   the internal field, three values of $B_{\rm dip,i} = 10^{15}, 3\times
   10^{14} {\rm ~and}~10^{14}$ G  are chosen.}
  \label{fig:Lx-vs-Hall-internal}
\end{center}
\end{figure*}
We calculated curves in the $L_{\rm X,\infty}$ vs. $\tau_{\rm c}$ plane in
this way choosing two different values of the initial internal field,
$B_{\rm int,i}$. For each value of $B_{\rm int,i}$, two values of the
initial dipole, $B_{\rm dip,i}$, were chosen, giving four curves in total 
for each value of the decay index $\alpha$. Two different 
values of $\alpha$ were chosen and results for either choice are shown in
the two panels of Fig. \ref{fig:Lx-vs-Hall-internal}.

Our main conclusions are summarized as follows:
\begin{itemize}
\item the NS surface emission saturates at a level close to,
  but lower than, the observed luminosity of young AXPs/SGRs
  ($\tau_{\rm c} \lesssim 10^4$ yrs). Given the uncertainties and
  approximations implicit in the derivation of the limit in
  Eq. \ref{eq:Tmax-surf}, the mismatch should  not be regarded as a major 
  issue.
  Further, as already noted, the gradual relaxation 
  of a magnetospheric twist could provide an additional channel for
  release of the energy of the internal field directly at the NS surface. 
  It is interesting to note that, despite not being limited by crustal
  neutrino processes in this case, a luminosity enhancement by just a factor of 2-3, 
  at most, is all {\rm that} is required to match the data, thus roughly confirming the
  overall energy budget estimated in our approximate calculations.
  A similar argument does not apply to the decaying parts of the
  curves, though, since those represent the evolution of $L_{\rm B,int}$, a
  limit that cannot be exceeded.

\item internal fields $\gtrsim 10^{16}$ G are required also in this 
scenario. The minimal value $B_{\rm int,i} \gtrsim 5\times10^{15}$ G provides 
a total magnetic power, $L^{(I)}_{\rm B,int}$, marginally consistent with 
$L_{\rm  X,\infty} \approx 2 \times 10^{35}{\rm reg\;s}^{-1}$ of the youngest 
sources. Properly accounting for the powerful bursting activity of young
sources and for realistic values of the radiative efficiency, would certainly
imply a significantly larger minimal field\footnote{A self-consistent lower
  limit is obtained by setting the neutrino luminosity marginally equal 
to $L^{\rm {(max)}}_{\rm X}$. This implies $L_{\rm B,int}$ twice as large 
and, thus,  $B_{\rm int,i} \gtrsim 7\times 10^{15}$G.}. A neat example of this 
is provided by the 27 December 2004  Giant Flare fron SGR~1806-20, which released 
$\sim 5 \times 10^{46}\;$erg in high-energy photons. 
Even assuming that this was a unique event during the whole lifetime
of the source, which we take to be $\tau_{\rm c} \simeq 1.4 \times
10^3\;$yr, 
it would correspond to an average power output of $\sim 1.5\times 10^{36}$ erg
s$^{-1}$, an order of magnitude larger than the
quiescent emission. 

\item the internal field must also be able to provide its large power output for a 
sufficiently long time, as to match the duration of the apparent plateau in 
the  $L_{\rm X,\infty}$ vs. $\tau_{\rm c}$ up to $\tau_{\rm c} \gtrsim 
10^4$ yrs. This is comparable to 
the decay time of initial internal fields 
$\sim 10^{16}$ G.

For a given value of $B_{\rm int,i}$, 
a strong initial dipole and/or a small $\alpha$ push the end of the plateau to
larger $\tau_{\rm c}$, as discussed in sec. \ref{sec:coredecay}. This is why the $\alpha =1$ model fails by 
overpredicting luminosities at $\tau_{\rm  c} \gtrsim 10^5$ yrs, 
unless dipole fields as low as $10^{14}$ G are assumed. Even for 
such a weak dipole, this model overpredicts the luminosity
of SGR 0418+5729 and is thus ruled out. The $\alpha =1.5$ model seems, on the other hand, to well
match the apparent bending at $\tau_{\rm c} \sim 10^5$ yrs, also remaining 
consistent with the position of SGR 0418+5729 (cfr. our estimate of its
minimal power output in sec. \ref{sec:two}).
  
\end{itemize}
Finally note that the value of $T^{\rm (max)}_{\rm s}$ used in this section
(Eq. \ref{eq:Tmax-surf}) would also apply in the case of core 
dissipation (sec. \ref{sec:coredecay}), where we ignored this effect  
and let the surface luminosity free to grow. Despite this, the surface 
luminosities calculated in sec. \ref{sec:coredecay} are close to the maximum 
implied by Eq. \ref{eq:Tmax-surf} and very close, indeed, to the observed 
luminosities of magnetar candidates. In fact, also in that case the surface 
emission is limited by neutrino-emitting processes, which take place in the
core rather than the crust. 

\subsubsection{Case $II$ Hall decay}
\label{sec:caseII}
Finally, we consider the Hall-driven decay of a purely crustal field. In this 
case, the field would be completely insensitive to conditions 
in the core and its decay time is correctly given by Eq. \ref{eq:cummingetal}. 
In a way completely analogous to Eq. \ref{eq:Lb-Hall}, it is possible 
to define the total power output in the crust for this case as
\begin{equation}
\label{eq:Lb-hall-II}
L^{(II)}_{\rm B,int} \approx 2.7 \times 10^{37} B^3_{16} R_{*,6}~~{\rm erg s}
^{-1}~~.
\end{equation}

\begin{figure*}
  \begin{center}
\leavevmode\includegraphics[width=8.cm, height=6.2cm]{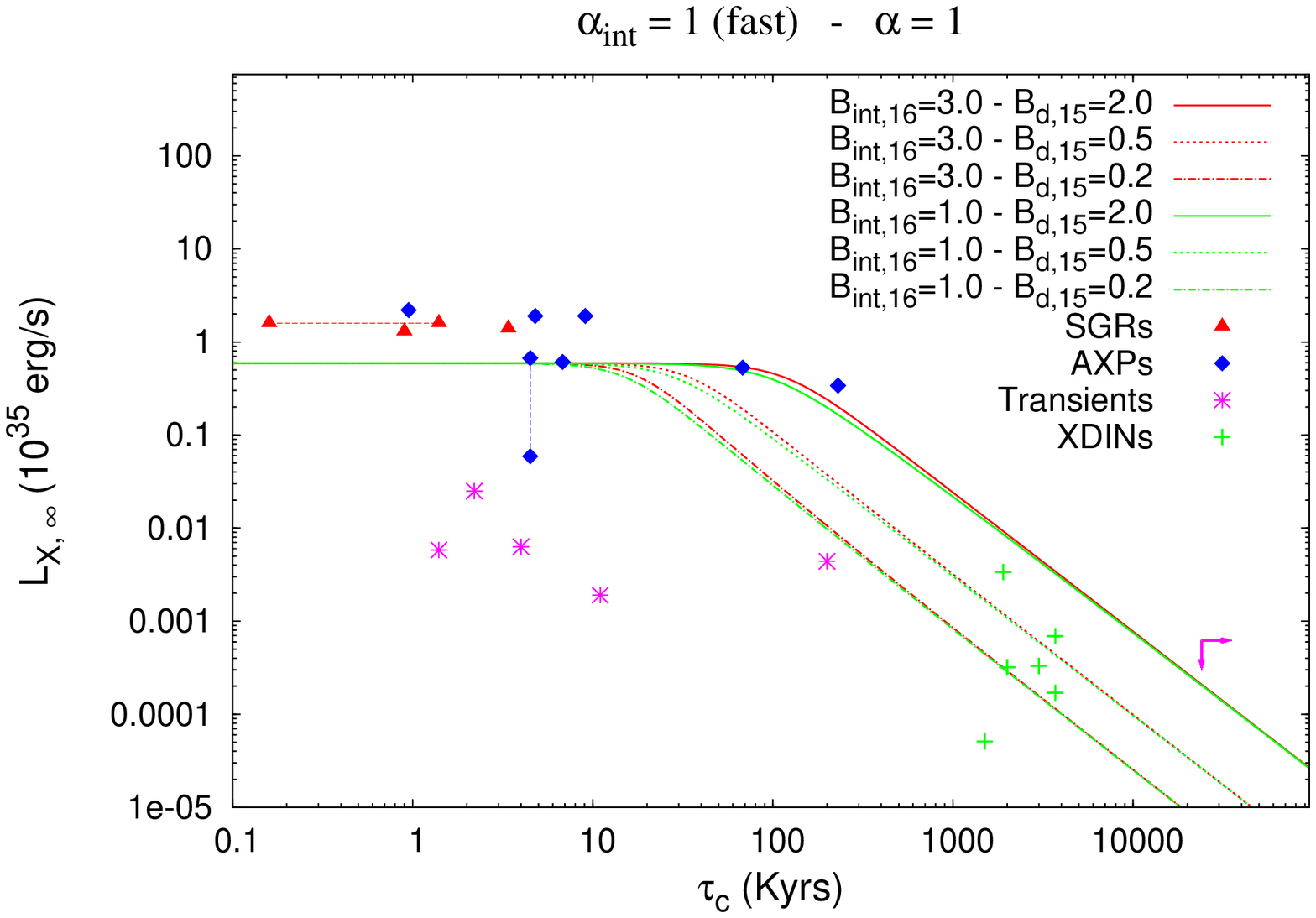}              
\hspace{3mm}
\leavevmode\includegraphics[width=8.cm, height=6.2cm]{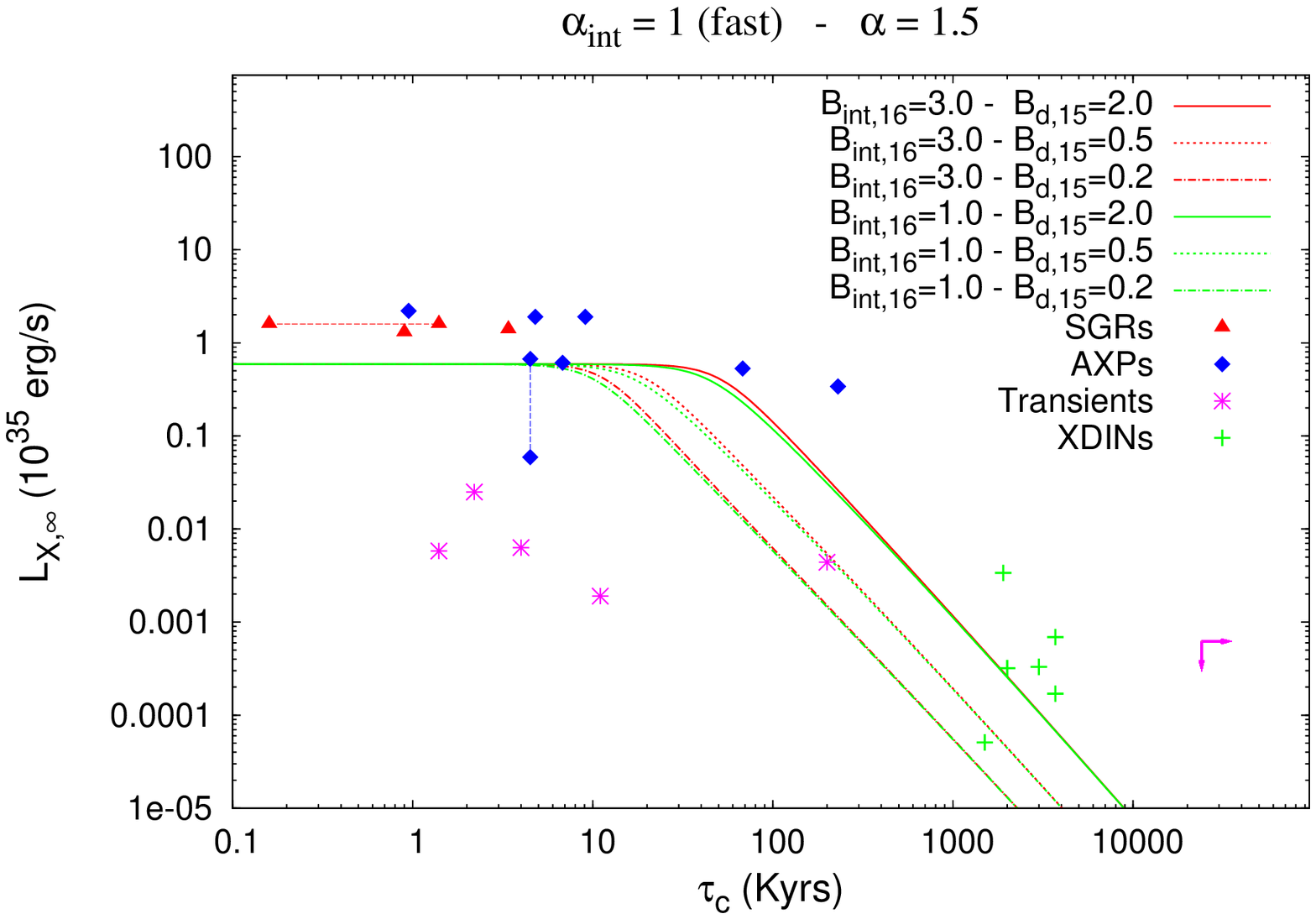}
  \caption{Curves show the evolution of the NS luminosity as a function 
   of $\tau_{\rm c}$, for Hall-driven decay of a purely crustal field. 
   The flat part of the curves corresponds to the maximum surface luminosity 
   of Eq. \ref{eq:Tmax-surf} and lasts as long as $L_{\rm B,int}$ is larger than this
   limit. Once it drops below that, curves switch to $L_{\rm B,int}$ and, thus,
   represent the maximum available power from field decay (\textit{i.e.} an
   upper limit to $L_{\rm X,\infty}$ in the (2-10) keV range). Points mark the measured 
   luminosities in the $(2-10)$  keV range of magnetar candidates. Different
   symbols are explained in the figure. The magenta arrows indicate the
   current upper limits to the spindown age and $L_{\rm X}$ of SGR 0418+5729.
   \textit{Left Panel:} the dipole field decays according to the case 
    $\alpha=1$ of sec. \ref{sec:two}, with the scaling $\tau_{\rm d} = 10^3
   B^{-1}_{15}$ yrs. Two values for the initial internal field are chosen, 
   $3\times 10^{16}$ G and $10^{16}$ G. For each value of the
    internal field, three curves are shown, corresponding to three different
    values of the initial dipole field, $2\times 10^{15}$ G, $5\times 10^{14}$
    G  and
    $2\times 10^{14}$ G. \textit{Right Panel:} same as left panel, but
    the dipole field is assumed to follow the $\alpha = 1.5$ case of 
    sec. \ref{sec:two} and different representative values of $B_{\rm dip,i}$
    are chosen.}
  \label{fig:Lx-vs-Hall-internal-II}
\end{center}
\end{figure*}

Fig. \ref{fig:Lx-vs-Hall-internal-II} depicts the results  the same 
calculation of the previous section, adopting the decay timescale 
Eq. \ref{eq:cummingetal} and the corresponding magnetic luminosity, Eq. \ref{eq:Lb-hall-II}.

One conclusion can be drawn also in this case, which is in common to all
other cases examined: an internal field of $\gtrsim 10^{16}$ G is required 
to explain the observed luminosities of active magnetars, in 
particular middle-aged ones ($\tau_{\rm c} \gtrsim 10^5$ yrs). Indeed, 
although the luminosity of the youngest sources is orders of magnitude less 
than the available $L_{\rm B,int}$ at that age, luminosities in excess 
of $10^{34}$ erg s$^{-1}$ are measured in middle-ages sources.
This turns out
to be a crucial property, since only fields larger than $10^{16}$ G have a
sufficiently large energy reservoir to account for it.

For $\alpha=1$, it seems impossible to account for the luminosity of the AXPs
1E2259+586, even assuming the highest value of $B_{\rm dip,i}$ that is
consistent with the distribution of sources of
Fig. \ref{fig:B-tau-fieldecay-models} and \ref{fig:B-P-fieldecay-models}.
The AXP 4U 0142+60 is, on the other hand,  just marginally consistent with
this upper curve. Note, however, that both sources correspond to a lower
initial dipole, for $\alpha=1$, as shown in Fig. \ref{fig:B-tau-fieldecay-models}. 
If they were both born with $B_{\rm dip,i} = 2\times 10^{15}$, their spin periods
would have to be $\lesssim 11$ s, as clearly seen in Fig. \ref{fig:B-P-fieldecay-models}. 
For $B_{\rm dip,i} \leq 10^{15}$ G the mismatch of model curves with
observations would be far too large. This discrepancy is thus very significant.

A wide range of quiescent luminosities is predicted for SGR 0418+5729, 
essentially depending on its initial dipole field. The latter is likely close
to $B_{\rm dip,i} \approx 10^{15}$ G for $\alpha=1$, (see
Fig. \ref{fig:B-tau-fieldecay-models} and \ref{fig:B-P-fieldecay-models}), for 
which a quiescent luminosity $\sim 2\times 10^{31}$ erg s$^{-1}$ would be
expected. A direct measurement of its quiescent emission may therefore provide
a conclusive answer about this scenario.

The case $\alpha=1$ would have to be considered at best marginal, given the
above considerations, essentially for the same reasons as the case $\alpha=1$
in sec. \ref{sec:explaining}. The additional general arguments we presented against
this small $\alpha$ value make this scenario, in conclusion, very unlikely. 
%

The case $\alpha=1.5$ falls short of the X-ray emission of several sources
beyond $\tau_{\rm c}\gtrsim 10^5$ yr and is, therefore, completely ruled out. 

Overall, unless a different scaling for $\tau_{\rm Hall}$ can be
provided, or unless an additional mechanism for decay of purely
crustal fields is proposed, this hypothesis appears inconsistent with
the observational properties of the source sample
considered. Accordingly, we are led to rule it out in its present
form.


\section{Summary and Conclusions}
\label{sec:conclusions}
In this work we have studied the implications of dipole field decay 
on the dynamical evolution of a spinning, magnetized NS, 
assuming a power-law scaling for the field decay time, $\tau_{\rm d} \propto 
B^{-\alpha}_{\rm dip}$ and $\dot{B}_{\rm dip} \propto B^{1+\alpha}_{\rm dip}$.
If the field decays sufficiently fast ($\alpha <2$)  the spindown time $\tau_{\rm c}$ 
grows faster than the decay time, $\tau_{\rm d}$. Once $\tau_{\rm c} >
\tau_{\rm d}$,  the dipole continues to decay while the NS spin 
period hardly changes and an asymptotic value of the spin period, $P_{\infty}$, is reached. 
For a slow ($\alpha > 2$) decay,  the above condition is
never met  and the NS continues spinning  down forever.

The distribution of magnetar candidates (SGRs, AXPs, transients) and XDINs 
in the $B_{\rm dip}$ vs. $\tau_{\rm c}$ ( or $B_{\rm dip}$ vs. $P$)
plane provides  a  strong evidence for a fast ($\alpha < 2$) dipole field decay. This conclusion
is based on two striking properties. First, the absence of
objects with large periods, well above 10~s, and in particular with
large dipole fields $B_{\rm dip}$ and large spindown ages $\tau_{\rm
c}$. We know that objects with large $B_{\rm dip}$ exist since we find
them at small $\tau_{\rm c}$. Similarly we know that objects with
large $\tau_{\rm c}$ also exist, but they are all found only with relatively
small $B_{\rm dip}$. If $B_{\rm dip}$ did not decay
then $\tau_{\rm c} = t$ and there would be no reason why there are no old (large $\tau_{\rm c}$) 
objects with large $B_{\rm dip}$. Note that such objects
should be more numerous that younger ones  and they should be
at least as detectable as  objects with a similar $\tau_{\rm c}$ and
smaller $B_{\rm dip}$.

Decay of the dipole field offers the most natural
interpretation of this fact, as it naturally produces bending
evolutionary tracks in the $B_{\rm dip}$ vs. $\tau_{\rm c}$ plane or, 
equivalently, the $\dot{P}$ vs.  $P$ and $B_{\rm dip}$ vs. $P$ planes
(see Figs. \ref{fig:B-tau-fieldecay-models}, \ref{fig:B-P-fieldecay-models} and
\ref{fig:p-pdot-tracks}). 
In particular, if the decay law has $\alpha<2$, an asymptotic spin
period is approached which corresponds to the scaling $B_{\rm dip,15}
\approx P_{\infty, 10} ~\tau^{-1/2}_{\rm c, kyr}$.  Sources that reach
their asymptotic period move along a trajectory with a constant
period, $P_{\infty}$, which, in turn, is related to the value of
$\alpha$ and of the initial magnetic dipole, $B_{\rm dip,i}$ (see
Eq. \ref{eq:def-pinfty}).

The second key feature is the fact that spin period periods of all (but one) sources with
$\tau_{\rm c} \geq 10^4\;$yr are within a very narrow
strip corresponding to $7\lesssim P_{\infty} \lesssim 12$ s. This
supports the existence of an asymptotic period and hence of a single decay
mechanism operating in all sources, with $\alpha <2$ and with a decay time $\tau_{\rm d} \approx 
10^3~B^{-\alpha}_{15}$ yr (sec. \ref{sec:two}).
The narrowness of the distribution in $P_{\infty}$ implies that
the decay index must be in the range $1 \lesssim 
\alpha <2 $. 

Hall decay in the crust, corresponding to $\alpha =1$, provides a
reasonable explanation for the observed $B_{\rm dip}\,$--$\,\tau_{\rm
c}$ distribution. It implies that all SGRs, AXPs, transients and XDINs
were drawn from a distribution of $B_{\rm dip,i} \in [0.2 - 2]
\times 10^{15}$ G.
A similarly good description of the data is provided by larger
values of $\alpha$, for which we consider $\alpha =1.5$
as representative. These allow for a wider distributions of initial 
dipoles, making them favored in this respect. As $\alpha$ approaches
2, asymptotic periods $>12$ s would necessarily be expected and the  current 
data  disfavor $\alpha$-values too close to 2. Note that all viable
  models imply that there cannot be a population of sources whose dipole field
  at birth is significantly larger than $\simeq 2\times 10^{15}$ G. These
  would indeed reach asymptotic spin periods significantly longer than observed.

An important implication of the dipole field decay with $\alpha <
2$ is that the sources are younger than what their spindown age
implies ($t < \tau_{\rm c}$).  The current value of $t/\tau_{\rm c}$
also depends on $B_{\rm dip,i}$. For magnetar candidates this ratio is
significantly smaller than unity and their real ages can be a factor
of $\sim$5--50 smaller than $\tau_{\rm c}$, depending on their initial
field, age, and exact value of $\alpha$.

The ratio $t/\tau_{\rm c}$ decreases for decreasing values of $\alpha$, at a
given $\tau_{\rm c}$ (see Fig. \ref{fig:summary}). For a fixed value of
$\alpha$, $t/\tau_{\rm c}$ decreases as the object ages.  

X-ray luminosities of XDINs, their clustering at $\tau_{\rm c}
\sim {\rm a\ few}\times 10^6\;$yr and the distribution of their
effective temperatures point to their ages being $t\gtrsim 10^5\;$yr.
This further constrains possible values of $\alpha$.  Specifically,
$\alpha=1$ decay implies that these sources are younger, or even much
younger, than this value. On the other hand, values of $\alpha \sim
1.5$ well match this age requirement for XDINs. Thus, the case
$\alpha=1$ is basically ruled out and a more restrictive condition,
$1.5\lesssim\alpha\lesssim 1.8$,
appears to be the most consistent with the whole available data.

Although the decaying dipole field dissipates energy, it turns out
that unlike the simplest expectations, this energy is
insufficient to power the strong X-ray emission of SGRs/AXPs (and,
possibly, other related classes).
While exponential decay is rapid enough that $L_{\rm B,dip}$ would
suffice for all sources, the current upper limit on the weak quiescent
emission of SGR 0418+5729 ($\lesssim 6\times 10^{31}$ erg s$^{-1}$)
already implies $\alpha \gtrsim 1$, as otherwise this source would be
too young to be so dim.

For $\alpha = 1$, the maximal available power, $L_{\rm B,dip}$, falls
short of the observed X-ray emission of persistent AXPs at $\tau_{\rm
c} \approx 10^5\;$yr and the current observed emission of
SGR~0418+5729 is far larger\footnote{This post burst luminosity might over estimate
the real quiescent emission. However even the emission from
the outbursts of SGR 0418+5729 suggests a minimal energy
output larger than $L_{\rm B,dip}$, for this source.} than its $L_{\rm B,dip}$ for all values of
$\alpha \ge 1$.  Hence,
$\alpha=1$ and energy provided just by magnetic dipole field decay
should be considered at best marginally consistent with the observed
distribution of sources in the $L_{\rm X}$ vs. $\tau_{\rm c}$ plane.
However, as we have discussed above, $\alpha = 1$ is ruled out by the
requirements on the ages of XDINs, which altogether imply
$1.5\lesssim\alpha\lesssim 1.8$. For such values of $\alpha$ (or even
for $\alpha>1$) it is impossible to power the X-ray emission of
SGRs/AXPs by the decay of the dipole field.
If magnetic energy powers this emission, the available reservoir must be 
larger than that of the dipole and must decay on a somewhat longer timescale. 
The most natural conclusion is that this is provided by the magnetic field in the 
NS interior (sec. \ref{sec:explaining}).

In order to identify at least the basic properties of the internal field,
we considered its decay either in the liquid core, through the solenoidal
mode of ambipolar diffusion, or the inner crust, through the Hall
term. In the latter case, we treat separately fields threading the
whole NS volume or fields confined to the crust alone.  
A  general conclusion in all cases is that the observed large $L_{\rm X}$ values  
at up to $\tau_{\rm c} \sim 10^5$ yrs require a very large initial internal field, 
$B_{\rm int,i} \gtrsim 10^{16}$ G . Combined with the maximum value of
  $\simeq 2\times 10^{15}$ G for the initial dipole field this implies that,
  at formation, the internal magnetic field in magnetars contains at least 20-30 times more
  energy than the dipole, likely one order of magnitude more than
    that. This raises a question about the stability of the implied
    configurations, which is a currently open field. Although it is
    difficult, at this stage, to have a precise answer on global field
    stability, recent investigations (cfr. \citet{Bra09}) have highlighted the key role of stable
    stratification of stellar interiors in providing stability to
    predominantly toroidal fields. As opposed to this, it was shown that a
    predominantly poloidal field would remain unstable, independent of the
    interior stratification. The maximum ratio of toroidal-to-poloidal field
    strength that can be stablized is estimated, conservatively, to be
    $\lesssim 10$, for a $\sim 10^{16}$ G toroidal field, roughly consistent
    with our results. Clearly, more work is needed to settle this important topic.

Fields threading the NS core have relatively long decay timescales ($\gtrsim 
10^4$ yr for $B_{\rm int} = 10^{16}$ G), whether their dissipation occurs 
primarily in the crust or the liquid core. 
This implies that it takes $\gtrsim 10^4\;$yr 
for $L_{\rm X}$ to decrease
appreciably. On the other hand, observations show that this decrease 
occurs at $\tau_{\rm c}\lesssim 10^5$ yrs, approximately 10 times larger than 
the real age. This favors values of $\alpha >1$ and, in particular, our choice 
$\alpha =1.5$ meets easily this requirement (cfr. Fig. \ref{fig:summary}). 
If $\alpha \simeq 1$, on the other hand,
$\tau_{\rm c}$ grows too fast with real time and when $L_{\rm X}$ eventually
starts decreasing,  $\tau_{\rm c}$ is already too large, in sharp contrast
with observations.

If the internal field is confined to the NS crust, on the other hand, it 
decays on a $\sim 10^3\;$yr timescale and the situation is reversed. The case 
$\alpha = 1.5$ largely fails because it implies 
that a given $\tau_{\rm c}$ is reached when the source is too old and 
its X-ray emission too weak. 
The case $\alpha=1$ is marginally consistent with $\tau_{\rm c} \sim
10^5$ yr-old sources, and only for an extreme choice of parameter values. 
Considering the strong independent arguments against such a low $\alpha$ value, we
conclude that purely crustal fields cannot account for the distribution of 
SGRs/AXPs in the $L_{\rm X}$ vs $\tau_{\rm c}$ plane. 

SGR 0418+5729 has the largest $\tau_{\rm c}$, weakest $B_{\rm dip}$ and 
$L_{\rm X}$ of our sample and thus provides crucial insight into the long-term 
evolution of magnetar candidates. It is thus of particular interest to summarize 
our main conclusions on its properties. For the reasons just mentioned, we
restrict attention to the two scenarios where the internal field threads the whole
NS  volume and, thus, to dipole field decay with $\alpha \approx 1.5$.  
We find very similar conclusions in the two cases, whether the internal field 
decays in the crust or liquid core. This is not surprising, given the very
similar values of $\alpha_{\rm int}$ and  $\tau_{\rm d,int}$ in both models.
From the decay of the dipole field with $\alpha=1.5$, this source was
likely born with $B_{\rm dip,i} \sim (3 - 5) \times 10^{14}$ G, which implies
its real age is $t_{\rm age} \approx (1 - 2)\times
10^6$ yrs, the youngest age corresponding to the larger initial field. With
these assumption on the initial dipole, its current dipole field would be
$\sim (4-7) \times 10^{12}$ G and the internal field $B_{\rm int} \sim (1.1
- 2) \times 10^{14}$ G,  independent of its initial value (as long as it
was $ \gtrsim 10^{16} $G). The average luminosity due to the decay of the
internal field is $L_{\rm B,int} \approx L_{\rm X}\sim (4 - 10) \times 10^{30}$ erg s$^{-1}$.

Note that the internal field is expected to be close to the lower limit for
being able to produce crust breaking (and, thus, bursts, according to the current
understanding of these events, cfr. TD96; Perna \& Pons 2011). Also note that 
the expected luminosity can be checked by continued monitoring of this source 
and is, incidentally, similar to the estimated average output from outbursts 
(cfr. sec. \ref{sec:two}).

\section{Discussion \& Evolutionary Sequence}
\label{sec:evol}
In the light of our results, we can sketch a tentative evolutionary picture 
for the different classes of objects considered here. The most obvious
link is between SGRs and AXPs, as already suggested by \cite{Kouve+1998}.

These sources  appear to form a continuous sequence, with persistent SGRs
 seen as the youngest, mostly magnetized and brightest sources. Persistent
AXPs are found at comparable field strength and similar, although slightly
older, ages.
A couple of evolved AXPs are known (1E2259+586 and 4U
0142+61), whose dipole fields have decayed substantially and whose X-ray
luminosity is somewhat weaker than in younger objects. The X-ray luminosity  of 
SGRs/AXPs is inconsistent with the decay of the dipole
field alone and needs to be powered by an additional, stronger field
component. The latter must thread the whole NS volume and its strength likely 
exceeds $10^{16}$ G at birth. At young ages the internal field likely decays 
through a combination of ambipolar diffusion in the core and Hall decay in the
crust. The released heat powers the strong thermal emission typical of these
sources. Neutrino-emitting processes, whether in the liquid core or in the
crust, limit the thermal luminosity of young magnetars to be $\lesssim
10^{35}$ erg s$^{-1}$ and provide a likely explanation for the observed
plateau in the distribution of $L_{\rm X}$ up to $\tau_{\rm c} \gtrsim 10^4$ yrs.

Transient SGRs/AXPs are not related in an obvious way to persistent sources. 
Their dipole fields are systematically weaker\footnote{It is remarkable that
most of them have $B_{\rm dip} \approx 2\times 10^{14}$ G.} and their spin 
periods also appear to be shorter, on average, suggesting that most of them  
have not yet reached the asymptotic spin period. The two oldest  transients 
 have significantly weaker dipole fields and somewhat 
slower spins (CXO J164710-455216 and SGR 0418+5729), thus supporting this
basic picture. 

The decay of $B_{\rm dip}$ appears consistent with having the same properties 
as in persistent sources. However, the X-ray luminosities of transients are much weaker 
and it is not possible, at present, to identify a decline at $\tau_{\rm c} \approx 100$
kyrs, as  seen in persistent sources.

The weak quiescent emission of transients is very striking, being much lower 
even than the power provided by their decaying dipole field. 
The relatively bright, short-time bursts they emit in outburst and the ensuing
large increase in X-ray luminosities, that make them temporarily similar to the 
persistent sources, supports, on the other hand, their connection to the
latter group\footnote{During outbursts, the X-ray luminosity of
    transients largely exceed their rotational energy losses, like in
    persistent
sources, which is not the case in quiescence (cfr. Tab. \ref{table}).}
If this is indeed the case, then, it is possible that transients also host a
decaying, strong internal field, despite the very low radiative efficiency of
their quiescent state. 
We can only speculate,
at this stage, that a strong azimuthal component of the crustal field provides
a possible explanation for their X-ray deficiency. If the energy released by 
magnetic dissipation is confined to a sufficient depth in the NS interior,
thermal insulation of the hot regions by such a field component could suppress
the surface X-ray emission by a large factor (cfr. Potekhin \& Yakovlev 2001,
Kaminker et al. 2009). The difference with persistent sources would then be 
attributed to the existence of a much more ordered field component in 
transients, which  globally suppresses the radiative efficiency $\epsilon_{\rm
  X}$. Only at outbursts does the internal field gain access to the more
efficient channel of X-ray radiation, probably due to flaring of smaller-scale 
field structures temporarily creating a preferential path for heat conduction, 
or a localized dissipation region close to the NS surface. This is
qualitatively consistent with the small (and shrinking) emitting areas 
and high (and decreasing) temperatures of the extra 
BB spectral components found in outburst emission from transients.  

\begin{figure}
\begin{center}
\includegraphics[width=12 cm, angle=90]{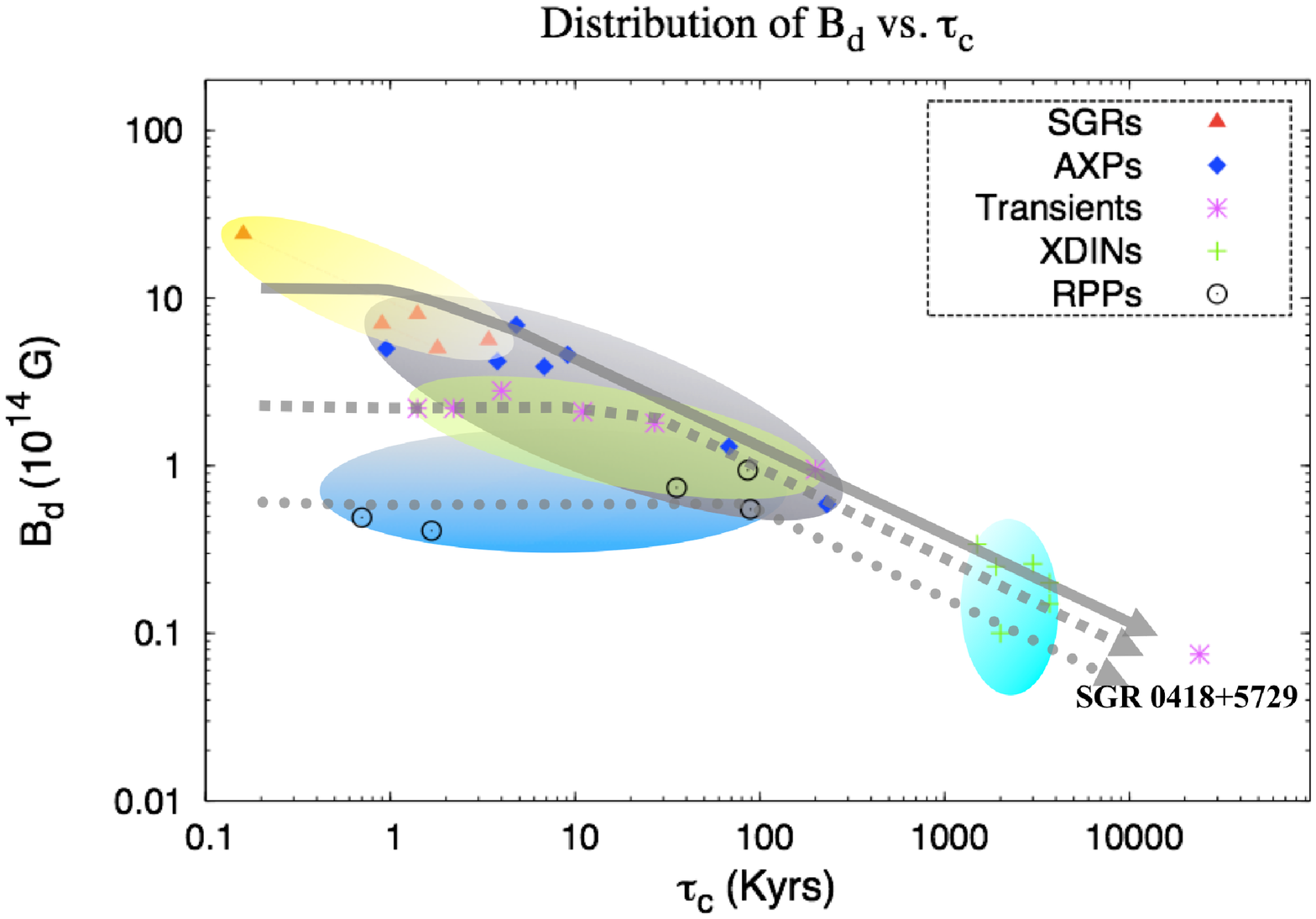}
\caption{A schematic diagram of the evolutionary sequences we infer for different high magnetic field NS source classes in the $B_{\rm dip}\,$--$\,\tau_{\rm c}$ plane (which can be thought of as a rotated version of the $P\,$--$\,\dot{P}$ plane since $P\propto B_{\rm dip}\tau_c^{1/2}$ and $\dot{P}\propto B_{\rm dip}\tau_{\rm c}^{-1/2}$). Members of different classes are denoted by different symbols, while the classes as a whole are indicated by colored ellipses. We clearly identify two distinct evolutionary tracks -- the SGR/AXP branch ({\it solid arrow}) and the transient branch ({\it short-dashed arrow}), which start with different initial dipole fields (and likely also different internal fields) but converge to similar values (at a given $\tau_{\rm c}$) after their fields decay significantly. The dimmest and fastest spinning XDIN (RX~J0420) must have had a significantly smaller initial dipole field than either of these two branches, and thus must have followed a different evolutionary track, which might be related to rotation-powered X-ray pulsars (RPP) or high-field radio pulsars (this tentative track is indicated by the {\it dotted arrow}). Such a significantly lower initial dipole field results in a somewhat faster asymptotic spin period.}
\label{fig:conclusions}
\end{center}
\end{figure}

In this context, the dichotomy in behaviour between transient and
persistent SGRs/AXPs
which likely represent two distinctevolutionary sequences (see
Fig.~\ref{fig:conclusions}), 
could point to a different mechanism for generation of their strong fields at
formation. One might speculate, based on the apparent properties of
their magnetic fields, that SGR/AXPs appear to be born with somewhat higher initial
dipole fields ($B_{\rm dip,i} \sim 10^{15}\;$G) compared to transients
(which are narrowly clustered around $B_{\rm dip,i} \sim 2\times
10^{14}\;$G).  SGR/AXPs also have a strong initial internal chaotic
field ($B_{\rm int,i}\gtrsim 10^{16}\;$G) that varies significantly on
length scales much smaller than the NS radius, which naturally
accounts for their good heat conduction from the core to the surface
and for the more frequent and more powerful bursting
activity. Transients, on the other hand, likely have a strong and
globally ordered tangential field, which efficiently suppresses heat
conduction from the core to the surface and also results in much fewer
and smaller bursts (that are likely powered by small scale
field structures). Thus, one might speculate that SGR/AXPs might
attain their fields at birth through the alpha-omega dynamo, if they
were initially sufficiently rapidly rotating, while NSs in the
transient branch might obtain their fields through a different
(unclear as of yet) mechanism.

The $B_{\rm dip}\,$ vs. $\,P$ distribution of XDINs is also consistent
with the same dipole field decay of other classes. We conclude that
these sources represent $\sim (1-6)\times 10^5$ year-old NSs born with
$B_{\rm dip,i} \sim (0.03-2)\times 10^{15}\;$G (for $\alpha = 1.5$; see bottom
left panel of Fig.~\ref{fig:B-tau-fieldecay-models}).
Their initial dipole fields have significantly decayed, making them
apparently look older than they really are. Hence, their weak X-ray
emission is likely dominated by their remnant heat, although a minor
contribution from the decay of their dipole field cannot be ruled out,
in general. As opposed to other classes, they show no evidence for an
additional, stronger field component in their interior. As such, their
most likely connection is with other high-field NSs not showing any
peculiar magnetar-like activity. The brighter XDINs had initial
dipole fields comparable to those of SGR/AXPs or transients, but they
likely have a weaker internal field. The dimmest and fastest spinning
of all XDINs, RX~J0420, was born with the weakest dipole field,
$B_{\rm dip,i} \sim 4\times 10^{13}\;$G (for $\alpha = 1.5$).
This is well below that of SGR/AXPs, or even of transients, and actually 
suggests a possible evolutionary link with either the high B-field radio pulsars
or the Rotation-Powered X-ray Pulsars (RPPs), which would depend on the
strength (and, possibly, the geometry) of both the initial dipole field and the
initial internal field.
This tentative option\footnote{Data for the RPPs are taken from Tab. 3 in Mereghetti (2011).}is also sketched in Fig.~\ref{fig:conclusions}, alongside the
clearer evolutionary tracks of the SGR/AXP branch and the transient branch,
that have been discussed above.

How does the ``weak field magnetar" SGR 0418+5729 fit in the general  picture? 
This object was likely born with a dipole field $\sim (3 - 5)
\times 10^{14}$ G that has now decayed  to $(4 - 7)\times 10^{12}$ G,
and an internal field $\gtrsim 10^{16}$ G, which has
now decayed to $\sim(1 - 2)\times 10^{14}\;$G. 
Despite its current appearance as a transient source, SGR
0418+5729 has most likely followed the track of persistent SGRs/AXPs
in the $L_{\rm X}$ vs.  $\tau_{\rm c}$ plane. The absence of more
luminous sources at similar $\tau_{\rm c}$ implies that the emission
of persistent sources must drop significantly in the range $\tau_{\rm
c} \sim (10^5 - 10^7$) yrs. 
It is not clear whether the decrease in luminosity of the persistent
sources passes through the upper tail of XDINs (like the lower curves
in the right panel of Fig. \ref{fig:Lx-vs-Hall-internal} may suggest)
or completely avoids them (like most curves in the right panels of
Figs. \ref{fig:Lx-vs-tau-solenoidal} and
\ref{fig:Lx-vs-Hall-internal}). In the former case XDINs would be a
less homogeneous class of sources, which included both passively
cooling, high-field NSs and some old magnetars. This would also imply
that SGR 0418+5729 follows the brightest XDINs along the same cooling
sequence. Its quiescent luminosity should thus be orders of magnitude
lower than that of the brightest XDINs. In the latter case, the XDINs
are just passively cooling NSs, with no particular relation to
persistent SGRs/AXPs, accidentally occupying a region in parameter
space that is between persistent and transient magnetars.

Overall we have demonstrated that the dipole magnetic field of
magnetars must decay on a time scale of $(10^3/B^{\alpha}_{\rm
dip,15})$ years. The most likely decay law has $1.5 \lesssim \alpha
\lesssim 1.8$.  At the same time we have shown that the power supplied by the
decaying dipole field is not sufficient to power the X-ray luminosity
of these objects.  Unless there is an external energy reservoir, like
a fallback disk (cfr. \citealt{Trumpetal+10, Alpetal+11}), this energy can arise
only from a stronger internal field whose decay is decoupled from the decay of
the dipole\footnote{It could actually be possible that it directly 
affects the decay of the dipole.}. Detailed analyses of the X-ray
emission depends on the structure of this internal field, its decay
mode as well as cooling and heat transfer within the NS. Further
observations of this exciting group of object could reveal some of the
hidden secrets of the NS interiors as well as provide clues on
magnetic field generation and amplification at their birth.

\section*{Acknowledgements}

TP thanks N. Kylafys for illuminating discussions on the nature of
magnetars. This research was supported by and ERC advanced research grant and 
by the Israeli Center of excellence for High Energy Astrophysics.

\bibliography{magnetar}
\label{lastpage}
\end{document}